\documentstyle[multicol,pre,eqsecnum,aps]{revtex}
\input{psfig.sty}

\def\eps{\varepsilon}
\def\Dm{\widetilde{\cal D}_{\mu}}
\def\n{{\bf n}}
\def\k{{\bf k}}
\def\x{{\bf x}}
\def\q{{\bf q}}
\def\p{{\bf p}}
\def\bfv{{\bf v}}
\def\I{\widetilde{I}_{1}}
\def\partt{\partial}
\def\btheta{\theta}

\def\nueff{\nu_{\mbox{\small\it eff}}\,}
\def\nunon{\nu_{\mbox{\small\it non}}\,}
\def\nuloc{\nu_{\mbox{\small\it loc}}}

\begin{document}
\draft

 \title{Anomalous scaling, nonlocality and anisotropy
 in a model of the passively advected vector field}

 \author{L. Ts. Adzhemyan, N. V. Antonov and A. V. Runov}

 \address{Department of Theoretical Physics, St~Petersburg University,
 Uljanovskaja 1, St~Petersburg, Petrodvorez, 198504 Russia}

\maketitle

 \begin{abstract}
A model of the passive vector quantity advected by the Gaussian velocity
field with the covariance
$\propto\delta(t-t')|{\bf x}-{\bf x'}|^{\varepsilon}$ is studied;
the effects of pressure and large-scale anisotropy are discussed.
The inertial-range behavior of the pair correlation function is described
by an infinite family of scaling exponents, which satisfy exact
{\it transcendental} equations derived explicitly in $d$ dimensions by means
of the functional techniques. The exponents are organized in a hierarchical
order according to their degree of anisotropy, with the spectrum unbounded
from above and the leading (minimal) exponent coming from the isotropic
sector. This picture extends to higher-order correlation functions.
Like in the scalar model, the second-order structure function appears
nonanomalous and is described by the simple dimensional exponent:
$S_{2} \propto r^{2-\eps}$. For the higher-order structure functions,
$S_{2n} \propto r^{n(2-\eps)+\Delta_{n}}$, the anomalous scaling
behavior is established as a consequence of the existence in the
corresponding operator product expansions of ``dangerous'' composite
operators, whose {\it negative} critical dimensions determine the anomalous
exponents $\Delta_{n}<0$. A close formal resemblance of the model with the
stirred Navier--Stokes equation reveals itself in the {\it mixing} of
relevant operators and is the main motivation of the paper.
Using the renormalization group, the anomalous exponents are calculated
in the $O(\varepsilon)$ approximation, in large $d$ dimensions, for the
even structure functions up to the twelfth order.
 \end{abstract}


 \section{Introduction} \label{sec:Intro}

The investigation of intermittency and anomalous scaling in fully developed
turbulence remains essentially an open theoretical problem. Much effort has
been invested recently into the understanding of the inertial-range behavior
of the passive scalar. Both the real experiments and numerical simulations
suggest that the breakdown of the classical Kolmogorov--Obukhov theory
\cite{Legacy} is even more strongly pronounced for a passively advected
scalar field than for the turbulent velocity itself. On the other hand, the
problem of passive advection appears easier tractable theoretically;
see Ref.~\cite{Nature} and references therein.

The most progress has been achieved for the so-called rapid-change model of
the passive scalar advection by a self-similar white-in-time velocity field
\cite{Kraich1}. The model is interesting because of the insight it offers
into the origin of intermittency and anomalous scaling in turbulence: for
the first time, anomalous exponents have been calculated on the basis of a
microscopic model and within controlled approximations
\cite{Falk1,GK,Pumir,VMF}. Within the ``zero-mode approach'' to the
rapid-change model, proposed in Refs.~\cite{Falk1,GK,Pumir}, nontrivial
anomalous exponents are related to the zero modes (homogeneous solutions) of
the closed exact differential equations satisfied by the equal-time
correlations. In this sense, the model appears exactly solvable. A recent
review and more references can be found in Ref.~\cite{Nature}.

In Ref. \cite{RG} and subsequent papers \cite{RG1,RG2,AH,cube,Novikov,RG3},
the field theoretic renormalization group (RG) and the operator
product expansion (OPE) were applied to the model \cite{Kraich1,Falk1,GK}.
The feature specific to the theory of turbulence is the existence in the
corresponding field theoretical models of the composite operators with
{\it negative} scaling (critical) dimensions. Such operators, termed
``dangerous'' in \cite{RG,RG1,RG2,AH,cube,Novikov,RG3}, give rise
to anomalous scaling, i.e., the singular dependence on the infrared (IR)
scale with certain nonlinear anomalous exponents.

The OPE and the concept of dangerous operators for the Navier--Stokes
(NS) turbulence were introduced in Ref.~\cite{JETP}; detailed review and
bibliography can be found in \cite{UFN,turbo}. The relationship between
the anomalous exponents and dimensions of composite operators was
anticipated in Ref.~\cite{Eyink2} for the stochastic
hydrodynamics and in \cite{Falk1,GK,Eyink} for the Kraichnan model
within certain phenomenological formulation of the OPE, the so-called
``additive fusion rules,'' typical to the models with multifractal
behavior \cite{DL}. A similar picture naturally arises within the context
of the Burgers turbulence and growth phenomena \cite{Burg,Burg1}.

Important advantages of the RG approach are its universality and
calculational efficiency: a regular systematic perturbation expansion
for the anomalous exponents was constructed, similar to the well-known
$\eps$-expansion in the theory of phase transitions, and the exponents were
calculated in the second \cite{RG,RG1,RG2,AH} and third \cite{cube}
orders of that expansion. For passively advected vector fields,
any calculation of the exponents for higher-order correlations calls
for the RG techniques already in the $O(\eps)$ approximation
\cite{RG1,Lanotte2,amodel}. Furthermore, the
RG approach is not related to the aforementioned solvability of the
rapid-change model and can also be applied to the case of finite
correlation time or non-Gaussian advecting field \cite{RG3}.

Recent research on the Kraichnan model and its descendants has mostly
been concentrated on the passive scalar advection. The large-scale
transport of vector quantities exhibits more interesting
behavior; see monograph \cite{Legacy} and references therein.
In this paper, we study the anomalous scaling and effects of anisotropy
and pressure, for the passive {\it vector} field advected by the
rapid-change velocity field. The model has already been introduced and
discussed independently in Refs. \cite{LA} and~\cite{AP}.

Before explaining our motivations, which follow the same lines as
those of Refs. \cite{LA,AP}, we shall discuss the definition of the
model in detail.

We shall confine ourselves with the case of transverse (solenoidal)
passive $\btheta(x)\equiv \{\theta_{i} (t, {\bf x})\}$ and advecting
${\bf v}(x)\equiv \{v_{i}(t, {\bf x})\}$ vector fields and
the advection-diffusion equation of the form
\begin{eqnarray}
\nabla _t\theta_{i} + \partial_{i} {\cal P} =
\nu_0\Delta \theta_{i} + f_{i},
\qquad
\nabla_t \equiv \partial _t + (v_{j}\partial_{j}) ,
\label{1}
\end{eqnarray}
where ${\cal P}(x)$ is the pressure, $\nu_0$ is the diffusivity, $\Delta$
is the Laplace operator and $f_{i}(x)$ is a transverse Gaussian stirring
force with zero mean and covariance
\begin{equation}
\langle  f_{i}(x)  f_{j}(x')\rangle
= \delta(t-t')\, C_{ij}({\bf r}/L), \qquad
{\bf r}\equiv{\bf x}-{\bf x'}.
\label{2}
\end{equation}
The parameter $L$ is an integral scale related to the stirring, and $C_{ij}$
is a dimensionless function finite as $L\to\infty$. Its precise form is not
essential; for generality, it is not assumed to be isotropic. Therefore, the
force maintains the steady state and is also a source of the large-scale
anisotropy in the system.

The velocity ${\bf v}(x)$ obeys a Gaussian distribution with zero
mean and covariance
\begin{equation}
\langle v_{i}(x) v_{j}(x')\rangle = D_{0}\, \delta(t-t')
\int {\cal D}{\bf p}\, P_{ij}({\bf p})\, p^{-d-\eps}\,
\exp \big[{\rm i}({\bf pr})\big].
\label{3}
\end{equation}

Here and below  ${\bf p}$ is the momentum, $p\equiv |{\bf p}|$,
${\cal D}{\bf p} = d{\bf p}/(2\pi)^{d}$,
$P_{ij}({\bf p}) \equiv \delta _{ij} - p_i p_j / p^2$ is the transverse
projector, $D_{0}>0$ is an amplitude factor
and $d$ is the dimensionality of the ${\bf x}$ space. The exponent
$0<\eps<2$ plays in the RG approach the same role as the parameter
$\eps=4-d$ does in the RG theory of critical behavior.
The IR regularization is provided by
the cut-off in integral (\ref{3}) from below at
$p=m$, where $1/m$ is another integral scale;
the precise form of the cut-off is not essential.
In what follows, we shall not distinguish the two IR
scales, setting $m\sim1/L$. The relations
\begin{equation}
D_{0}/\nu_0 = g_{0} = \Lambda^{\eps}
\label{Lambda}
\end{equation}
define the coupling constant $g_{0}$ (i.e., the formal expansion parameter
in the ordinary perturbation theory) and the characteristic ultraviolet
(UV) momentum scale $\Lambda$.

Due to the transversality conditions,
$\partial_{i}\theta_{i} = \partial_{i}v_{i}=0$, the pressure can be
expressed as the solution of the Poisson equation,
\begin{equation}
\Delta{\cal P} = - \partial_{i} v_{j} \, \partial_{j} \theta_{i}.
\label{Poisson}
\end{equation}

The issue of interest is, in particular, the behavior of the
equal-time structure functions
\begin{equation}
S_{n} ({\bf r}) \equiv \big\langle
[ \theta_{r} (t,{\bf x}) - \theta_{r} (t,{\bf x'})]^{n} \big\rangle
\label{struc}
\end{equation}
in the inertial range, specified by the inequalities
$1/\Lambda \ll r\ll L \sim 1/m$.
Here $\theta_{r} \equiv \theta_{i} r_{i} /r$ is the component of the
passive field along the direction ${\bf r}={\bf x}-{\bf x'}$,
an analog of the streamwise component of the turbulent velocity field
in real experiments.

The general symmetry of the vector problem permits one to add
on the left-hand side of the advection-diffusion equation
the ``stretching term'' of the form
$(\theta_{j}\partial_{j}) v_{i}$. This general vector model is studied
in Ref. \cite{amodel}, and its special magnetic case, where
the pressure term disappears, was studied earlier in a number of papers
in detail; see Refs. \cite{Lanotte2,V96,RK97,Lanotte,Arad}.

From the physics viewpoints, the model (\ref{1})--(\ref{3})
can be considered as the linearized NS equation
with the prescribed statistics of the background field $\bfv$
and the additional convention that the field $\bfv$ is ``soft''
and the perturbation $\theta$ is ``hard,'' that is,
$\partial\theta\gg\partial\bfv$. Our motivation, however, is different:
a close formal resemblance of the model (\ref{1})--(\ref{3})
with the stirred NS turbulence.

Although the model (\ref{1}) is formally a special case of the general
``${\cal A}$-model'' \cite{amodel}, it appears in a sense exceptional and
requires special investigation. In this case, the stretching term is absent
and the analog of the kinetic energy $\theta^{2}(x)$ is conserved, as for
the passive scalar and stochastic NS equations. An important consequence
is the existence of a constant flux solution, characteristic of the
real NS turbulence \cite{Legacy}. The second-order structure function
appears nonanomalous and is described by the simple ``dimensional''
exponent, $S_{2} \propto r^{2-\eps}$.

Furthermore, since only {\it derivatives} of the field $\theta$ enter
into Eq. (\ref{1}), the latter possesses additional symmetry
$\theta\to\theta+{\rm const}$. The leading anomalous exponents
in the magnetic and general cases are determined by the composite operators
built of the field $\theta$ without derivatives \cite{Lanotte2,amodel};
in our case they become trivial and the leading nontrivial exponents are
related to the composite operators built solely of the {\it gradients}
of $\theta$.

A similar distinction exists also between the density and tracer scalar
fields advected by a compressible velocity, as discussed in Ref. \cite{RG1}
in detail. But in contrast with the scalar case, where the leading
exponent for a given-order structure function is determined by an individual
composite operator even in the presence of the $\theta\to\theta+{\rm const}$
invariance \cite{RG,RG1,RG2}, in the vector model (\ref{1}) the
inertial-range behavior of any structure function is determined by a
{\it family}
of composite operators with the same symmetry and dimension, and in order
to find the corresponding {\it set} of exponents and to identify the leading
contribution, one has to consider the renormalization of the whole family,
which implies the {\it mixing} of individual operators.

In the scalar case, the anomalous exponents for all structure functions
are given by a single expression which includes $n$, the order of a
function, as a parameter \cite{Falk1,GK,RG}. This remains true for the
vector models with the stretching term \cite{Lanotte2,amodel}. In the
special vector model (\ref{1}), the number and the form of the operators
entering into the relevant family depend essentially on $n$, and different
structure functions should be studied separately. As a result, no general
expression valid for all $n$ exists in the model, in contrast with
the scalar \cite{Falk1,GK}, magnetic \cite{Lanotte2} and general
vector \cite{amodel} models.

In this respect, the model (\ref{1}) is one step closer to the nonlinear
NS equation, where the inertial-range behavior of structure
functions is believed to be related with the Galilean-invariant (and hence
built of the velocity gradients) operators, which mix heavily in
renormalization; see \cite{UFN,turbo} and references therein.

Another important question recently addressed is the effects of
large-scale anisotropy on the inertial-range statistics;
see, e.g., Ref. \cite{Nature} and references therein.
In particular, it was shown that in the presence of anisotropic
forcing, the exponents describing the inertial-range scaling
of the passively advected scalar \cite{RG3,CLMV} and vector
\cite{Lanotte2,Lanotte,Arad} fields and the turbulent velocity field itself
\cite{Arad99,Arad991,LP1,KS} are organized in a hierarchical order according
to their degree of anisotropy, with the leading contribution
coming from the isotropic sector. The consistency of this picture
with the presence of nonlocal terms in the  equations for the
correlation functions, caused by the pressure contributions, has been
addressed recently \cite{WL} and answered positively in Ref. \cite{AP}
on the example of the model (\ref{1})--(\ref{3}), for the pair correlation
function in three dimensions, and in Ref. \cite{amodel} on the example of
the general ${\cal A}$-model, for the correlation functions of arbitrary
order in $d$ dimensions.

The plan of the paper and the main results are the following.
In Sec. \ref{sec:QFT}, we give the field theoretic formulation of the model,
its diagrammatic techniques, and derive exact equations for the response
function and pair correlation function: the so-called Dyson--Wyld equations.

In Sec. \ref{sec:PAIR}, we study the inertial-range behavior of
the pair correlation function in the presence of the large-scale
anisotropic forcing. This issue for the model (\ref{1})--(\ref{3})
was already discussed in Ref. \cite{AP}, where the numerical solutions
for the scaling exponents were presented in three dimensions for the
isotropic sector and low-order anisotropic sectors.

In this paper, starting from the Dyson--Wyld equations, we give the
general recipe of deriving nonperturbative exact equations and obtain
explicitly transcendental equations for the scaling exponents, related
to different irreducible representations, in $d$ dimensions. This allows
us to give general description of the behavior of the full set of
solutions in isotropic and anisotropic sectors, and to derive analytical
results in all sectors to order $O(\eps)$ in $d$ dimensions. Some details
are given in Appendix~A. These results are illustrated by a few
nonperturbative solutions obtained numerically in two and three dimensions
for the isotropic and low-order anisotropic sectors; later, in
Secs.~\ref{sec:Ashells} and~\ref{sec:hi-fi} they are confirmed using
the RG and OPE techniques and extended to higher-order structure functions.

In Sec. \ref{sec:RGE}, we perform the ultraviolet (UV) renormalization of
the model and derive the corresponding RG equations with exactly known RG
functions ($\beta$ functions and anomalous dimensions of the basis fields
and parameters). For $d^{2}>3$, these equations possess an IR stable fixed
point, which establishes the existence of IR scaling with exactly known
scaling dimensions of the basis fields and parameters of the model.

In Sec. \ref{sec:OPE}, we discuss the operator product expansion and
its relationship to the issue of anomalous scaling. We show that
nontrivial exponents describing the inertial-range behavior of the
$2n\,$th order even structure function are related to critical dimensions
of scalar composite operators built of $2n$ derivatives of the advected
field. Explicit calculation of the dimensions of relevant operators is given
in Sec.~\ref{sec:Operators}. The case $n=1$ can be treated exactly using
certain functional Schwinger equation
(in the case at hand, it has the meaning of the energy balance equation).
Like in the scalar case \cite{Kraich1}, the function $S_{2} \propto
r^{2-\eps}$ appears nonanomalous with the simple dimensional exponent;
Sec.~\ref{sec:Operators2}. The case $n=2$ is studied in detail;
Sec.~\ref{sec:Operators4}. The critical dimensions of the relevant family
of operators are calculated to order $O(\eps)$ in $d$ dimensions; they
include a negative dimension, and the function $S_{4}$ shows anomalous
scaling. The families of the anomalous exponents related to the higher-order
functions $S_{2n}$ are calculated in Sec.~\ref{sec:Operators6}, in the
limit of large $d$, for $n$ as high as 6; some technical details are
given in Appendix~B.

Generalization to the case of anisotropic sectors is discussed in
Secs.~\ref{sec:Ashells} and~\ref{sec:hi-fi}. There, the RG and OPE
techniques confirm the general picture established earlier for the
pair correlation function (infinite sets of exponents, hierarchy, absence
of saturation) and extend it to the case of the higher-order
structure functions.

The results obtained are reviewed and discussed in the Conclusion,
where the lessons one can learn regarding the stirred NS equation and
possible generalization to this nonlinear problem are also briefly outlined.

\section{Field theoretic formulation and the Dyson--Wyld equations}
\label {sec:QFT}

The stochastic problem (\ref{1})--(\ref{3}) is equivalent to the field
theoretic model of the extended set of three fields
$\Phi\equiv\{\btheta', \btheta, {\bf v}\}$ with action functional
\begin{equation}
{\cal S}(\Phi)=\btheta' D_{\theta}\btheta' /2+
\btheta' \left[ - \nabla_{t} + \nu _0\Delta \right] \btheta
-{\bf v} D_{v}^{-1} {\bf v}/2.
\label{action}
\end{equation}
The first three terms in  Eq. (\ref{action}) represent the
Martin--Siggia--Rose-type action for the stochastic problem (\ref{1}),
(\ref{2}) at fixed ${\bf v}$ (see, e.g., \cite{UFN,turbo} and references
therein), while the last term represents the Gaussian averaging over
${\bf v}$. Here $D_{\theta}$ and $D_{v}$ are the correlation functions
(\ref{2}) and (\ref{3}), respectively, $\btheta'$ is an auxiliary transverse
vector field, the required integrations over $x=(t,{\bf x})$ and summations
over the vector indices are implied, for example,
\[ \btheta' \partial_{t} \btheta \equiv \int dt\, d{\bf x}\, \theta'_{i}
(t,{\bf x})\, \partial_{t}\, \theta_{i} (t,{\bf x}). \]
The pressure term can be omitted in the functional (\ref{action})
owing to the transversality of the auxiliary field:
\[ \int dx \theta_{i}' \partial_{i} {\cal P} =
- \int dx {\cal P} \partial_{i} \theta_{i}' =0 .\]
Of course, this does not mean that the pressure contribution can simply be
neglected: the field $\btheta'$ acts as the transverse projector and
selects the transverse part of the expression in the square brackets in
Eq.~(\ref{action}).

The formulation (\ref{action}) means that statistical averages
of random quantities in stochastic problem (\ref{1})--(\ref{3}) can be
represented as functional averages with the weight $\exp{\cal S}(\Phi)$,
so that the generating functionals of total [$G(A)$] and connected
[$W(A)$] Green functions of the problem are given by the functional integral
\begin{equation}
G(A)=\exp  W(A)=\int {\cal D}\Phi \exp [{\cal S}(\Phi)+A\Phi ]
\label{field}
\end{equation}
with arbitrary sources $A\equiv A^{\theta'},A^{\theta},A^{\bfv}$
in the linear form
\[A\Phi \equiv \int dx\,
[A^{\theta'}_{i}(x)\theta'_{i} (x)+A^{\theta }_{i}(x)
\theta_{i} (x) + A^{\bfv}_{i}(x)v_{i}(x)].\]

The model (\ref{action}) corresponds to a standard Feynman
diagrammatic technique with the triple vertex
$ -\btheta' ({\bf v}\partt) \btheta \equiv
\theta_{i}'\theta_{j} v_{k} V_{ijk}$
with vertex factor
\begin{equation}
V_{ijk}({\bf p})=  {\rm i}\, \delta_{ij}\, p_{k},
\label{vertex}
\end{equation}
where ${\bf p}$ is the momentum flowing into the vertex via the
field $\theta'$. The bare propagators in the frequency--momentum
($\omega$--$\p$) representation have the forms
\begin{mathletters}
\label{lines}
\begin{equation}
\bigl\langle \theta_{i} (\omega,\p) \theta'_{j}(-\omega,-\p) \bigr\rangle _0 =
\bigl\langle
\theta'_{i}(\omega,\p) \theta_{j}(-\omega,-\p) \bigr\rangle _0^*=
\frac{P_{ij}(\p)} {(-{\rm i}\omega +\nu _0 p^2)} ,
\label{lines1}
\end{equation}
\begin{equation}
\bigl\langle \theta_{i}(\omega,\p) \theta_{j}(-\omega,-\p)\bigr\rangle _0
= \frac {C_{ij}(\p)} {(\omega^{2}+\nu _0 ^2 p^{4})},
\label{lines2}
\end{equation}
\begin{equation}
\bigl\langle \theta_{i}'(\omega,\p) \theta_{j}'(-\omega,-\p) \bigr\rangle _0 = 0,
\label{lines3}
\end{equation}
\end{mathletters}
where $C_{ij}(\p)$ is the Fourier transform of the function
$C_{ij}({\bf r})$ from (\ref{2}); the bare propagator
$\langle v_{i} v_{j}  \rangle _0$ is given by Eq.~(\ref{3}).

The action functional (\ref{action}) is invariant with respect to the
dilatation $\theta\to\lambda\theta$, $\theta'\to\theta'/\lambda$,
$C\to\lambda^{2} C$, where $C$ is the correlation function (\ref{2}).
It then follows
that any total or connected Green function with $n$ fields $\theta$ and
$p$ fields $\theta'$ is proportional to $C^{(n-p)/2}$. Since the
function $C$ appears in the bare propagator (\ref{lines2}) only in the
numerators, the difference $n-p$ is an even non-negative integer for any
nonvanishing function; the Green functions with $n-p<0$ vanish
identically. On the contrary, the 1-irreducible
function $\langle\theta(x_{1})\cdots\theta(x_{n})\, \theta'(y_{1})
\cdots\theta'(y_{p})\rangle_{\rm 1-ir}$ contains the factor
$C^{(p-n)/2}$ and therefore vanishes for $n-p>0$; this
fact will be relevant in the analysis of the renormalizability of
the model (see Sec.~\ref{sec:RGE}).

The pair correlation functions $\langle\Phi\Phi\rangle$ of the
multicomponent field $\Phi$ satisfy standard Dyson equation,
which in the component notation reduces to the system of two
nontrivial equations for the exact correlation function
${\cal D} _{ij}(\omega,\p)=\langle \theta_{i}(\omega,\p)
\theta_{j}(-\omega,-\p)  \rangle$ and the exact response function
${\cal G}_{ij}(\omega,\p)=\langle \theta_{i}(\omega,\p)
\theta'_{j}(-\omega,-\p) \rangle$. We shall see below
that the latter function does not include the correlation function
(\ref{2}), therefore it is isotropic and can be written as
${\cal G}_{ij}(\omega,\p)= P_{ij}(\p) {\cal G}(\omega,p)$
with certain isotropic scalar function ${\cal G}(\omega,p)$.
Thus the component equations, usually referred to as the Dyson--Wyld
equations, in our model take on the form (cf. Refs. \cite{RG1} for
the scalar and \cite{Lanotte2} for the magnetic models)
\begin{mathletters}
\label{Dyson}
\begin{equation}
{\cal G}^{-1}(\omega, p) P_{ij}(\p)
= \bigl[-{\rm i}\omega +\nu_0 p^{2}\bigr] P_{ij}(\p) -
\Sigma_{ij}^{\theta'\theta} (\omega, \p),
\label{Dyson1}
\end{equation}
\begin{equation}
{\cal D}_{ij}(\omega, \p)= |{\cal G}(\omega, p)|^{2}\, \Bigl[
C_{ij} (\p) +\Sigma _{ij}^{\theta'\theta'} (\omega, \p)\Bigr],
\label{Dyson2}
\end{equation}
\end{mathletters}
where $\Sigma^{\theta'\theta}$ and $\Sigma^{\theta'\theta'}$ are self-energy
operators represented by the corresponding 1-irreducible diagrams;
the other functions $\Sigma^{\Phi\Phi}$ vanish identically. It is
also convenient to contract Eq. (\ref{Dyson1}) with the projector
$P_{ij}(\p)$ in order to obtain the scalar equation:
\begin{mathletters}
\label{Dyson'}
\begin{equation}
{\cal G}^{-1}(\omega, p) = -{\rm i}\omega +\nu_0 p^{2} -
\Sigma^{\theta'\theta} (\omega, p),
\label{Dyson3}
\end{equation}
where we have written
\begin{equation}
\Sigma^{\theta'\theta} (\omega, p) \equiv
\Sigma_{ij}^{\theta'\theta} (\omega, \p) P_{ij}(\p) / (d-1).
\label{Dyson4}
\end{equation}
\end{mathletters}

The feature characteristic of the rapid-change models like
(\ref{action}) is that all the skeleton multiloop diagrams entering
into the self-energy operators $\Sigma^{\theta'\theta}$ and
$\Sigma^{\theta'\theta'}$ contain effectively closed circuits
of retarded propagators $ \langle\theta\theta'\rangle_{0}$ and
therefore vanish; it is crucial here that the velocity propagator
(\ref{3}) contains the $\delta$ function in time and the bare
propagator (\ref{lines3}) vanishes. Therefore the self-energy
operators in (\ref{Dyson}) are given by the
one-loop approximations exactly and have the forms
\begin{mathletters}
\label{Dyson9}
\begin{equation}
\Sigma^{\theta'\theta}=
\raisebox{-2.6ex}{\psfig{file=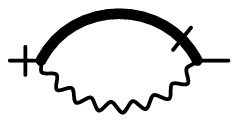,width=5em}}\,\, ,
\label{Dyson7}
\end{equation}
\begin{equation}
\Sigma^{\theta'\theta'}=
\raisebox{-2.6ex}{\psfig{file=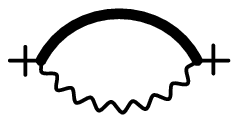,width=5em}}\,\, .
\label{Dyson8}
\end{equation}
\end{mathletters}
The thick solid lines in the diagrams denote the {\it exact\,} propagators
$\langle\theta\theta'\rangle$ and $\langle\theta\theta\rangle$;
the ends with a slash correspond to the field $\theta'$, and
the ends without a slash correspond
to $\theta$; the wavy lines denote the velocity propagator
(\ref{3}); the vertices correspond to the factor (\ref{vertex}).
The first equation does not include the correlation function (\ref{2}),
which justifies the isotropic form of the function ${\cal G}_{ij}$.
The analytic expressions for the diagrams in Eq. (\ref{Dyson9}) have
the forms
\begin{mathletters}
\label{sigma}
\begin{eqnarray}
\Sigma^{\theta'\theta} (\omega, p) = \frac{P_{ij}(\p)}{(d-1)}
\int {\cal D}\omega' \int {\cal D}{\bf k} \,
V_{ii_{3}i_{1}} (\p)\, P_{i_{3}i_{4}}(\p-\k) {\cal G}(\omega',\p-\k) \,
\frac{D_{0}\,P_{i_{1}i_{2}}(\k)} {k^{d+\eps}}\, V_{i_{4}ji_{2}} (\p) ,
\label{sigma1}
\end{eqnarray}
\begin{eqnarray}
\Sigma^{\theta'\theta'}_{ij} (\omega, \p)=
\int {\cal D}\omega' \int {\cal D}{\bf k} \,
V_{ii_{3}i_{1}} (\p )\,
{\cal D}_{i_{3}i_{4}} (\omega',\p-\k) \,
\frac{D_{0}\,P_{i_{1}i_{2}}(\k)} {k^{d+\eps}}\,
V_{ji_{4}i_{2}} (-\p ) .
\label{sigma2}
\end{eqnarray}
\end{mathletters}
Here we have denoted ${\cal D} \omega'\equiv d \omega'/(2\pi)$, used the
explicit form (\ref{3}) of the velocity covariance and the relation
$ P_{i_{1}i_{2}}(\k)\, V_{i_{4}ji_{2}} (\p-\k) =
P_{i_{1}i_{2}}(\k)\, V_{i_{4}ji_{2}} (\p)$
for the vertex factor in Eq. (\ref{vertex}). We also recall that the
integrations over $\k$ should be cut off from below at $k=m$.

The integrations with respect to $\omega'$ on the right-hand sides of
Eqs. (\ref{sigma}) give the equal-time response function
${\cal G}(k)=\int {\cal D} \omega' {\cal G}(\omega',k)$
and the equal-time pair correlation function ${\cal D}_{ij}(\k)=
\int {\cal D} \omega' {\cal D}_{ij}(\omega',\k)$;
note that both the self-energy operators appear independent of
$\omega$. The only contribution to ${\cal G}$ comes from the bare
propagator (\ref{lines1}), which in the $t$ representation is
discontinuous at coincident times. Since the correlation function
(\ref{3}), which enters into the one-loop diagram
for $\Sigma ^{\theta'\theta}$, is symmetric in $t$ and $t'$, the response
function must be defined at $t=t'$ by half the sum of the limits,
which is equivalent to the convention $ {\cal G}(k)=\int {\cal D} \omega'
(-i\omega'+\nu _0 k^2)^{-1}=1/2 $. This allows one to write
the equation (\ref{Dyson3}) in the form
\begin{equation}
{\cal G}^{-1}(\omega, p) = [-{\rm i}\omega + \nueff(p) p^{2}] ,
\label{Dyson101}
\end{equation}
where the $p$-dependent effective ``eddy diffusivity'' is given by
\begin{equation}
2 p^{2} \nueff(p) \equiv  2\nu_0 p^{2} + D_{0} \,\int
\frac{{\cal D}{\bf k}}{k^{d+\varepsilon}}\,
\left[ 1- \frac{(\p\k)^2}{p^2k^2} \right]
\left[  p^{2} - \frac { p^{2}k^{2} - (\p\k)^{2} }
{ (d-1) |\p-\k|^{2} } \right].
\label{Dyson102}
\end{equation}

It follows from Eq. (\ref{Dyson102}) that the eddy diffusivity can be written
as the sum of two parts: $\nueff(p) =\nuloc + \nunon(p)$, where the
local part is $p$-independent and coincides with the expression for the
effective diffusivity known in the scalar and magnetic cases, while the
nonlocal contribution has a finite limit at $m=0$ but retains a nontrivial
dependence on the momentum:
\begin{mathletters}
\label{Dyson103}
\begin{eqnarray}
\nuloc=\nu_0+\frac{D_{0}}{2}\int \frac{{\cal D}{\bf k}}{k^{d+\varepsilon}}\,
\left[1-\frac{(\p\k)^2}{p^2k^2}\right]=\nu_0+D_{0}\,C_{d}\,m^{-\eps}
\frac{(d-1)} {2d\eps},
\label{Dyson103A}
\end{eqnarray}
\begin{eqnarray}
\nunon(p)=\frac{D_{0}}{2}\int \frac{{\cal D}{\bf k}}{k^{d+\varepsilon}}\,
\left[1-\frac{(\p\k)^2}{p^2k^2}\right]\frac{(\p\k)^2-p^{2}k^{2}}{(d-1)p^{2}
|\p-\k|^{2}}.
\label{Dyson103B}
\end{eqnarray}
\end{mathletters}
Here and below $C_{d} \equiv  {S_{d}}/{(2\pi)^{d}}$ and
$S_d\equiv 2\pi ^{d/2}/\Gamma (d/2)$ is the surface area of the
unit sphere in $d$-dimensional space and $\Gamma(z)$ is the Euler
Gamma function. The parameter $m$ in
$\nuloc$ has arisen from the lower limit in the integral over $\k$.
For $m=0$, equation (\ref{Dyson103B}) gives
\begin{eqnarray}
\nunon(p)= - D_{0}p^{-\eps} J / (4\pi)^{d/2} ,
\label{J2}
\end{eqnarray}
where
\begin{eqnarray}
J= \frac{(d+1)\,\Gamma\left(\displaystyle\frac{\eps}{2}\right)
\Gamma\left(1-\displaystyle\frac{\eps}{2}\right)
\Gamma\left(1+\displaystyle\frac{d}{2}\right)}
{8\,\Gamma\left(1+\displaystyle\frac{d+\eps}{2}\right)
\Gamma\left(2+\displaystyle\frac{d-\eps}{2}\right)} ,
\label{J}
\end{eqnarray}
while for $p=0$ one obtains
\begin{eqnarray}
\nunon= -C_{d}D_{0} \,m^{-\eps} \frac{(d+1)}{2d(d+2)\eps}, \qquad
\nueff= \nu_0 + C_{d}D_{0} \,m^{-\eps} \frac {(d^{2}-3)} {2d(d+2)\eps}.
\label{nonlocal}
\end{eqnarray}

Equations (\ref{Dyson101})--(\ref{nonlocal}) give the explicit exact
expression for the response function in our model; it will be used in Sec.
\ref{sec:RGE} for the exact calculation of the RG functions. The integration
of Eq. (\ref{Dyson2}) over the frequency $\omega$ gives a closed equation for
the equal-time correlation function; it is important here that the $\omega$
dependence of the right-hand side is contained only in the prefactor
$|{\cal G}(\omega,p)|^{2}$:
\begin{eqnarray}
2 \nueff(p) p^{2} {\cal D}_{ij} (\p) = C_{ij} (\p) +
\Sigma ^{\theta'\theta'}_{ij} (\p).
\label{9}
\end{eqnarray}
Using Eqs. (\ref{sigma2}), (\ref{Dyson102}) and (\ref{Dyson103}),
the equation for ${\cal D}_{ij}$ can be rewritten in the form
\begin{eqnarray}
2\nu_0p^{2}\, {\cal D}_{ij}(\p)=C_{ij} (\p)
+p^{2}\,D_{0}\, \int \frac{{\cal D}{\bf k}}
{k^{{d+\eps}}} \left[ 1- \frac{(\p\k)^2}{p^2k^2} \right]
\biggl\{ P_{ii_{1}}(\p) {\cal D} _{ii_{2}}(\p-\k) P_{i_{2}j}(\p)
- {\cal D} _{ij}(\p) \biggr\} -2 \nunon (p)p^{2}\, {\cal D}_{ij} (\p).
\nonumber
\end{eqnarray}
\vskip -0.5cm
\begin{eqnarray}
{}
\label{10}
\end{eqnarray}

The subtracted term in the curly brackets is the contribution of
$\nuloc$ in Eq. (\ref{9}) written in its integral form (\ref{Dyson103A}).
For $0< \eps <2$, the IR cutoff in Eq. (\ref{10}) can be removed.
Indeed, owing to the subtraction, the integral over ${\bf k}$ is
finite for $m\to0$: the possible IR
divergence at ${\bf k}=0$ is suppressed by the expression in the
curly brackets. In what follows we set $m=0$ in Eq. (\ref{10}).

It is instructive to compare expression (\ref{nonlocal}) for the effective
diffusivity in our model with its analogs for the scalar \cite{Falk1} and
magnetic \cite{Lanotte2} cases. For those, the nonlocal part of the effective
diffusivity vanishes identically, while its local part coincides with Eq.
(\ref{Dyson103A}). Therefore the ratio $\nuloc/\nueff=(d-1)(d+2)/(d^{2}-3)$
[we have put $p=0$ and neglected the bare diffusivity $\nu_0$] can be
considered as a measure of the nonlocality contribution into the turbulent
diffusion. It tends to unity as $d\to\infty$, increases monotonically as
$d$ decreases and diverges at $d^{2}=3$. This means that the nonlocality
contribution is negligible at large $d$ (see also Ref. \cite{amodel} for
the general vector model), becomes comparable with the local contribution
as $d$ is reduced, and dominates the diffusion in low dimensions (in
particular, $\nuloc/\nueff=5/3$ and 4 for $d=3$ and 2, respectively).

We also notice that, according to Eq. (\ref{nonlocal}), the effective
diffusivity $\nueff(p)$ becomes {\it negative} for small $\p\ll m$ and
$d^{2}<3$. Therefore the response function in the time-momentum
representation,
\[ {\cal G}_{ij}(t,\p)=\Theta(t)\,P_{ij}(\p)\,
{\rm exp}\bigl\{-\nueff(p)p^{2}t\bigr\}, \]
where $\Theta(t)$ is the step function,
grows with time for small $\p$, thus signalling that the steady-state
solution cannot be stable for $d^{2}<3$: any small perturbation would lead
to the exponential in-time growth of the mean field $\langle\theta\rangle$.
Indeed, one can easily see that Eq. (\ref{9}) for the correlation
function has no solutions at small $\p$ and $d^{2}<3$: its left-hand side
is negative, while the right-hand side is strictly positive for all $d$.
We shall see below in Sec.~\ref{sec:Iso} that the inertial-range solutions
of Eq. (\ref{9}) also become singular and disappear in the limit $d^{2}\to3$.

Although the instability occurs for unphysical value of $d$, it deserves
some attention as a result of the competition between the local and nonlocal
contributions: from Eqs. (\ref{Dyson103}) it follows that $\nuloc$ is
strictly {\it positive} for all $d$, while $\nunon$ is strictly
{\it negative}. In a few respects, such an instability differs from that
established in Ref. \cite{V96} for the magnetic (local) case, in three
dimensions, where $\nunon=0$, the effective diffusivity coincides with its
local part (\ref{Dyson103A}) and is always positive. Thus the response
function, and hence the mean $\langle\theta\rangle$, show no hint of
misbehaviour at the threshold of the instability; the latter reveals itself
on the level of the pair correlation function and can be related to the
complexification of the inertial-range exponents;
see Refs.~\cite{V96,RK97,Lanotte,Arad}.

\section{Inertial-range behavior and scaling exponents for the pair
correlation function in the presence of large-scale anisotropy}
\label {sec:PAIR}

It is well known \cite{Falk1,GK,Pumir} that nontrivial
inertial-range exponents are determined by zero modes, i.e., the solutions
of Eq. (\ref{10}) neglecting both the forcing [$C({\bf r} )=0$] and the
dissipation ($\nu_0=0$). Whatever be the forcing, equations for zero modes
are linear and $SO(d)$ covariant, and their solution can be sought in the
form of decomposition in irreducible representations of the rotation group.
Equation (\ref{10}) then falls into independent equations for the
coefficient functions. In three dimensions, such decompositions were
used in Refs.~\cite{AP,Arad}.

Below we use more elementary derivation, which allows one to obtain
explicitly transcendental equations for the scaling exponents, related
to different irreducible representations, in $d$ dimensions. We
restrict ourselves with the case of {\it uniaxial} anisotropy, specified
by an unit vector ${\bf n}$, which is sufficient to find {\it all}
independent exponents, and use explicit expressions for the
basis functions in the {\it momentum representation}. Then the
transversality condition can be easily imposed from the very
beginning, and there is no need to check it {\it a posteriori}.

In contrast with the real-space Legendre decomposition, employed
in Refs. \cite{Lanotte2,Lanotte}, our representation is consistent
with the rotational symmetry: it can be considered as the projection of
the complete $SO(d)$ decomposition onto the subspace of the functions
with uniaxial symmetry. Therefore, no additional assumptions,
like the hierarchy of exponents, are needed to disentangle the
equations for different anisotropic sectors.

We start with the isotropic case and then discuss the general situation.

\subsection{Solution in the isotropic shell} \label {sec:Iso}

In the isotropic case, the inertial-range solution is sought in the form
\begin{equation}
{\cal D}_{ij}(\p) = A\, P_{ij} (\p)\, p^{-d-\gamma}.
\label{isol}
\end{equation}
The zero-mode analog of Eq. (\ref{10}) takes on the form
\begin{equation}
D_{0} \int \frac{{\cal D} \k}{k^{d+\varepsilon}}
\left[ 1- \frac{(\p\k)^2}{p^2k^2} \right] \,
\left\{\left[ \frac{1}{|\p-\k|^{d+\gamma}} - \frac{1}{p^{d+\gamma}}
\right] -\frac{k^{2}\left[ 1- {(\p\k)^2}/{p^2k^2} \right] }
{(d-1)|\p-\k|^{d+\gamma+2} } \right\} = 2 \nunon (p)\, p^{-d-\gamma} .
\label{isol2}
\end{equation}

In the following, we shall need the standard reference integrals
\begin{mathletters}
\label{convolu}
\begin{eqnarray}
\int {\cal D} \k
\frac{\left[ 1- {(\p\k)^2}/{p^2k^2} \right]^{n}}
{k^{d+\alpha} |\p-\k|^{d+\beta}} \equiv
I_{n}(\alpha,\beta) \, \frac
{p^{-d-\alpha-\beta}}{(4\pi)^{d/2}},
\label{convolu1}
\end{eqnarray}
where
\begin{eqnarray}
I_{n}(\alpha,\beta) =
\frac {\Gamma\left(n+\displaystyle\frac{d-1}{2}\right)
\Gamma\left(-\displaystyle\frac{\alpha}{2}\right)
\Gamma\left(n-\displaystyle\frac{\beta}{2}\right)
\Gamma\left(\displaystyle\frac{\alpha+\beta+d}{2}\right) }
{\Gamma\left(\displaystyle\frac{d-1}{2}\right)
\Gamma\left(n+\displaystyle\frac{\alpha+d}{2}\right)
\Gamma\left(\displaystyle\frac{\beta+d}{2}\right)
\Gamma\left(n-\displaystyle\frac{\alpha+\beta}{2}\right) } \, .
\label{convolu2}
\end{eqnarray}
\end{mathletters}

The integral (\ref{convolu1}) is finite in the region of parameters
specified by the inequalities $\alpha<0$ (convergence at $\k\to0$),
$\beta<2n$ (convergence at $\k-\p\to0$, improved by the factor
$[ 1- {(\p\k)^2}/{p^2k^2} ]^{n}\propto |\k-\p|^{2n} $) and
$\alpha+\beta >-d$ (convergence at $\k\to\infty$). However,
expression (\ref{convolu2}) is meaningful in a wider range of parameters
and, in the spirit of analytical and dimensional regularizations
\cite{Collins,book3}, it can be considered as the analytical continuation
of the integral (\ref{convolu1}) from the region in which it converges.
The precise meaning of such a continuation is that Eq. (\ref{convolu})
gives the correct value of the integral with proper subtractions which
ensure its convergence.
For example, if the factor  $1/|\p-\k|^{d+\beta}$ is replaced with the
difference $1/|\p-\k|^{d+\beta}-1/p^{d+\beta}$ (that is, the zeroth
term of the Taylor expansion in $\k$ is subtracted), the integral
becomes convergent for $\alpha<2$ and expression (\ref{convolu})
gives the correct answer for this ``improved'' integral.

One can easily see that Eq. (\ref{isol2}) involves such a subtraction, which
improves its convergence at small $\k$. Therefore, one can use the formal
expression (\ref{convolu}) in (\ref{isol2}) and simultaneously omit the
subtracted term. Then Eq. (\ref{isol2}) takes on the form
\begin{equation}
I_{1}(\eps,\gamma)-I_{2}(\eps-2,\gamma+2)/(d-1)=-2 J,
\label{isol3}
\end{equation}
with $J$ from Eq.~(\ref{J}). We shall see below that the {\it leading
admissible} solution of this equation is $\gamma=2-\eps$, so that the
integrals entering into Eq. (\ref{isol2}) are convergent for all $0<\eps<2$
and the above procedure is internally consistent.
In Eq. (\ref{isol3}), we omit the overall nonvanishing factor
\[ \frac {\Gamma \left( 1 -\displaystyle\frac{\eps}{2} \right)
 \Gamma \left( 1 -\displaystyle\frac{\gamma}{2} \right)
 \Gamma \left( \displaystyle\frac{\gamma+d+\eps}{2} \right) }
{\eps(d+1)\,\Gamma \left( 1 +\displaystyle\frac{d+\eps}{2} \right)
 \Gamma \left( 1 +\displaystyle\frac{d+\gamma}{2} \right)
 \Gamma \left(2- \displaystyle\frac{\gamma+\eps}{2} \right) } \]
and obtain the desired equation for the
zero-mode exponents in the isotropic case:
\begin{equation}
(d-1)(d+\gamma)(2-\gamma-\eps)/(d+1)+\eps = 2 \, \frac {
\Gamma \left( 1 +\displaystyle\frac{d}{2} \right)
\Gamma \left( 1 +\displaystyle\frac{\eps}{2} \right)
\Gamma \left( 1 +\displaystyle\frac{d+\gamma}{2} \right)
\Gamma \left( 2 -\displaystyle\frac{\eps+\gamma}{2} \right)}
{\Gamma \left( 2 +\displaystyle\frac{d-\eps}{2} \right)
\Gamma \left( 1 -\displaystyle\frac{\gamma}{2} \right)
\Gamma \left( \displaystyle\frac{d+\eps+\gamma}{2} \right)}.
\label{isol4}
\end{equation}

The transcendental equation (\ref{isol4}) has infinitely many solutions.
This means that the inertial-range behavior of the correlation function is
given by an infinite sum of powerlike contributions of the form
(\ref{isol}); the leading term is given by the minimal exponent $\gamma$.

Some solutions can be ruled out as not admissible \cite{Falk1,GK};
admissible solutions are non-negative for $\eps=0$ (see, e.g.,
\cite{Lanotte2,Arad}). The remaining solutions, all having the forms
$\gamma=-d-2k+O(\eps)$ with non-negative integer $k$, are also meaningful
and describe the behavior of the correlation function at large scales
$r\gg L$ \cite{Falk1}; we shall not discuss them in the following.

It then follows from (\ref{isol4}) that the leading admissible inertial-range
solution is $\gamma=2-\eps$ (no corrections of order $\eps^{2}$ and higher).
In the coordinate representation, this corresponds to
$S_{2}\propto r^{2-\eps}$, that is, the second-order structure function
is nonanomalous like in the scalar model \cite{Kraich1}.

For small $\eps$, all the subleading exponents can be written in the form
\begin{equation}
\gamma_{k}=2k - \eps\,  \frac {(d-1)(d+2)}{(d^{2}-3)} + \eps^{2}\,
\frac {(d+1)(d+2)}{2(d^{2}-3)^{2}} \left\{(d-1)\, {\cal K} -
\frac{(d+1)} {(k-1)(d+2k)} \right\} +O(\eps^{3})
\label{isol5}
\end{equation}
with $k=2,3,4$ and so on. The functions $\psi(z) = d \ln \Gamma(z) /dz$ can
be eliminated from the coefficient
\[ {\cal K}= \psi(k+d/2)-\psi(2+d/2)+\psi(k)-\psi(1) \]
using the relation $\psi(z+1) = \psi(z) + 1/z$. In order to prevent the
appearance of ambiguities in the zeroth order in $\eps$, it is convenient
to rewrite the right-hand side of Eq.~(\protect\ref{isol4}) such that the
Gamma functions have no poles at $\gamma_{k}=2k$. This is easily done using
the relation $\Gamma(1+z)\Gamma(1-z) = \pi z/\sin(\pi z)$ and is also
useful for the numerical solution.

\begin{figure}
\centerline{\psfig{file=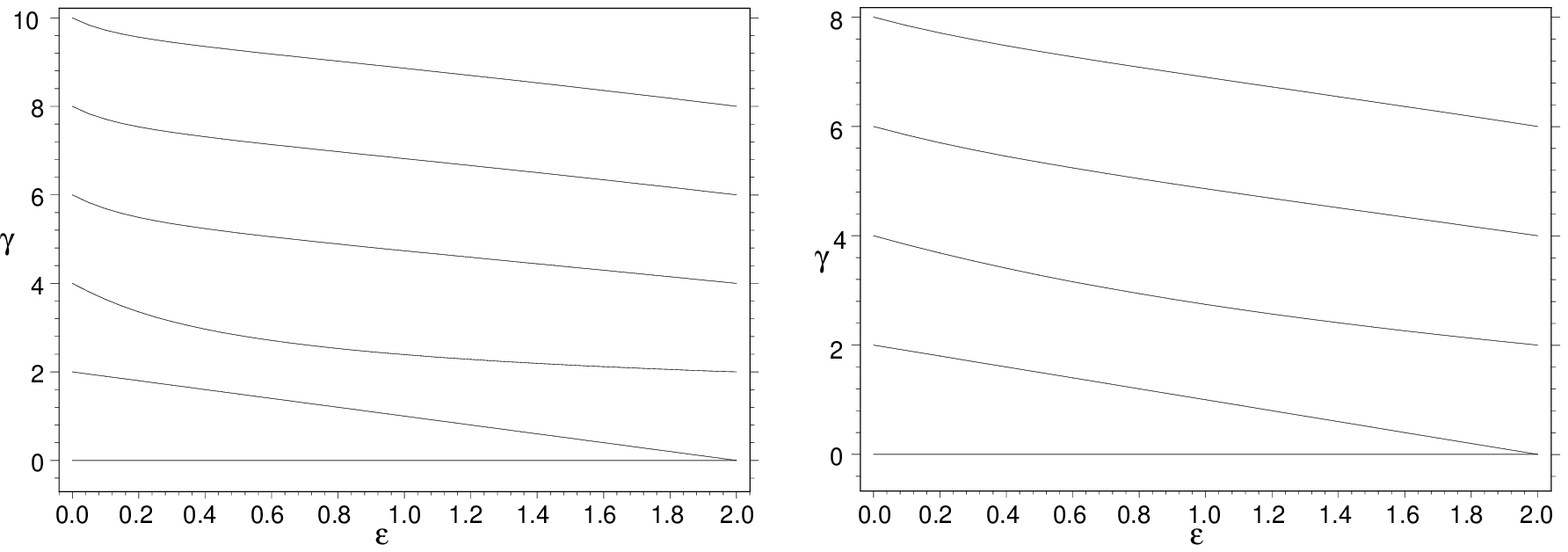,width=\textwidth}}
\caption{Leading scaling exponents for the isotropic sector $l=0$
in $d=2$ (left) and $d=3$ (right).}
\label{EXPO}
\end{figure}

Nonperturbative solutions of Eq. (\ref{isol4}) can only be obtained
numerically. They are illustrated by Fig.~\ref{EXPO} for $d=2$ (left) and
$d=3$ (right); the latter is in agreement with Fig.~2 from Ref. \cite{AP}.
One can see that solutions (\ref{isol5}) exist for all $0\le\eps\le2$,
decrease monotonically as $\eps$ grows and turn to $\gamma_{k}=2k-2$ at
$\eps=2$. The exponent $\gamma=0$ corresponds to the solution
$\delta_{ij} \delta(\bf p)$ which exists for all $d$ (see below).

It is easily seen that the integrals entering into Eq. (\ref{isol2}) are
divergent on these solutions at $\k-\p \to0$. One can say that
the knowledge of the exponents is insufficient to discuss the convergence:
it is necessary to know the behavior of the entire solution in the region
of small momenta, $|\k-\p| \ll m$, where it no longer reduces to a sum of
power terms with the exponents (\ref{isol5}). However, it is intuitively
clear that the form of the solution at such small momenta is irrelevant
for the calculation of the inertial-range exponents. Indeed, we made no
assumptions about the form of the solution in that range, sought them
in purely powerlike form and managed to derive closed equations for the
exponents using the prescriptions of the analytical regularization.
A simple justification of this procedure follows.

In coordinate representation, the solution is sought in the form
${\cal D} ({\bf r}) \propto r^{\gamma}{\cal C}(m{\bf r})$ (for simplicity,
here and below we omit its vector indices). The convergence problems
arise if $\gamma>0$, which is implied in what follows. It is natural
to assume that the scaling function ${\cal C}(m{\bf r})$ is such that
${\cal D} ({\bf r})$ vanishes at ${\bf r}=0$ along with all its
derivatives up to the $n\,$th order, where $n$ is the maximal integer
satisfying the inequality $\gamma>n$. This gives the set of
integral relations
\begin{equation}
\int d \q\, q_{i_{1}}\dots q_{i_{k}}\, {\cal D} ({\bf q}) =0,
\qquad k=0,1,\dots, n
\label{isol6}
\end{equation}
for the correlation function
${\cal D} ({\bf q}) \propto q^{-d-\gamma} \widetilde {\cal C}({\bf q}/m)$
in momentum representation. The integral entering into
Eq. (\ref{isol2}) can symbolically be written as
\begin{equation}
\int d \q\, F(\q, \p)\, {\cal D} ({\bf q}),
\label{isol7}
\end{equation}
where we have introduced the new variable $\q=\p-\k$. The form of the
kernel $F(\q, \p)$ is clear from the comparison with Eq. (\ref{isol2});
the divergence can arise from the region $\q\to0$, where the solution
behaves as ${\cal D} ({\bf q}) \propto q^{-d-\gamma}$. Owing to
relations (\ref{isol6}), the value of integral (\ref{isol7}) does not
change if one subtracts from $F(\q, \p)$ the first terms of its Taylor
expansion up to the $n\,$th order:
\begin{equation}
\int d \q \left\{
F(\q, \p) - F(0, \p) - q_{i} \frac{\partial F(0, \p)}{\partial q_{i}} -
\dots - \frac{1}{n!} q_{i_{1}}\dots q_{i_{n}}
\frac{\partial^{n} F(0, \p)}{\partial q_{i_{1}}\cdots
\partial q_{i_{n}}} \right\} {\cal D} ({\bf q}) .
\label{isol8}
\end{equation}
Now one can set $m=0$ in the function ${\cal D} ({\bf q})$, that is,
replace the exact solution with its inertial-range asymptotic expression
${\cal D} ({\bf q}) \propto q^{-d-\gamma}$: the possible divergence at
${\bf q}=0$ is suppressed by the expression in the curly brackets, which
behaves as $q^{n+2}$ for ${\bf q}\to0$.
Therefore, one can use the formal rules of analytical regularization
and simultaneously omit the subtracted terms \cite{book3}: this gives
the correct answer for the convergent integral with proper subtractions
in Eq.~(\ref{isol8}).

We thus conclude that the exponents (\ref{isol5}) may appear in the full
solution as correction terms; in the corresponding integrals exact solution
can be replaced with its powerlike asymptote and the resulting integrals
are properly given by the rules of analytical regularization.

Our conclusions are in agreement with those drawn in Ref. \cite{AP}
for the equation in coordinate space, although the analysis in
momentum space appears rather different. In particular, the
momentum-space analysis reveals the close resemblance between the
scalar and vector models: for the former, the correlation function
in momentum representation also satisfies an {\it integral}
equation and the above discussion is needed to fix the convergence
problem. It is also worth noting that the procedure employed in Ref.
\cite{AP} for the calculation of divergent integrals involves analytical
continuation from the region of convergence, and is therefore close to
the concept of analytical regularization.

Furthermore, the ``realizability'' of solutions (\ref{isol5}) is
guaranteed by the fact that in the RG approach they are identified
with the critical dimensions of composite operators entering into the
corresponding operator product expansions; see Sec.~\ref{sec:OPE}.

Besides powerlike solutions, Eq. (\ref{isol2}) also possesses the solution
of the form $\delta _{ij}\delta(\p)$. To demonstrate this, we use the
well-known representation of the $d$-dimensional delta function
\begin{equation}
\delta(\p) = \lim_{\sigma\to0} \int {\cal D}\x\, (\Lambda x)^{-\sigma} \,
\exp \bigl[ {\rm i} (\p\x)\bigr] = S_{d}^{-1}\, p^{-d}\, \lim_{\sigma\to0}
[\sigma (p/\Lambda)^{\sigma}]
\label{pode}
\end{equation}
and substitute it into Eq. (\ref{isol2}). For small $\sigma$, the integral
on its left-hand side is finite. Therefore it vanishes as $\sigma \to 0$
in Eq. (\ref{pode}), and the equation is satisfied. Another such
solution, $[\delta_{ij} -d n_{i}n_{j}]\, \delta(\p)$, belongs to the first
anisotropic sector. The both solutions are automatically transverse owing
to the relation $p_{i}\delta(\p)=0$. In coordinate representation, they
correspond to constant terms, so that the exponent $\gamma=0$ can formally
be assigned to them. They are indeed present in the pair correlation
function, but disappear from the structure function (\ref{struc}) owing
to the relation $S_{2}=0$ for ${\bf r}=0$.

We have already established in the end of Sec.~\ref{sec:QFT} that the model
(\ref{1})--(\ref{3}) becomes unstable at $d=\sqrt3$ and no steady-state
solution for the correlation function exists in the range of small momenta,
$p\ll m$. It is easily seen that expressions (\ref{isol5}) for the
inertial-range exponents diverge in this limit. The same singularity occurs
in the first-order expressions for another exponents in our model; see,
e.g., Eqs. (\ref{corre3})--(\ref{corre9}), (\ref{anal}) and (\ref{mnogo})
in subsequent sections. Moreover, the RG analysis shows that the actual
expansion parameter is $g_{*} \propto \eps/ (d^{2}-3)$ rather than $\eps$
itself; see Eq. (\ref{fixed}) in Sec.~\ref{sec:RGE}. Therefore, the
higher-order terms of the $\eps$ expansions become more and more singular
as $d$ approaches $\sqrt3$ from above and any finite-order approximation
cannot be trusted.

Numerical solution shows that for $d$ close to the threshold, the behaviour
of exponents $\gamma_{k}$ consists of two pronounced stages. At the beginning,
the exponents decrease rapidly as $\eps$ increases in agreement with the
first-order expression in Eq. (\ref{isol5}). Then, for a very small value of
$\eps$ (which tends to zero as $d\to\sqrt3$) the instantaneous stabilization
takes place at an almost constant value $\gamma_{k}\approx 2k-2$, and for
$\eps=2$ one has $\gamma_{k}=2k-2$ exactly.

For $d=\sqrt3$ the solutions do not exist. This fact can be naturally
explained, and extended to the other exponents, on the basis of the
$\eps$ expansion. The series in $\eps$ can be rewritten as series in
$g_{*} \propto \eps/ (d^{2}-3)$, with coefficients regular at $d^2=3$.
Then they can be inverted into expansions of $g_{*}$ in $\gamma$ [more
precisely, in $\gamma-2k=O(\eps)$], where the coefficients are also
regular. In other words, the inverted $\eps$ series can be written as
$\eps= (d^{2}-3) f(\gamma)$, where $f$ is a function regular as
$d\to\sqrt3$. For $d=\sqrt3$ this gives $\eps=0$, that is, no solution
exists for $\gamma$ if $\eps>0$.

\subsection{Scaling behavior in anisotropic sectors} \label {sec:Kartina}

Now let us turn to the case of uniaxial anisotropy, specified by an unit
vector ${\bf n}$ in the correlation function (\ref{2}), and
introduce the following irreducible $l$th rank tensors
\begin{equation}
{\cal N}^{(l)}_{i_{1}\cdots i_{l}} \equiv {\cal IRP} \bigl[
n_{i_{1}} \cdots n_{i_{l}} \bigr].
\label{tensorN}
\end{equation}
Here and below, ${\cal IRP}$ denotes the irreducible part, obtained by
subtracting the appropriate expression involving the Kronecker delta
symbols, such that the resulting tensor is traceless with respect to any
pair of indices. In particular, ${\cal N}^{(1)}_{i}=n_{i}$,
${\cal N}^{(2)}_{ij} = n_{i}n_{j} - \delta_{ij} /d$,
${\cal N}^{(3)}_{ijk}= n_{i} n_{j} n_{k} -
(\delta _{ij} n_{k} + \delta _{ik} n_{j}+
\delta _{jk} n_{i})/(d+2)$ and so on.

It is easy to see that structures (\ref{tensorN})
are orthogonal on the sphere:
\begin{equation}
\int d {\bf n}\, {\cal N}^{(l)} \, {\cal N}^{(s)} =0
\qquad {\rm for} \quad l\ne s,
\label{tensorN3}
\end{equation}
where $d {\bf n}$ is the area element of the unit sphere in
the $d$-dimensional space.

The contraction of the tensor (\ref{tensorN}) with a fixed vector ${\bf p}$
gives
\begin{equation}
 {\cal N}^{(l)}_{i_{1}\cdots i_{l}}\, p _{i_{1}}  \dots p _{i_{l} } =
  p^{l} \, P_{l} (z) , \qquad z \equiv ({\bf np}) /p,
\label{zviloga}
\end{equation}
where $P_{l} (z) = z^{l}+O(z^{l-2}) $ are the Gegenbauer polynomials, which
reduce to the Legendre polynomials and trigonometrical functions in three
and two dimensions, respectively  \cite{G}.

The relation (\ref{tensorN3}) implies that these polynomials are
orthogonal on the sphere even if their arguments are different:
\begin{equation}
\int d{\bf n}\,P_{l}(z)P_{s}(z')=0\qquad{\rm for}\quad l\ne s,
\label{tensorN4}
\end{equation}
where $z \equiv ({\bf np}) /p$ and
$z' \equiv ({\bf n p'}) /p'$ with any fixed vectors
${\bf p}$ and ${\bf p'}$.

In terms of the structures (\ref{tensorN}) and transverse projectors
$P_{ij}$ the desired decomposition of the pair correlation function
can be written as follows:
\begin{equation}
{\cal D}_{ij} ({\bf p}) = \sum_{l=0}^{\infty} \,
{\cal D}_{ij}^{(l)} ({\bf p}),
\label{decomposition}
\end{equation}
where the summation runs over all even values of $l$ and the coefficient
functions have the forms
\begin{equation}
{\cal D}_{ij}^{(l)} ({\bf p}) = P_{i i_{1}} ({\bf p}) \, \Bigl[
A^{(l)}(p) \, \delta_{i_{1} i_{2}} \, \Bigl(
{\cal N}^{(l)}_{i_{3}i_{4}\cdots i_{l+2}}\, p _{i_{3}}p _{i_{4}}
\dots p _{i_{l+2}} \Bigr) + \Bigl( B^{(l)}(p) \, p^{2}\,
{\cal N}^{(l)}_{i_{1}i_{2}i_{3}i_{4}\cdots i_{l}}\,
p _{i_{3}}p _{i_{4}} \dots p _{i_{l}} \Bigr) \Bigr] \,
P_{ i_{2}j} ({\bf p}).
\label{decomposition2}
\end{equation}
Whatever be the coefficient functions $A^{(l)}$ and $B^{(l)}$,
dependent only on $p=|{\bf p}|$, the expression (\ref{decomposition2})
is symmetric in the tensor indices $i$, $j$ and orthogonal to the
vector ${\bf p}$: $p_{i}\, {\cal D}_{ij}^{(l)} ({\bf p})=0$.

The first tensor structure in Eq. (\ref{decomposition2}) can be expressed
in terms of the Gegenbauer polynomials using the relation (\ref{zviloga}),
while the second structure can be expressed in terms of the polynomials
$P_{l}(z)$ and their derivatives using the relation
\[ l(l-1)\, {\cal N}^{(l)}_{i_{1}i_{2}i_{3}i_{4}\cdots i_{l}}\,
p _{i_{3}}p _{i_{4}} \dots p _{i_{l}} = \frac {\partial^{2}}
{\partial p_{i_{1}}\, \partial p_{i_{2}}} \, \Bigl[ p^{l} \, P_{l} (z)
\Bigr], \]
which is obvious from the definition (\ref{tensorN}) and the
the relation (\ref{zviloga}).

Substituting the series (\ref{decomposition}) into the zero-mode analog of
Eq.~(\ref{10}) then gives the equation for the coefficient functions
$A^{(l)}$, $B^{(l)}$, which can be symbolically written as $\Lambda_{ij}
({\bf p})=0$. Its left-hand side can be decomposed using the same tensor
structures (\ref{tensorN}), so that the equation can be reduced to an
infinite family of the scalar equations
\begin{equation}
\int d {\bf n}\,\Lambda_{ii}({\bf p})\,  ({\bf p}{\bf n})^{l}  =0 , \qquad
\int d {\bf n}\, \Lambda_{ij}({\bf p}) \n_i \n_j \,({\bf p}{\bf n})^{l-2} =0,
\label{skalars}
\end{equation}
with the integration over the unit sphere; cf. (\ref{tensorN3}). Owing
to the generalized orthogonality relation (\ref{tensorN4}), the $l$th
pair of equations (\ref{skalars}) involves only the coefficient functions
$A^{(l)}$, $B^{(l)}$ with the same index $l$, so that the equations for
different values of $l$ have decoupled.

In the inertial range, the solution in the $l\,$th sector is sought
in the form $A^{(l)} = a_{l} p^{-d-l-\gamma_{l}}$,
$B^{(l)}= b_{l}  p^{-d-l-\gamma_{l}}$, which corresponds to
${\cal D}^{(l)} ({\bf r}) \propto r^{\gamma_{l}}$ in
coordinate representation; for the isotropic shell, $l=0$, this reduces
to the single term (\ref{isol}). For $l\ge2$, one obtains a pair of
linear equations for each pair of coefficients $a_{l}$, $b_{l}$:
\begin{eqnarray}
C_{11}^{(l)} a_{l}+ C_{12}^{(l)}b_{l}=0, \qquad
C_{21}^{(l)} a_{l}+ C_{22}^{(l)}b_{l}=0,
\label{system}
\end{eqnarray}
where the coefficients $C_{\alpha\beta}^{(l)}$ depend on $\eps$, $d$
and the unknown exponent $\gamma_{l}$. They all can be expressed in
terms of the basic integrals (\ref{convolu}). In particular, for $l=2$ one
obtains
\begin{eqnarray}
 C_{11}^{(2)}&=& (d-1)^{2} \widetilde{I}_{1} - (d^{2}-1) I_{2} + dI_{3},
\nonumber \\
C_{12}^{(2)}&=& C_{21}^{(2)}=
-(d-1)\widetilde{I}_{1} + (d+1) I_{2} -dI_{3},
\nonumber  \\
C_{22}^{(2)}&=& (d-1)(d^{2}-2)\widetilde{I}_{1}/2 - (d^{2}-1) I_{2} +dI_{3},
\label{coefficients}
\end{eqnarray}
where $\widetilde{I}_{1}\equiv I_{1} +2J$ with $J$ from (\ref{J}) and
$I_{n}\equiv I_{n}(\gamma_2-2,\eps+2)$. The $l\,$th pair of equations
involves integrals $I_{n}\equiv I_{n}(\gamma_l-2,\eps+2)$
with $n$ as high as $l/2+2$. The coefficients $C_{\alpha\beta}^{(l)}$
for higher values of $l$ up to $l=12$ are given in Appendix~A.

The desired equation for $\gamma_{l}$ is obtained as the requirement that
the linear homogeneous system (\ref{system}) have nontrivial solutions:
\begin{eqnarray}
\det |C_{\alpha\beta}^{(l)}|=0.
\label{determinant}
\end{eqnarray}
For any given $l$, the determinant can easily be written down in terms of
standard integrals, provided the coefficients are known
[see Eq. (\ref{coefficients}) for $l=2$ and Appendix~A for $l\le12$];
then using the explicit expressions (\ref{convolu}) for $I_{n}$ one obtains
transcendental equations similar to Eq. (\ref{isol4}) but much more
cumbersome. We shall not write them down explicitly for the sake of brevity
and turn to the corresponding solutions.

It is also worth noting that for $l\ge4$ the formal divergence
of the integrals in Eq. (\ref{isol2}) occurs already for the leading
exponents and the discussion, similar to that given in Sec.~\ref{sec:Iso},
is needed to justify the use of analytic regularization.

For $d=2$, the two structures in Eq. (\ref{decomposition2}) coincide
and the determinant $\det|C_{\alpha\beta}^{(l)}|$ vanishes identically.
We shall return to this case in the end of the Section, and for now on we
assume $d\ne2$.

For any given $l\ge2$ and small $\eps$, all possible solutions for the
exponents $\gamma_{l}$ can be written in the form
$\gamma_{l} = (l-2+2k)+ O(\eps)$ with $k=0,1,2,\dots$.
The leading exponent corresponds to $k=0$; it is unique for
any $l\ge2$ and has the form
\begin{eqnarray}
\gamma_{l}= (l-2) \left\{ 1+ \varepsilon \,
\frac {(d+2)(l-3)(d^2-4d+2ld+l^2-5l+4)}
{(d^2-3)(d+2l-6)(d+2l-4)(d+2l-2)} + O(\varepsilon^{2}) \right\} .
\label{corre3}
\end{eqnarray}
For all $k\ge1$, there are exactly two solutions. For $k=1$, they
can be written as
\begin{eqnarray}
\gamma_{l} = l-\eps\, \frac {(d+2)\,x_{\pm}}
{(d^{2}-3)(d+2l-2)} +O(\eps^{2}),
\label{slopes3}
\end{eqnarray}
where the slopes $x_{\pm}$ satisfy the quadratic equation
\begin{eqnarray}
x^{2} (d+2l-4)(d+2l) -  \bigl[ d^{4} +(4l-5) d^{3} +
4(l^{2}-4l+2)d^{2} + (-4l^{3}-2l^{2}+14l-4) d -
2l(l-1)(l-2) (3l+1)\bigr] x -
\nonumber \\
-l(l-1)(d+l-1) \bigl[d^{3}+(3l-5) d^{2} + (2l^{2}-11l+8)d +
2(-l^{3} +l^{2}+3l-2)\bigr] =0.
\nonumber
\end{eqnarray}

In particular, for $l=2$ this gives
\begin{equation}
x_{\pm}= \left\{ d^{3}+3d^2-8d -16 \pm \sqrt { (d+4)
(d^5+2d^4-7d^3-4d^2+8d+16)} \right\} \big /\, 2(d+4) .
\label{gam2}
\end{equation}

For $k=2$, the solutions have the form
\begin{eqnarray}
\gamma_{l} &=& (l+2) + \varepsilon \frac{ (d+2) [
-d^{4}+ (2-6l)d^{3}+(3+9l-11l^{2}) d^{2} + (-6+6l+16l^{2}-6l^{3})d
+l(l^{2}-1) (l+10)]} {(d^{2}-3)(d+2l-2)(d+2l)(d+2l+2)}+O(\eps^{2}),
\nonumber \\
\gamma_{l} &=& (l+2) - \varepsilon \frac{(d-1)(d+2)}{(d^{2}-3)}+O(\eps^{2}).
\label{corre8}
\end{eqnarray}

The situation simplifies for all $k\ge3$: then the solutions in order
$O(\eps)$ are degenerate and have the form
\begin{eqnarray}
\gamma_{l}= (l-2+2k) - \varepsilon \frac{(d-1)(d+2)}{(d^{2}-3)}
+O(\eps^{2}),
\label{corre9}
\end{eqnarray}
that is, the slope is the same as for the second solution in Eq.
(\ref{corre8}) and the standard slope (\ref{isol5}) for the isotropic sector.
However, the degeneracy is removed by the $O(\eps^{2})$ terms, and the two
solutions (\ref{corre9}) do not coincide identically.

At the opposite edge, $\eps=2$, all the solutions can also be found
analytically for any given $l\ge2$. They take on the values $l-2$ (single),
$l$, $l+2$, $l+4$, and so on (two-fold degeneracy for $d\ne2$
and single otherwise). In addition to these ``standard'' values, for all
$l\ge2$ there are exactly two $d$-dependent solutions $\gamma_{l}^{\pm}$;
they satisfy certain quadratic equations and have the forms
\begin{eqnarray}
\gamma_{l}^{+} &=&- \frac{d}{2} + \frac{1}{2} \sqrt {
\frac{ d^{3}+4ld^{2}-d^{2}-4ld+4l^{2}d -8l+4l^{2}}{(d-1)} },
\nonumber \\
\gamma_{l}^{-}&=& - \frac{d}{2} + \frac{1}{2} \sqrt {
\frac{ d^{3}+4ld^{2}-9d^{2}-4ld+4l^{2}d +8 -8l+4l^{2}}{(d-1)} }
\label{nonstandard}
\end{eqnarray}
(only one solution of each equation is admissible). Note that
$\gamma_{l}^{+}>\gamma_{l}^{-}$ for all $l$ and $d>1$. For large $d$,
exponents (\ref{nonstandard}) behave as $\gamma_{l}^{+}=l+O(1/d)$ and
$\gamma_{l}^{-}=l-2+O(1/d)$, respectively, so that all solutions become
``standard'' at $d=\infty$.

\begin{figure}
\centerline{\psfig{file=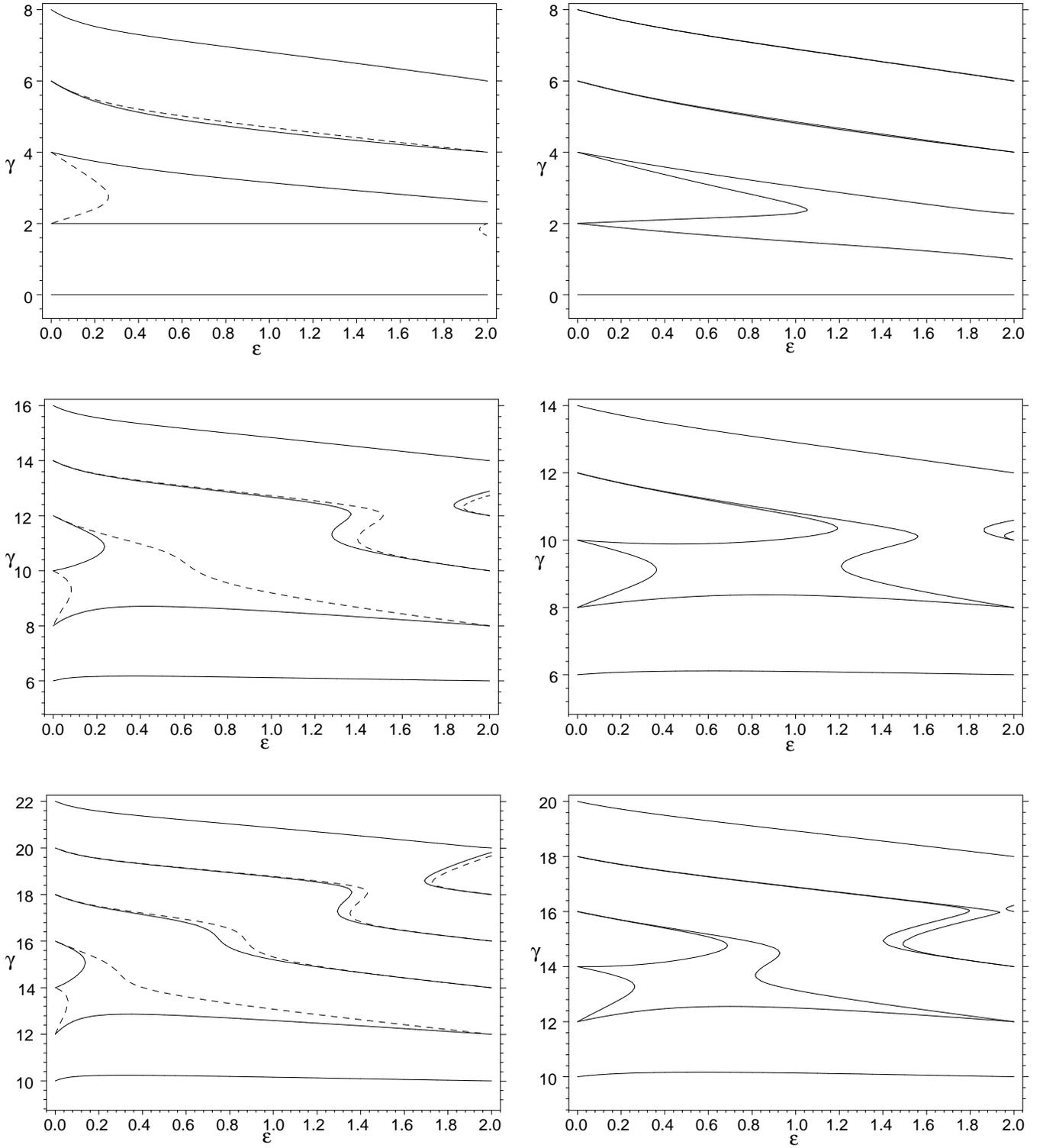,width=\textwidth}}
\caption{Leading scaling exponents for the sectors $l=2$, 8 and 12
(from above to below) in $d=2$ (left) and $d=3$ (right).
Dashed lines denote solutions that exist as limits $d\to2$ but
disappear in two dimensions.}
\label{EXPO2}
\end{figure}

Nonperturbative solutions for intermediate values of $\eps$ between
0 and 2 can only be obtained numerically. We have performed the
calculation in two and three dimensions for $l\le12$; the results
are illustrated by Fig.~\ref{EXPO2} for $l=2$, 8 and 12. For $d=3$ and
$l\le10$, our solutions are in agreement with the results presented in
Fig.~2 of Ref. \cite{AP}, except for the case $l=2$: the behaviors of the
solutions $\gamma_{2}=2+O(\eps)$ and $\gamma_{2}=4+O(\eps)$ are different.
We believe that this disagreement is not conceptual and is explained by
calculational errors in \cite{AP}. For the sake of brevity, we do not
give the solutions for $l=4$, 6 and 10, which in three dimensions are in
agreement with \cite{AP}. The exponent $\gamma=0$ for $l=2$ corresponds
to the solution $[\delta_{ij}-dn_{i}n_{j}]\,\delta(\bf p)$, which exists
for all $d$ (see Sec.~\ref{sec:Iso}).

The figures illustrate the following qualitative behavior of the solutions,
which holds for all $d\ne2$ and $l\ge2$. The leading solution (\ref{corre3})
exists for all $\eps$ and turns to $l-2$ for $\eps=2$. In fact, it is hardly
distinguishable from a constant, $\gamma_{l}\approx l-2$, for all values of
$\eps$ ($\gamma_{2}\equiv 0$, see Sec.~\ref{sec:Iso}).

For any $l$, some critical value $k_{c}=k_{c}(l,d)$ exists such that
for all $k > k_{c}$ the behavior of the solutions $\gamma_{l}=l-2+2k+
O(\eps)$ is simple: the both solutions exist for all $\eps$, decrease
slowly as $\eps$ grows and turn to $\gamma_{l}=l-4+2k$ for $\eps=2$.
(In fact, the both solutions corresponding to given $l$ and $k$ are
very close to each other for all values of $\eps$.)

For fixed $l$, the critical value $k_{c}$ decreases as $d$ increases, so
that all solutions with $k=1,2,\dots$ become simple (in the above sense)
provided $d$ is large enough. For fixed $d$, the critical value $k_{c}$
increases with $l$; in particular, in three dimensions $k_{c}=2$ for
$l=2$, 4, 6, $k_{c}=3$ for $l=8$, 10 and $k_{c}=4$ for $l=12$.

An interesting interaction between the solutions with a fixed $l$ and
different $k$ occurs for $1\le k \le k_{c}$. Two branches starting at
$\eps=0$ with different values of $k$ can coalesce and disappear for
some value of $\eps$ between 0 and 2. Another possible process is the
creation of a pair of solutions for some $0<\eps<2$. A solution that
starts at $\eps=0$ can annihilate with a solution from a pair that was
created for some finite value of $\eps$. The interplay between these
creation-annihilation processes can produce a very complicated pattern,
as illustrated by Fig.~\ref{EXPO2} for the sectors $l=2$, 8 and 12
(see also Fig.~2 in Ref. \cite{AP}).

It turns out, however, that the creation and annihilation of solutions
eventually compensate each other in the sense that the number of branches
starting at $\eps=0$ with $0< k <k_{c}$ is equal to the number of branches
arriving at $\eps=2$ and confined between the leading solution
($\gamma_{l}=l-2$ for $\eps=2$) and the lowest ``simple'' solution
($\gamma_{l}=l-4+2k_{c}$ for $\eps=2$).  The balance is possible owing to
the existence of two ``nonstandard'' solutions (\ref{nonstandard}) at the
edge $\eps=2$. They also determine the boundary between the solutions with
``simple'' and ``interesting'' behavior: the uppermost solution with the
interesting behavior, $\gamma_{l}=l-2+2k_c+O(\eps)$, turns to
$\gamma_{l}^{+}$ at $\eps=2$. [Some reservations are needed if a
``standard'' solution at $\eps=2$ lies between the roots $\gamma_{l}^{\pm}$
or coincides with one of them. In particular, for $d=3$ and $l=2$,
the standard solution $\gamma_{2}=2$ exists but is isolated in the sense
that no {\it real\,} branches attach it from the region $\eps<2$. For $d=3$
and $l=12$, one obtains $\gamma_{12}^{-}=16$, and this standard value
acquires three-fold degeneracy].

The behavior eventually simplifies in the limit $d\to\infty$. All the
solutions become simple in the above sense and they are described by
straight lines: $\gamma_{l} = l-2$ for the leading solution and
$\gamma_{l} = l-2+2k -\eps$ for all $k\ge1$.

The annihilation of coalescing solutions actually means that they become
complex: the effect known for the magnetic model \cite{V96}, where it
occurs in the isotropic shell. It was argued in Refs. \cite{V96,Lanotte}
that the complexification leads to the instability of the steady state
(exponential growth of the pair correlation function).
We shall not discuss this important issue here and only stress an essential
distinction between the two cases. In the magnetic model, the {\it leading
admissible} exponent $\gamma=O(\eps)$ coalesces with the solution
$\gamma=-d+O(\eps)$, which is not admissible and describes the large-scale
behavior at $r\gg L$ [see the remark and references below Eq. (\ref{isol4})].
In model (\ref{1})--(\ref{3}) the coalescence occurs only in anisotropic
sectors and only for {\it nonleading admissible} exponents. If the steady
state remains stable, the inertial-range behavior in the corresponding
sectors will include oscillations on the powerlike background; cf. the
discussion in Ref.~\cite{Novikov}.

In two dimensions, the tensor structures in decomposition
(\ref{decomposition2}) become coincident, and the determinant
$\det|C_{\alpha\beta}^{(l)}|$ in Eq. (\ref{determinant}) vanishes
identically. All the coefficients $C_{\alpha\beta}^{(l)}$ for $d=2$ become
equal up to the sign [see Eq. (\ref{coefficients}) for $l=2$]. Therefore,
the equation for the exponents $\gamma_{l}$ can simply be written as
\begin{eqnarray}
C_{11}^{(l)}=0.
\label{dett}
\end{eqnarray}
All solutions of the $d$-dimensional equation (\ref{determinant})
have well-defined limits as $d\to2$, and all true two-dimensional
solutions are indeed recovered in this limit. However, this limit gives
more solutions than the correct two-dimensional equation: one half of the
solutions obtained in the limit $d\to2$ from the $d$-dimensional case do
not satisfy Eq. (\ref{dett}) and should be discarded. This behavior is
illustrated by Fig.~\ref{EXPO2}, where the solid lines on the diagrams
with $d=2$ and $l\ge2$ denote
solutions obtained both from the two-dimensional equation (\ref{dett})
and as limits $d\to2$ from $d$-dimensional solutions, and the dashed lines
denote spurious solutions which are obtained in the limit $d\to2$ from Eq.
(\ref{determinant}), but do not satisfy equation (\ref{dett}). We shall see
in Secs.~\ref{sec:Operators6} and~\ref{sec:Ashells} that similar effect
is encountered in the RG and OPE approach: all critical dimensions have
well-defined limits as $d\to2$, but some of them should be ruled out due to
linear relations between composite operators which hold in two dimensions.

\section{Renormalization, RG functions and RG equations}
\label {sec:RGE}

The analysis of the UV divergences is based on the analysis of canonical
dimensions \cite{Collins,book3}. Dynamical models of the type (\ref{action}),
in contrast to static models, have two scales, so that the canonical
dimension of some quantity $F$ (a field or a parameter in the action
functional)
is described by two numbers, the momentum dimension $d_{F}^{k}$ and
the frequency dimension $d_{F}^{\omega}$. They are determined such that
$[F] \sim [L]^{-d_{F}^{k}} [T]^{-d_{F}^{\omega}}$, where $L$ is the
length scale and $T$ is the time scale. The dimensions are found
from the obvious
normalization conditions $d_k^k=-d_{\bf x}^k=1$, $d_k^{\omega}
=d_{\bf x}^{\omega}=0$, $d_{\omega}^k=d_t^k=0$,
$d_{\omega}^{\omega}=-d_t^{\omega}=1$, and from the requirement
that each term of the action functional be dimensionless (with
respect to the momentum and frequency dimensions separately).
Then, based on $d_{F}^{k}$ and $d_{F}^{\omega}$,
one can introduce the total canonical dimension
$d_{F}=d_{F}^{k}+2d_{F}^{\omega}$ (in the free theory,
$\partial_{t}\propto\Delta$), which plays in the theory of
renormalization of dynamical models the same role as the conventional
(momentum) dimension does in static problems \cite{book3}.

The dimensions for the model (\ref{action}) are given in
Table \ref{table1}, including the parameters
which will be introduced later on.
From the Table it follows that the model is
logarithmic (the coupling constant $g_{0}$ is dimensionless)
at $\eps=0$, so that the UV divergences have the form of
the poles in $\eps$ in the Green functions.

The total canonical dimension of an arbitrary
1-irreducible Green function $\Gamma = \langle\Phi \cdots \Phi
\rangle _{\rm 1-ir}$ is given by the relation
\begin{equation}
d_{\Gamma }=d_{\Gamma }^k+2d_{\Gamma }^{\omega }=
d+2-N_{\Phi }d_{\Phi},
\label{deltac}
\end{equation}
where $N_{\Phi}=\{N_{\theta'},N_{\theta},N_{\bf v}\}$ are the
numbers of corresponding fields entering into the function
$\Gamma$, and the summation over all types of the fields is
implied.
The total dimension $d_{\Gamma}$ is the formal index of the UV divergence.
Superficial UV divergences, whose elimination requires
counterterms, can be present only in those functions $\Gamma$ for
which $d_{\Gamma}$ is a non-negative integer.

Analysis of the divergences should be based on the following auxiliary
considerations \cite{UFN,turbo}:

(i) From the explicit form of the vertex and bare propagators in
the model (\ref{action}) it follows that $N_{\theta'}- N_{\theta}=2N_{0}$
for any 1-irreducible Green function, where $N_{0}\ge0$
is the total number of bare propagators $\langle \theta \theta
\rangle _0$ entering into the function (see Sec.~\ref{sec:QFT}).
Therefore, the difference
$N_{\theta'}- N_{\theta}$ is an even non-negative integer for
any nonvanishing function.

(ii) If for some reason a number of external momenta occurs as an
overall factor in all the diagrams of a given Green function, the
real index of divergence $d_{\Gamma}'$ is smaller than $d_{\Gamma}$
by the corresponding number (the Green function requires
counterterms only if $d_{\Gamma}'$  is a non-negative integer).
In the model (\ref{action}), the derivative $\partt$ at the
vertex $\theta'({\bf v}\partt)\theta$ can be moved onto the
field $\theta'$ by virtue of the transversality of the field
${\bf v}$. Therefore, in any 1-irreducible diagram it is always
possible to move the derivative onto any of the external
``tails'' $\theta$ or $\theta'$, which decreases the real index
of divergence: $d_{\Gamma}' = d_{\Gamma}- N_{\theta}-N_{\theta'}$.
The fields $\theta$, $\theta'$ enter into the counterterms only
in the form of derivatives $\partial\theta$, $\partial\theta'$.

From the dimensions in Table \ref{table1} we find
$d_{\Gamma} = d+2 - N_{\bfv} + N_{\theta}- (d+1)N_{\theta'}$
and $d_{\Gamma}'=(d+2)(1-N_{\theta'}) - N_{\bf v}$.
It then follows that for any $d$, superficial
divergences can only exist in the 1-irreducible functions
$\langle\theta'\theta\dots\theta\rangle_{\rm 1-ir}$ with $N_{\theta'}=1$
and arbitrary value of $N_{\theta}$, for which $d_{\Gamma}=2$,
$d_{\Gamma}'=0$. However, all the functions with $N_{\theta}>
N_{\theta'}$ vanish (see above) and obviously do not require
counterterms. We are left with the only superficially
divergent function $\langle\theta'\theta\rangle_{\rm 1-ir}$,
which does not depend on the correlation function (\ref{2}) and
therefore is isotropic;  see Sec. \ref{sec:QFT}.
The corresponding counterterm must contain two symbols
$\partial$, and owing to the isotropy and transversality conditions
reduces to the only structure $\btheta'\Delta\btheta$.

Inclusion of this counterterm is reproduced by the multiplicative
renormalization of the parameters $g_{0}$, $\nu_0$ in the action
functional (\ref{action}) with the only
independent renormalization constant $Z_{\nu}$:
\begin{equation}
\nu_0=\nu Z_{\nu},\quad g_{0}=g\mu^{\eps}Z_{g}, \quad
Z_{g}=Z_{\nu}^{-1}.
\label{reno}
\end{equation}
Here $\mu$ is the reference mass in the minimal subtraction (MS)
scheme, which is always used in what follows, $g$ and $\nu$
are renormalized analogs of the bare parameters $g_{0}$ and $\nu_0$,
and $Z=Z(g,\eps,d)$ are the renormalization constants. Their
relation in Eq. (\ref{reno}) results from the absence of renormalization
of the last term in Eq. (\ref{action}). No renormalization of the fields
and the ``mass'' $m$ is required, i.e., $Z_{\Phi}=1$ for all $\Phi$ and
$m_{0}=m$. The renormalized action functional has the form
\begin{equation}
{\cal S}_{R}(\Phi)=\btheta' D_{\theta}\btheta' /2+
\btheta' \left[ - \nabla_{t} + \nu Z_{\nu} \Delta \right] \btheta
-{\bf v} D_{v}^{-1} {\bf v}/2,
\label{renact}
\end{equation}
where the amplitude $D_{0}$ from Eq. (\ref{3})
expressed in renormalized parameters
using Eqs. (\ref{reno}): $D_{0}\equiv g_{0}\nu_0 = g\mu^{\eps} \nu$.

The explicit form of the constant $Z_{\nu}$ is determined by
the requirement that the 1-irreducible function
$\langle\theta'\theta\rangle_{{\rm 1-ir}}$ expressed in renormalized
variables be UV finite (i.e., be finite for $\eps\to0$). This
requirement determines $Z_{\nu}$ up to an UV finite contribution; the
latter is fixed by the choice of the renormalization scheme. In the MS
scheme all renormalization constants have the form ``1 + only poles
in  $\eps$.''  The function $\langle\theta'\theta\rangle_{{\rm 1-ir}}$
in our model is known exactly; see Eqs. (\ref{Dyson101})--(\ref{nonlocal})
in Sec. \ref{sec:QFT}. We substitute Eqs. (\ref{reno}) into it and choose
$Z_{\nu}$ to cancel the pole in $\eps$ in the resulting expression. This is
equivalent to the requirement that $\nu_{eff}(p)$ be finite; its pole part
is independent of $p$ and is therefore contained in Eq. (\ref{nonlocal}).
This gives:
\begin{equation}
Z_{\nu} = 1 - g\, C_{d}\, \frac{(d^{2}-3)}{2d(d+2)\eps},
\label{Z}
\end{equation}
with coefficient $C_{d}$ from (\ref{Dyson103}). The result (\ref{Z})
is exact, i.e., it has no corrections of order $g^{2}$, $g^{3}$, and so on;
this is a consequence of the fact that the one-loop approximation
(\ref{Dyson7}) for the response function is exact. Also note that
expression (\ref{Z}) differs from the exact expression for
$Z_{\nu}$ in the scalar \cite{RG} and magnetic \cite{Lanotte2} cases.

The relation $ {\cal S}(\Phi,e_{0})={\cal S}_{R}(\Phi,e,\mu)$ (where
$e_{0}=\{g_{0}, \nu_{0}, m\} $ is the complete set of bare parameters,
and $e =\{g, \nu, m\}$ is the set of their renormalized analogs) implies
$W(A,e_{0})=W_{R}(A,e,\mu)$, where $W$ is the functional (\ref{field})
and $W_{R}$ is its renormalized counterpart obtained by the replacement
${\cal S}\to{\cal S}_{R}$. We use $\widetilde{\cal D}_{\mu}$
to denote the differential operation $\mu\partial_{\mu}$ for fixed
$e_{0}$ and operate on both sides of this relation with it. This
gives the basic RG differential equation:
\begin{equation}
{\cal D}_{RG}\,W_{R}(e,\mu) = 0,
\label{RG1}
\end{equation}
where ${\cal D}_{RG}$ is the operation $\widetilde{\cal D}_{\mu}$
expressed in the renormalized variables:
\begin{equation}
{\cal D}_{RG}\equiv {\cal D}_{\mu} + \beta(g)\partial_{g}
-\gamma_{\nu}(g){\cal D}_{\nu}.
\label{RG2}
\end{equation}
In Eq. (\ref{RG2}), we have written ${\cal D}_{x}\equiv x\partial_{x}$ for
any variable $x$, and the RG functions (the $\beta$ function and
the anomalous dimension $\gamma$) are defined as
\begin{mathletters}
\label{RGF}
\begin{equation}
\gamma_{F}(g)\equiv \Dm \ln Z_{F}\qquad  {\rm for\ any\ } Z_{F},
\label{RGF1}
\end{equation}
\begin{equation}
\beta(g)\equiv\Dm  g=g[-\eps+\gamma_{\nu}(g)].
\label{RGF2}
\end{equation}
\end{mathletters}
The relation between $\beta$ and $\gamma$ in Eq. (\ref{RGF2}) results from
the definitions and the last relation in Eq. (\ref{reno}). From the relations
(\ref{Z}) and (\ref{RGF}) one obtains explicit expressions for the RG
functions:
\begin{equation}
\gamma_{\nu}(g)=\frac{-\eps {\cal D}_g \ln Z_{\nu}}{1-{\cal D}_g
\ln Z_{\nu}}= gC_{d}\, \frac {(d^{2}-3)}{2d(d+2)} .
\label{gammanu}
\end{equation}

From Eq. (\ref{RGF2}) it follows that the RG equations of the model
have an IR stable fixed point [$\beta(g_{*})=0$, $\beta'(g_{*})>0$]
with the coordinate
\begin{equation}
g_{*}= \frac{2d(d+2)\,\eps}{C_{d}(d^{2}-3)}.
\label{fixed}
\end{equation}
From the relation between the RG functions in Eq. (\ref{RGF2})
the value of $\gamma_{\nu}(g)$ at the fixed point is found exactly:
$\gamma_{\nu}^{*} \equiv \gamma_{\nu}(g_*)= \eps$.

For $d^{2}<3$, the fixed point is negative and therefore not accessible
for the RG flow with physical (positive) initial data for $g$. This is in
agreement with the conclusion of Sec. \ref{sec:QFT} that no stable
steady state exists in the model for $d^{2}<3$ (see the discussion
below Eq. (\ref{10})).

For $d^{2}>3$, the fixed point is positive;
this establishes the existence of scaling behavior in the IR region
($\Lambda r \gg 1$ and any fixed $mr$) for all correlation functions
of the model. Let $F$ be some
multiplicatively renormalized quantity (say, a correlation function
involving composite operators), i.e., $F=Z_{F}F_{R}$ with
certain renormalization constant $Z_{F}$. It satisfies the RG
equation of the form $[{\cal D}_{RG} + \gamma_{F}] F_{R}=0$ with
$\gamma_{F}$ from (\ref{RGF1}) and ${\cal D}_{RG}$ from (\ref{RG2}).
The solution of the RG equation then shows that in
the IR region $F$ takes on the scaling form
\begin{equation}
F \simeq  \Lambda ^{-\gamma_{F}^{*}}\,
D_{0}^{d^{\omega}_{F}} \, r^{-\Delta_F}
\xi_{F}(mr),
\label{RGR}
\end{equation}
where
\begin{equation}
\Delta_{F} = d_{F}^{k}+ \Delta_{\omega}
d_{F}^{\omega}+\gamma_{F}^{*}, \quad \Delta_{\omega}=2-\gamma_{\nu}^{*}
\label{32B}
\end{equation}
is the critical dimension of the function $F$, $d^{\omega}_{F}$ and
$d_{F}$ are its frequency and total canonical dimensions,
$\gamma_{F}^{*} = \gamma_{F}(g_{*})$ is the value of its anomalous dimension
at the fixed point, $\Delta_{\omega}=2-\gamma_{\nu}^{*}=2-\eps$
is the critical dimension of the frequency, and $\xi_{F}(mr)$
is the scaling function whose form is not determined by the RG
equation itself. Derivation of Eq. (\ref{32B}) and more detail can
be found in Refs. \cite{RG3,UFN,turbo,book3}. In particular, for
the structure functions (\ref{struc}) with  $d_{F}=0$,
$d_{F}^{\omega}=-n/2$ (see Table \ref{table1}) and  $\gamma_{F}^{*}=0$
(see Sec. \ref{sec:Operators}) one obtains
\begin{equation}
S_{n}({\bf r})= D_{0}^{-n/2}\,  r^{n(1-\eps/2)}\, \xi_{n}(mr),
\label{100}
\end{equation}
so that the dependence on the UV scale $\Lambda$ disappears, while
the dependence on the IR scale $m$ is contained in the scaling
functions $\xi_{n}(mr)$.

\section{Operator product expansion and anomalous scaling}
  \label {sec:OPE}

Representations (\ref{RGR})--(\ref{100}) for any scaling
functions $\xi(mr)$ describe the behavior of the correlation functions
for $\Lambda r \gg 1$ and any fixed value of $mr$.
The inertial range corresponds to the additional condition $mr \ll 1$.
The form of the functions $\xi(mr)$ is not determined by the RG equations
themselves; in analogy with the theory of critical phenomena, their
behavior for $mr\to0$ is studied using the operator product expansion (OPE);
see Refs. \cite{Collins,book3}. Below we concentrate on the equal-time
structure functions (\ref{struc}), (\ref{100}).

According to the OPE, the behavior of the quantities entering into
the right-hand side of Eq. (\ref{struc}) for
${\bf r} = {\bf x} - {\bf x'} \to 0$  and fixed
$ {\bf x} + {\bf x'} $ is given by the infinite sum
\begin{equation}
\Bigl[ \theta_{r} (t,{\bf x}) - \theta_{r} (t,{\bf x'})\Bigr]^{n}
= \sum_{F} C_{F} ({\bf r}) F\left(t,\, \frac{{\bf x}+{\bf x'}}{2} \right),
\label{OPE}
\end{equation}
where $C_{F}$ are coefficients regular in $m^{2}$ and $F$ are all possible
renormalized local composite operators allowed by the symmetry. More
precisely, the operators entering into the OPE are those which appear in the
naive Taylor expansion, and all the operators that admix to them
in renormalization.

In what follows it is assumed that the expansion is made in irreducible
tensors (scalars, vectors and traceless tensors); the possible tensor indices
of the operators $F$ are contracted with the corresponding indices of the
coefficients $C_{F}$. With no loss of generality, it can also be assumed that
the expansion is made in ``scaling'' operators, i.e., those having definite
critical dimensions $\Delta_{F}$ (see Sec.~\ref{sec:Operators}).

The structure functions (\ref{struc}) are obtained by averaging Eq.
(\ref{OPE}) with the weight $\exp {\cal S}_{R}$, the mean values
$\langle F\rangle$ appear on the right-hand side. Their asymptotic
behavior for $m\to0$ is found from the corresponding RG equations
and has the form $\langle F\rangle \propto  m^{\Delta_{F}}$.

From the RG representation (\ref{100}) and the operator product expansion
(\ref{OPE}) we therefore find the following expression for the structure
function in the inertial range ($\Lambda r\gg1$, $mr\ll1$)
\begin{equation}
S_{n}({\bf r})= D_{0}^{-n/2}\, r^{n(1-\eps/2)} \,
\sum_{F} A_{F}(m{\bf r})\, (mr)^{\Delta_{F}} ,
\label{OR}
\end{equation}
where the coefficients $A_{F}$ are regular in $(mr)^{2}$.

Some general remarks are now in order.

Owing to the translational invariance, the operators having the
form of total derivatives give no
contribution to Eq. (\ref{OR}):  $\langle \partial F(x) \rangle =
\partial\langle  F(x) \rangle =\partial\times\, {\rm const}=0$
(these operators become relevant if the stirring force in Eq. (\ref{1})
violates the translational invariance, like in the problem
discussed in Ref. \cite{Wiese}).

In the model (\ref{1})--(\ref{3}), the operators with an odd number of
fields $\theta$ also have vanishing mean values; their contributions
vanish along with the odd structure functions themselves (they will
be ``activated'' in the presence of a nonzero mixed correlation function
$\langle{\bfv}f\rangle$; we shall not discuss this possibility here).

If the tensor $C_{ij}({\bf r})$ in Eq. (\ref{2}) is taken to be isotropic,
the model becomes $SO(d)$ covariant and only the contributions of the scalar
operators survive in (\ref{OR}). Indeed, in the isotropic case the mean value
of a tensor operator depends only on scalar parameters, its tensor indices
can only be those of Kronecker delta symbols. It is impossible, however, to
construct nonzero irreducible (traceless) tensor solely of the delta symbols.

In the presence of anisotropy, irreducible tensor operators acquire
nonzero mean values and their contributions appear on the right-hand
side of Eq. (\ref{OR}). Like in Sec.~\ref{sec:PAIR}, consider the case of
the uniaxial anisotropy, specified by an unit vector ${\bf n}$ in the
correlation function (\ref{2}). In this case, the mean value of a $l\,$th rank
traceless operator involves the vector ${\bf n}$ along with the delta symbols
and is necessarily proportional to the $l\,$th rank symmetric traceless
tensor ${\cal N}^{(l)}_{i_{1}\cdots i_{l}}$ from Eq. (\ref{tensorN}).
The contraction with the corresponding coefficient $C_{F}$ gives rise
to the $l\,$th order Gegenbauer polynomial $P_{l}(z)$ with
$z= ({\bf r}{\bf n})/r$; see Eq. (\ref{zviloga}). In general, the
expansion in irreducible tensors in (\ref{OPE}) after the averaging
leads to the $SO(d)$ decomposition employed in Refs. \cite{AP,Arad},
the $l\,$th shell corresponding to the contribution of the $l\,$th
rank composite operators.

The feature characteristic of the models describing turbulence is the
existence of the so-called ``dangerous'' composite operators with
{\it negative} critical dimensions; see Refs.~\cite{UFN,turbo}.
Their contributions into the OPE give rise to singular behavior of
the scaling functions for $mr\to0$, that is, the anomalous scaling.
The leading term in the $l\,$th anisotropic sector is given by the
$l\,$th rank tensor operator with minimal (not necessarily negative)
dimension $\Delta[F]$.

Since $\Delta_F=d_{F}+O(\eps)$, see Eq. (\ref{32B}), the operators with
minimal $\Delta_F$ are those involving maximum possible number of fields
$\theta$ and minimum possible number of derivatives (at least for small
$\eps$). Both the problem (\ref{1})--(\ref{3}) and the quantities
(\ref{struc}) possess the symmetry $\theta\to\theta+{\rm const}$.
It then follows that the expansion (\ref{OPE}) involves only operators
invariant with respect to this shift and therefore built of the
{\it gradients} of $\theta$.

As already mentioned above, the operators entering into the right-hand side
of Eq. (\ref{OPE}) are those which appear in the Taylor expansion, and those
that admix to them in renormalization. The leading term of the
Taylor expansion for $S_{n}$ is the $2n\,$th rank operator which can
symbolically be written as $(\partial\theta)^{n}$; its decomposition in
irreducible tensors gives rise to operators of lower ranks. These
contributions exist in the OPE (before averaging) even if the stirring
force in not included into Eq. (\ref{1}); in the language of
Refs. \cite{Falk1,GK,Pumir} it is then tempting to identify them with
zero modes, i.e., the solutions of the homogeneous (unforced) analogs
of the closed exact equations satisfied by the equal-time correlations.
In the presence of the stirring force, operators of the form
$(\partial\theta)^{k}$ with $k <n$ admix to them in renormalization
and appear in the OPE; their contributions correspond to solutions of
the inhomogeneous equations. Owing to the linearity of problem (\ref{1}),
operators with $k >n$ (whose contributions would be more important)
do not admix in renormalization to the terms of the Taylor expansion
for $S_{n}$ and do not appear in the corresponding OPE. All these
operators have minimal possible canonical dimension $d_{F}=0$ (see
Table \ref{table1}) and determine the leading terms of the $mr\to0$ behavior
in the sectors with $j\le 2n$. Operators involving more derivatives
than fields $\theta$ (and thus having canonical dimensions $d_{F}=1$,
2 and so on) determine correction terms for $j\le 2n$ and leading terms
for higher anisotropic sectors with $j>2n$. The renormalization and
dimensions of the most important operators are studied in the next Section.

\section{Renormalization and critical dimensions of
 composite operators} \label {sec:Operators}

We recall that the term ``composite operator'' refers to any local (unless
stated to be otherwise) monomial or polynomial built of primary fields and
their derivatives at a single spacetime point $x\equiv (t,{\bf x})$;
see Refs. \cite{Collins,book3}.
Since the arguments of the fields coincide, correlation functions with
such operators contain additional UV divergences, which are removed
by additional renormalization procedure. For the renormalized
correlation functions the RG equations are obtained, which describe
IR scaling of certain ``basis'' operators  $F$ with definite critical
dimensions $\Delta_{F}\equiv\Delta[F]$. Due to the renormalization,
$\Delta[F]$ does not coincide in general with the naive sum of
critical dimensions of the fields and derivatives entering into $F$.
As a rule, composite operators ``mix'' in renormalization,
i.e., an UV finite renormalized operator $F^{R}$ has the form
$F^{R}=F+$ counterterms, where the contribution of the
counterterms is a linear combination of $F$ itself and,
possibly, other unrenormalized operators which ``admix'' to $F$.

Let $F\equiv\{F_{\alpha}\}$ be a closed set, all of whose
monomials mix only with each other in renormalization.
The renormalization matrix $Z_{F}\equiv\{Z_{\alpha\beta}\}$
and the  matrix of anomalous dimensions
$\gamma_{F}\equiv\{\gamma_{\alpha\beta}\}$
for this set are given by
\begin{equation}
F_{\alpha }=\sum _{\beta} Z_{\alpha\beta}
F_{\beta }^{R},\qquad \gamma _F=Z_{F}^{-1}\Dm Z_{F},
\label{2.2}
\end{equation}
and the corresponding matrix of critical dimensions
$\Delta_{F}\equiv\{\Delta_{\alpha\beta}\}$ is given by Eq. (\ref{32B}),
in which $d_{F}^{k}$ and $d_{F}^{\omega}$ are understood as
the diagonal matrices of canonical dimensions of the operators in
question (with the diagonal elements equal to sums of corresponding
dimensions of all fields and derivatives constituting $F$) and
$\gamma^{*}_{F}\equiv\gamma_{F} (g_{*})$ is the
matrix (\ref{2.2}) at the fixed point (\ref{fixed}).

Critical dimensions of the set $F\equiv\{F_{\alpha}\}$ are
given by the eigenvalues of the matrix $\Delta_{F}$. The ``basis''
operators that possess definite critical dimensions have the form
\begin{equation}
F^{bas}_{\alpha}=\sum_{\beta} U_{\alpha \beta}F^{R}_{\beta}\ ,
\label{2.5}
\end{equation}
where the matrix $ U_{F} =  \{U_{\alpha \beta} \}$ is such that
$\Delta'_{F}= U_{F} \Delta_{F} U_{F}^{-1}$ is diagonal.

In general, counterterms to a given operator $F$ are
determined by all possible 1-irreducible Green functions
with one operator $F$ and arbitrary number of primary fields,
$\Gamma=\langle F(x) \Phi(x_{1})\dots\Phi(x_{2})\rangle_{\rm 1-ir}$.
The total canonical dimension (formal index of UV divergence)
for such a function is given by
\begin{equation}
d_\Gamma = d_{F} - N_{\Phi}d_{\Phi},
\label{index}
\end{equation}
with the summation over all types of fields entering into
the function. For superficially divergent diagrams,
$d_\Gamma$ is a non-negative integer; cf. Sec. \ref{sec:RGE}.

Let us begin with the simplest operators of the form $\theta^{n}(x)$,
with free tensor indices or involving any contraction.
From Table \ref{table1}
in Sec. \ref{sec:RGE} and Eq. (\ref{index}) we obtain $d_{F}=-n$,
$d_\Gamma = -n+N_{\theta}-N_{\bf v} -(d+1)N_{\theta'}$.

From the analysis of the diagrams it follows that the total number
$N_{\theta}$ of the fields $\theta$ entering into the 1-irreducible
function $\Gamma = \langle \theta^{n}(x) \theta(x_{1})\cdots
\theta(x_{n_{\theta}}) \rangle_{\rm 1-ir}$
cannot exceed the number of the fields $\theta$
in the operator $\theta^{n}$ itself, i.e., $N_{\theta}\le n$
[cf. item (i) in Sec. \ref{sec:RGE}].
Therefore, the divergence can only exist in the functions
with $N_{\bf v}= N_{\theta'}=0$, and arbitrary value of
$n=N_{\theta}$, for which the formal index vanishes, $d_\Gamma =0$.
However, at least one of $N_{\theta}$ external ``tails''
of the field $\theta$ is attached to a vertex
$\btheta'({\bf v}\partt)\btheta$ (it is impossible to construct
nontrivial, superficially divergent diagram of the desired
type with all the external tails attached to the vertex $F$),
at least one derivative $\partial$ appears as an extra
factor in the diagram, and, consequently, the real index of
divergence $d_\Gamma'$ is necessarily negative.

This means that the operators $\theta^{n}$ require no counterterms
at all, i.e., they are in fact UV finite, $\theta^{n}=Z\,[\theta^{n}]^{R}$
with $Z=1$. It then follows that the critical dimension of
$\theta^{n}(x)$ is simply given by the expression (\ref{32B})
with no correction from $\gamma_{F}^{*}$ and is therefore reduced
to the sum of the critical dimensions of the factors:
$\Delta [\theta^{n}] = n\Delta[\theta] =n (-1+\eps/2)$.
Since the structure functions (\ref{struc}) are linear combinations of
pair correlation function involving the operators $\theta^{n}$,
this relation shows that they indeed satisfy the homogeneous
RG equation (\ref{RG1}), discussed in Sec. \ref{sec:RGE}.

In the OPE for the pair correlation function, analogous to Eq. (\ref{OR}),
the operators $\theta^{2}$ and $\theta_{i}\theta_{j}$ with the dimensions
$2(-1+\eps/2)$ give rise to constant terms. They correspond to the
solutions with $\delta(\p)$, discussed in the end of Sec.~\ref{sec:PAIR}.
Such terms, caused by various operators of the form $\theta^{n}$, are also
present in higher-order correlation functions. They disappear from the
structure functions (\ref{struc}), whose inertial-range behavior is
determined by operators built only of gradients (see Sec.~\ref{sec:OPE}).

\subsection{Scalar operators of the form $(\partial\theta)^{2}$
and the scaling of $S_{2}$} \label {sec:Operators2}

The leading terms of the inertial-range behavior of the second-order
structure function $S_{2}$ are determined by the critical dimensions
of the composite operators built of two gradients:
\begin{equation}
F_{1} = \partial_{i}\theta_{j}\partial_{i}\theta_{j}, \qquad
F_{2} = \partial_{i}\theta_{j}\partial_{j}\theta_{i}.
\label{F2}
\end{equation}
For the transverse field $\theta$, the second operator reduces to a
total derivative,
$F_{2} = \partial_{i} \partial_{j} (\theta_{j}\theta_{i})$,
and its dimension $\Delta_{2} = 2+2 \Delta_{\theta} = \eps$
does not appear on the right-hand side of Eq.~(\ref{OR}).

The dimension of the first operator is found exactly: $\Delta_{1} = 0$.
This can be demonstrated using the Schwinger equation of the form
\begin{equation}
\int{\cal D}\Phi\, \frac{\delta}{\delta\theta'_{i}(x)} \,
\biggl\{ \theta_{i}(x) \exp \bigl[ {\cal S}_{R}( \Phi) + A \Phi \bigr]
\biggr\}  =0
\label{Schwi}
\end{equation}
(in the general sense of the term, Schwinger equations are any relations
stating that any functional integral of a total variational derivative is
equal to zero, see, e.g., Refs. \cite{Collins,book3}). Here ${\cal S}_{R}$ is
the renormalized analog of the action (\ref{action}), and the notation
introduced in Eq. (\ref{field}) is used.
Equation (\ref{Schwi}) can be rewritten in the form
\begin{equation}
\langle\langle \theta' D_{\theta} \theta - \nabla_{t}[\theta^{2}/2]
-\partial_{i} [\theta_{i} {\cal P}] +
\nu Z_{\nu}\triangle[\theta^{2}/2]-\nu Z_{\nu} F_{1}\rangle\rangle
_{A}=-A_{\theta'} \delta W_{R}(A)/\delta A_{\theta}.
\label{Schwi2}
\end{equation}
Here $D_{\theta}$ is the correlation function (\ref{2}),
$\theta^{2} \equiv \theta_{i} \theta_{i}$,
$\langle\langle \dots\rangle\rangle _{A}$
denotes the averaging with the weight $ \exp [{\cal S}_{R}( \Phi) +
A \Phi]$, $W_{R}$ is determined by Eq. (\ref{field}) with the replacement
${\cal S}\to {\cal S}_{R}$, and the argument $x$ common to all the quantities
is omitted. The contribution with the pressure ${\cal P}$ from
Eq. (\ref{Poisson}) arises due to the fact that the differentiation in
Eq. (\ref{Schwi}) is performed with respect to a transverse field;
see the remark below Eq.~(\ref{action}).

The quantity $\langle\langle F\rangle\rangle _{A}$ is the
generating functional of the
correlation functions with one operator $F$ and any number of
fields $\Phi$, therefore the UV finiteness of the operator $F$ is
equivalent to the finiteness of the functional
$\langle\langle F\rangle\rangle _{A}$.
The quantity on the right-hand side of Eq. (\ref{Schwi2}) is
finite (a derivative of the renormalized functional
with respect to a finite argument), and so is the operator on the
left-hand side. Our operator $F_{1}$  does not admix in
renormalization to the operator $\theta' D_{\theta}\theta$
($F_{1}$ contains too many fields $\theta$), and to the
operators $\nabla_{t}[\theta^{2}/2]$, $\partial_{i} [\theta_{i} {\cal P}]$
and $\triangle[\theta^{2}/2]$
(they have the form of total derivatives, and $F_{1}$ does not
reduce to this form). On the other hand, the operators
$\theta' D_{\theta}\theta$ and $\partial_{i} [\theta_{i} {\cal P}]$
do not admix to $F_{1}$ (they are nonlocal, and $F_{1}$ is local),
while the derivatives
$\nabla_{t}[\theta^{2}/2]$ and $\triangle[\theta^{2}/2]$
do not admix to $F_{1}$ owing to the fact that each field
$\theta$ enters into the counterterms of the operators
$F_{n}$ only in the form of derivative $\partial\theta$
(see above). Therefore, all three types of operators entering into
the left-hand side of Eq. (\ref{Schwi2}) are independent,
and they must be UV finite separately.

Since the operator $\nu Z_{\nu} F_{1}$ is UV finite, it coincides
with its finite part, i.e., $\nu Z_{\nu}F_{1}=\nu F_{1}^{R}$,
which along with the relation $F_{1}=Z_{1}F_{1}^{R}$ gives
$Z_{1}=Z_{\nu}^{-1}$ and therefore $\gamma_{1}=-\gamma_{\nu}$.
For the critical exponent $\Delta_{1}=\eps + \gamma_{1}^{*}$
we then obtain $\Delta_{1}=0$ exactly (we recall that
$\gamma_{\nu}^{*}=\eps$; see the discussion below Eq.
(\ref{fixed}) in Sec.~\ref{sec:RGE}).

It then follows from (\ref{OR}) that the leading term of the inertial-range
behavior of the second-order structure function has the form
$S_{2} \propto D_{0}^{-1} \, r^{2-\eps}$,
in agreement with the solution $\gamma=2-\eps$ obtained in
Sec. \ref{sec:PAIR} from the exact equation. Therefore, this function is
not anomalous, like its analog for the scalar model
\cite{Kraich1,Falk1,GK}, and the anomalous scaling reveals itself
only on the level of the fourth-order structure function.

\subsection{Scalar operators of the form $(\partial\theta)^{4}$ and
the anomalous scaling of $S_{4}$} \label {sec:Operators4}

Let us turn to the scalar composite operators built of four gradients
$\partial\theta$, which cannot be reduced to the form of total derivatives.
This family includes six independent monomials,
all of which can be obtained from the fourth rank operator
$ \Phi^{mnps}_{ijkl} \equiv \partial_i\theta_m\,\partial_j\theta_n\,
\partial_k\theta_p\, \partial_l\theta_s $ by various contractions of
the tensor indices:
\begin{equation}
F_1=\Phi^{ijkl}_{jilk}, \quad   
F_2=\Phi^{iikk}_{jjll}, \quad   
F_3=\Phi^{ijkk}_{jill}, \quad   
F_4=\Phi^{iijk}_{jkll}, \quad   
F_5=\Phi^{iijj}_{klkl}, \quad   
F_6=\Phi^{iijk}_{jlkl}.         
\label{FF}
\end{equation}
At first glance, it seems that one can add another independent monomial,
$F_7=\Phi^{ijkl}_{lijk}$, but in fact it reduces to $F_1$ up to total
derivatives:
\begin{equation}
 3F_{1}-6F_{7} = \partial_{i} \Bigl[ -6 \theta_{k} \Phi^{spi}_{ksp}
+ 3 \theta_{k} \Phi^{pis}_{skp} +2 \theta_{i} \Phi^{kps}_{skp} \Bigr],
\label{FFR}
\end{equation}
where the notation is analogous to that in Eq. (\ref{FF});
see Ref.~\cite{Eight}.

Now let us turn to the calculation of the renormalization constants for the
family (\ref{FF}) in the one-loop approximation.
Let $\Gamma_{\alpha}(x;\theta)$ be the generating functional of the
1-irreducible Green functions with one composite operator $F_{\alpha}$
from Eq. (\ref{FF}) and any number of fields $\theta$. Here $x\equiv
(t,{\bf x})$ is the argument of the operator and $\theta$ is
the functional argument, the ``classical counterpart'' of the random
field $\theta$. We are interested in the fourth term of the
expansion of $\Gamma_{\alpha}(x;\theta)$ in $\theta$, which we denote
$\Gamma^{(4)}_{\alpha}(x;\theta)$. It has the form
\[ \Gamma^{(4)}_{\alpha}(x;\theta) =
\frac{1}{4!} \int dx_{1} \cdots \int dx_{4}\,
\theta_{i_{1}} (x_{1})\cdots\theta_{i_{4}}(x_{4})\,
\langle F_{\alpha}(x)
\theta_{i_{1}}(x_{1})\cdots\theta_{i_{4}}
(x_{4})\rangle_{\rm 1-ir}. \]
In the one-loop approximation this function is represented
diagrammatically as follows:
\begin{equation}
\Gamma^{(4)}_{\alpha}(x;\theta)= F_{\alpha} + \frac{1}{2} \,
\raisebox{-2.6ex}{\psfig{file=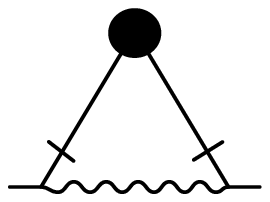,width=4em}}.
\label{Gamma2}
\end{equation}
Here the thin solid lines denote the {\it bare\,} propagator
$\langle\theta\theta'\rangle_{0}$ from Eq. (\ref{lines1}),
the ends with a slash correspond to the field $\theta'$, and the
ends without a slash correspond to $\theta$; the wavy line
denotes the velocity propagator (\ref{3}); the vertices correspond
to the factor (\ref{vertex}).
The first term is the ``tree'' approximation, and the black circle with
two attached lines in the diagram denotes the variational derivative
\begin{equation}
V_{i_{1} i_{2}}^{(\alpha)} (x;\, x_{1}, x_{2}) \equiv \delta^{2}
F_{\alpha}(x) / {\delta\theta_{i_{1}}(x_{1})\delta\theta_{i_{2}}(x_{2})}.
\label{BigV}
\end{equation}
The diagram is written analytically in the form
\begin{equation}
\int dx_{1} \cdots \int dx_{4}
V_{i_{1} i_{2}}^{(\alpha)} (x;\, x_{1}, x_{2})
\langle \theta_{i_{1}} (x_{1}) \theta'_{i_{3}} (x_{3}) \rangle_{0}
\langle \theta_{i_{2}}(x_{2}) \theta'_{i_{4}}(x_{4}) \rangle_{0}
\langle v_{i_{5}}(x_{3}) v_{i_{6}} (x_{4}) \rangle_{0}
\partial_{i_{5}} \theta_{i_{3}} (x_{3})
\partial_{i_{6}} \theta_{i_{4}} (x_{4}),
\label{Coor}
\end{equation}
with the bare propagators from Eqs. (\ref{3}) and (\ref{lines});
the derivatives appear from the ordinary vertices (\ref{vertex}).
It is convenient to represent the vertex (\ref{BigV}) in the form
\begin{equation}
V_{i_{1} i_{2}}^{(\alpha)}(x;\, x_{1}, x_{2}) =
\frac {\partial^{2} F_{\alpha} (a)} {\partial a_{i_{3} i_{4}}
\partial a_{i_{5} i_{6}}} \,
\partial_{i_{1}} \delta_{i_{3} i_{4}}
(x-x_{1})\, \partial_{i_{2}} \delta_{i_{5} i_{6}}(x-x_{2}),
\label{Vertex}
\end{equation}
where
\[ \delta_{ij} (x-x') \equiv \delta(t-t') P_{ij} ({\bf x}-{\bf x'}) =
\delta(t-t') \int {\cal D} {\bf k} P_{ij} ({\bf k})
\exp [{\rm i}{\bf k} \cdot ({\bf x}-{\bf x'})] \]
is the delta function on the transverse subspace. The first (combinatorial)
factor in Eq. (\ref{Vertex}) is understood as follows: the gradients
$\partial_{i}\theta_{j}(x)$ in the operator $F_{\alpha}$ are replaced
with a constant tensor $a_{ij}$, the differentiation is performed with
respect to its elements, and {\it after the differentiation} they are
replaced back with the gradients, $a_{ij}\to\partial_{i}\theta_{j}(x)$.

Using the identity $\partial_{k} \delta_{ij} (x-x') =
- \partial_{k}' \delta_{ij} (x-x')$ and the integration by parts,
the derivatives can be moved from the vertex onto the propagators,
and the integrations with respect to $x_{1}$ and $x_{2}$ are then
easily performed:
\begin{equation}
\frac {\partial^{2} F_{\alpha}(a)} {
\partial a_{i_{7} i_{1}}\partial a_{i_{8} i_{2}}} \,
\int dx_{3} \int dx_{4} \,
\langle \partial_{i_{7}}
\theta_{i_{1}} (x) \theta'_{i_{3}} (x_{3}) \rangle_{0} \,
\langle \partial_{i_{8}}
\theta_{i_{2}}(x) \theta'_{i_{4}}(x_{4}) \rangle_{0} \,
\langle v_{i_{5}}(x_{3}) v_{i_{6}} (x_{4}) \rangle_{0}  \,
\partial_{i_{5}} \theta_{i_{3}} (x_{3}) \,
\partial_{i_{6}} \theta_{i_{4}} (x_{4}).
\label{Coor_2}
\end{equation}

In order to find the renormalization constants, we need not the entire
exact expression (\ref{Coor_2}), rather we need its UV divergent part.
The latter is proportional to a polynomial built of four factors
$\partial\theta$ at a single spacetime point $x$. The needed four
gradients have already been factored out from the expression (\ref{Coor}):
two factors from the vertex (\ref{Vertex}) and two factors
from the ordinary vertices (\ref{vertex}). Therefore, we can neglect the
spacetime inhomogeneity of the gradients and replace them with their
values at the point $x$. Expression (\ref{Coor_2}) can therefore be written,
up to an UV finite part, in the form
\begin{equation}
\frac {\partial^{2} F_{\alpha}(a)} {
\partial a_{i_{7} i_{1}}\partial a_{i_{8} i_{2}}} \,
\partial_{i_{5}} \theta_{i_{3}} (x) \,
\partial_{i_{6}} \theta_{i_{4}} (x) \,
X_{i_{7}i_{1}i_{3}i_{8}i_{2}i_{4}i_{5}i_{6}},
\label{Coor2}
\end{equation}
where we have denoted
\begin{equation}
X_{i_{7}i_{1}i_{3}i_{8}i_{2}i_{4}i_{5}i_{6}} \equiv
\int dx_{3} \int dx_{4} \,
\langle \partial_{i_{7}}
\theta_{i_{1}} (x) \theta'_{i_{3}} (x_{3}) \rangle_{0} \,
\langle \partial_{i_{8}}
\theta_{i_{2}}(x) \theta'_{i_{4}}(x_{4}) \rangle_{0} \,
\langle v_{i_{5}}(x_{3}) v_{i_{6}} (x_{4}) \rangle_{0}  \,
\label{X}
\end{equation}
or, in the momentum-frequency representation, after the integration over
the frequency,
\begin{equation}
X_{i_{7}i_{1}i_{3}i_{8}i_{2}i_{4}i_{5}i_{6}}  =  \frac{D_{0}}{2\nu_0} \,
\int \frac{{\cal D}{\bf k}}{k^{d+\eps}} \,
P_{i_{1}i_{3}}  ({\bf k}) \,
P_{i_{2}i_{4}}  ({\bf k}) \,
P_{i_{5}i_{6}}  ({\bf k}) \,
\frac{k_{i_{7}} k_{i_{8}} }{k^{2}} ,
\label{X2}
\end{equation}
with $D_{0}$ from Eq. (\ref{3}). Using the isotropy relations
\begin{equation}
\int {\cal D} {\bf k} \, f(k) \,  \frac {k_{i_{1}} \cdots k_{i_{2n}} }
{k^{2n}} = \frac {\delta_{i_{1}i_{2}}\delta_{i_{3}i_{4}}\cdots
\delta_{i_{2n-1}i_{2n}}+ {\rm all\ possible\ permutations}}
{d(d+2)\cdots(d+2n-2)} \, \int {\cal D} {\bf k} \, f(k)
\label{isotropy}
\end{equation}
the integral (\ref{X2}) can be reduced to the simple scalar integral
\begin{equation}
\int \frac{{\cal D}{\bf k}}{k^{d+\eps}} = C_{d}\,\frac{m^{-\eps}}{\eps},
\label{logint}
\end{equation}
with $C_{d}$ from Eq.~(\ref{Dyson103}); the parameter $m$ has arisen from
the lower limit in the integral over $\k$. The explicit answer for the
quantity (\ref{X2}) is given in Appendix~B.

Contraction of the delta symbols with the first factors in Eq. (\ref{Coor2})
gives rise to various monomials built of four gradients of $\theta$; up to
total derivatives, they reduce to the operators from the family~(\ref{FF}).
Then the function $\Gamma^{(4)}_{\alpha}(x;\theta)$ from Eq. (\ref{Gamma2})
in the one-loop approximation of the renormalized perturbation theory
(i.e., to the first order in $g$) up to an UV finite part can be written in
the form
\begin{equation}
\Gamma^{(4)}_{\alpha}(x;\theta)= F_{\alpha} +
\frac{gC_{d}}{\eps} \,(\mu/m)^{\eps}\,  R_{\alpha},
\label{Gamma3}
\end{equation}
where $\mu$ has appeared from the relation
$D_{0} = g_{0} \nu_{0} = g\nu\mu^{\eps}$ and
\begin{equation}
 R_{\alpha} = \sum_{\beta} A_{\alpha\beta}  F_{\beta}
\label{GammaR}
\end{equation}
are linear combinations of the monomials (\ref{FF}) with the coefficients
$A_{\alpha\beta}$ dependent only on $d$:
\begin{eqnarray}
R_{1} &=&
- \frac { (3d^2+22d+36) F_{1}} {d(d+2)(d+4)(d+6)}
+\frac {2(d+3)F_{2}}{ d(d+2)(d+4)(d+6)}
+\frac {6 F_{3}} {d(d+2)(d+6)}
\nonumber
\\&&
+\frac {2 (d+5) F_{4}}{(d+2)(d+6)}
+ \frac {4(d+3)F_{5}}{d(d+2)(d+4) (d+6)}
-\frac { 4F_{6}}{ (d+2)(d+6)},
\nonumber
\\
R_{2}&=&
- \frac {12 F_{1}}{ d (d+2)(d+4)(d+6)}
+ \frac { ({d}^{4}+ 10{d}^{3}+19{d}^{2} -44 d-90) F_{2}}
{d (d+2)(d+4)(d+6)}
+ \frac {(d+12)F_{3}}{ d(d+2) (d+6)}
\nonumber
\\&&
+ \frac { 12F_{4}} {d(d+2)(d+6)}
+\frac {2 ({d}^{3}+8{d}^{2}+10d-18)F_{5}}
{d( d+2 )(d+4)(d+6)}
-\frac {4 F_{6}} {(d+2)(d+6)},
\nonumber
\\
R_{3}&=&
 \frac { ({d}^{2}+18d+48)F_{1}} {2d (d+ 2 ) (d+4 )(d+6)}
+\frac {2 (2d+9 )F_{2}} {d (d+2)(d+4)(d+6 )}
+\frac {({d}^{2}+4 d-15)F_{3}} {2d(d+6)}
\nonumber
\\&&
-\frac {2F_{4}}{ (d+2)(d+6)}
-\frac {2({d}^{2}+6d+6 )F_{{5}}}{ d(d+2) (d+ 4)(d+6 )}
+ \frac {2(d+4)F_{6}}{(d+2)(d+6)},
\nonumber
 \\
R_{{4}}&=&
\frac {2 (d+3)F_{{1}}}{ d(d+2)(d+4)(d+6)}
+ \frac { (d+3)({d}^{2}+8d+14) F_{{2}}}
{2d (d+2 )(d+4)(d+6)}
+\frac { 3F_{{3}}} {d (d+2)(d+6)}
\nonumber
\\&&
+ \frac { ({d}^{3}+5{d}^{2}-14d-36 ) F_{4}} {2d (d+2)(d+6)}
-\frac {2 (d+3) (d+5) F_{5}} {d(d+2)(d+4)(d+6)}
-\frac {2F_{{6}} }{  (d+2)(d+6)},
 \nonumber
 \\
R_{5}&=&
-\frac {6F_{1}} {d(d+2)(d+4)(d+6)}
+\frac { ({d}^{3}+10{d}^{2}+ 27d+15)F_{2}}{d(d+2)(d+4)(d+6)}
+\frac {3F_{3}} {d (d+2) (d+6 )}
 \nonumber
 \\&&
-\frac { 2(d+3)F_{4}} {d (d+2) (d+6 )}
+ \frac {({d}^{4}+10{d}^{3}+18{d}^{2}-54d-114) F_{5}}{d(d+2)(d+4)(d+6)}
+\frac {12 F_{6}} {d (d+2) (d+6 )},
 \nonumber
 \\
R_{6}&=&
\frac {({d}^{2}+18d+48)F_{1}} {4d(d+2)(d+4)(d+6)}
+\frac { (2d+9)F_{2}}{d(d+2)(d+4)(d+6)}
+\frac {({d}^ {2}+6d+6) F_{3}} {2d (d+2)(d+6)}
 \nonumber
 \\&&
-\frac {F_{4}}{ (d+2)(d+6)}
-\frac {({d}^{2}+6d+6)F_{5}} {d(d+2)(d+4)(d+6)}
+\frac {({d}^{3}+6{d}^{2}-11 d-42)F_{6}}{2d (d+2)(d+6)}.
\label{Rs}
\end{eqnarray}

\begin{figure}
\centerline{\psfig{file=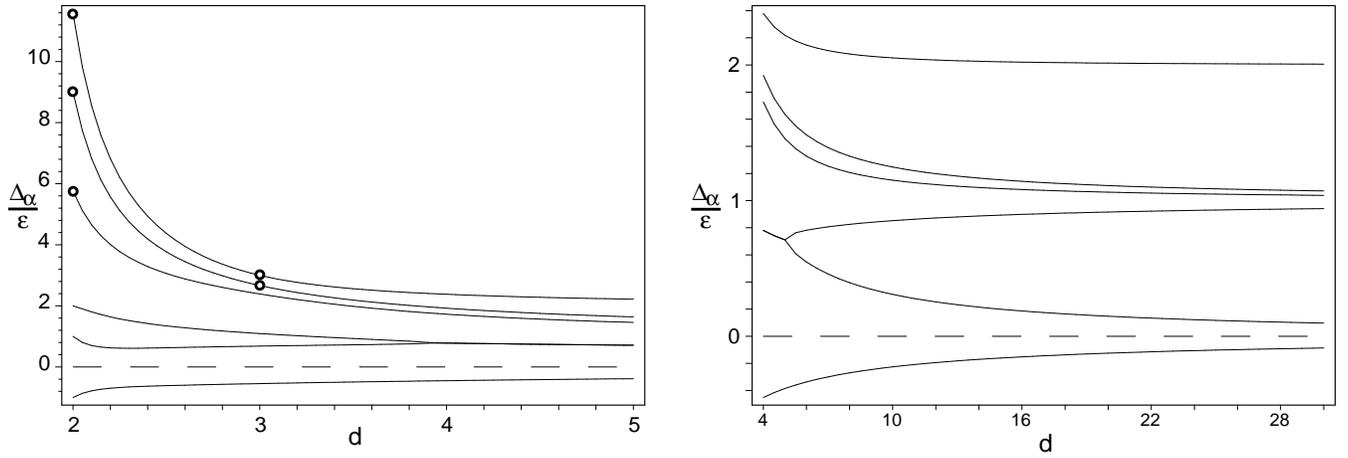,width=\textwidth}}
\caption{Critical dimensions $\Delta_{1}$--$\Delta_{6}$ (from below to above)
of the operators (\protect\ref{FF}) in the order $O(\varepsilon)$ {\em vs}
the space dimensionality $d$ for $2\le d\le5$ (left) and $4\le d\le30$
(right). The empty circles denote the operators which become trivial in
$d=2$ and 3. The dimensions tend to 0 ($\Delta_{1,2}$), $\eps$
($\Delta_{3}$--$\Delta_{5}$) and $2\eps$ ($\Delta_{6}$) for $d\to\infty$.}
\label{Eigenvalues}
\end{figure}

The constants $Z_{\alpha\beta}$ are found from the requirement that
the functions (\ref{Gamma2}) for the renormalized analogs of
operators (\ref{FF}), defined by the relation
$F_{\alpha}=Z_{\alpha\beta}F_{\alpha}^{R}$, be UV finite, i.e., be finite
for $\eps\to0$. In the MS scheme this gives:
\begin{equation}
 Z_{\alpha\beta} = \delta _{\alpha\beta} + gC_{d}\,
A_{\alpha\beta}/\eps +O(g^{2})
\label{Delta_44}
\end{equation}
with the coefficients $A_{\alpha\beta}$ from (\ref{Rs}).
For the matrix of anomalous dimensions (\ref{2.2})
at the fixed point (\ref{fixed}) one has
$\gamma_{\alpha\beta}^{*} = -g_{*}C_{d} A_{\alpha\beta} +O(\eps^{2})$,
and for the matrix of critical dimensions from Table~\ref{table1}
and Eq. (\ref{32B}) one obtains
\begin{equation}
\Delta_{\alpha\beta} = 2\eps\, \delta_{\alpha\beta}
+ \gamma_{\alpha\beta}^{*}.
\label{Delta_4}
\end{equation}
Critical dimensions associated with the family of operators (\ref{FF}) are
given by the eigenvalues $\Delta_{\alpha}$ of the matrix (\ref{Delta_4}).
In particular, for $d=3$ one obtains:
\begin{equation}
\Delta_1 \approx -0.55 \varepsilon, \quad
\Delta_2 \approx 0.68 \varepsilon, \quad
\Delta_3 \approx 1.1 \varepsilon, \quad
\Delta_4 \approx 2.4 \varepsilon, \quad
\Delta_5=8 \varepsilon/3, \quad
\Delta_6=3 \varepsilon .
\label{eigenvalues}
\end{equation}
For general $d$, only one of the eigenvalues is found analytically:
\begin{equation}
\Delta_5=\frac{(d+1)^2}{(d^2-3)}\,\eps
\label{anal}
\end{equation}
(we recall that all these eigenvalues have corrections of order $O(\eps^{2})$
and higher). The critical dimensions $\Delta_{\alpha}$ as functions of $d$
are presented in Fig.~\ref{Eigenvalues}. They are always real, except for
the pair $\Delta_{2,3}$ which becomes complex conjugate in the interval
$4<d<5$ (in Fig.~\ref{Eigenvalues} the real part is shown). One can also
see that for all $d$, exactly one of the dimensions, denoted by
$\Delta_1$ in Eq. (\ref{eigenvalues}), is negative, and the others are
positive. Existence of a negative dimension implies that the fourth-order
structure function in model (\ref{1})--(\ref{3}) exhibits the
inertial-range anomalous scaling; for small $mr$ it has the form
\begin{equation}
S_{4}({\bf r})= D_{0}^{-2}\, r^{4-2\eps} \,
\sum_{\alpha} A_{\alpha} (mr)^{\Delta_{\alpha}} + \dots,
\label{OR4}
\end{equation}
see Eq.~(\ref{OR}).
The dots stand for the corrections of the form $(mr)^{2+O(\eps)}$
and higher, which arise from the operators including more derivatives
than fields, and possible anisotropic contributions, related to
nonscalar operators. The leading term, singular for $mr\to0$, is
determined by the negative dimension $\Delta_1$.

The dimensions diverge for $d\to\sqrt{3}$ as a result of the
divergence of $g_{*}$ in Eq.~(\ref{fixed}). For $d\to\infty$,
they simplify and form three groups which tend to 0, $\eps$
and $2\eps$. More precisely, in the first group there are two
dimensions with the behavior $\Delta = \pm2\sqrt{2}\eps/d+O(1/d^{2})$.
This means that for $d=\infty$, the anomalous scaling of $S_{4}$
vanishes: the phenomenon known for the scalar Kraichnan model
\cite{Infty} and questioned for the NS turbulence \cite{FFR,AR}.
In the next Section we shall see that the simplification of the
exponents and vanishing of the anomalous scaling for $d\to\infty$
also holds for the higher structure functions.

Although the above expressions for the eigenvalues are well-defined for
any $d>\sqrt{3}$, low integer dimensions require special care because
of additional linear relations between the operators. For $d=3$, there
are two such relations:
$\Phi_5 -  \Phi_1/2 - \Phi_2 + 2\Phi_4=0$ and $ \Phi_6 - \Phi_3/2=0$.
For $d=2$, one more relation arises: $\Phi_4 - \Phi_1/2=0$.
(It is also noteworthy that the derivative on the right-hand side of
Eq. (\ref{FFR}) in two and three dimensions vanishes identically.)

Using these relations, one can check that two basis operators (\ref{2.5})
for $d=3$ and three basis operators for $d=2$ vanish. Therefore, the
corresponding eigenvalues are in fact meaningless and should be discarded
(in particular, this happens with the dimension (\ref{anal})). The
eigenvalues that survive for $d=2$ and 3 belong to the three and four
{\it lowest} branches in Fig.~\ref{Eigenvalues}, respectively (they are
denoted by the thick dots). In particular, the dangerous operator remains
nontrivial, so that our model exhibits the anomalous scaling also in two
and three dimensions. We recall that similar behavior was demonstrated by
the zero-mode solutions: all $d$-dimensional expressions for the exponents
have well-defined limits for $d\to2$, but a part of them becomes in fact
spurious owing to the vanishing of the corresponding amplitudes;
see the discussion in the end of Sec.~\ref{sec:Kartina}.

\subsection{Scalar operators of the form $(\partial\theta)^{2n}$ and
the anomalous scaling of $S_{2n}$} \label {sec:Operators6}

The leading terms of the inertial-range behavior of a higher-order structure
function $S_{2n}$ are related to the scalar composite operators of the form
$(\partial\theta)^{2k}$ with $0<k \le n$. In the previous section we have
established the anomalous scaling behavior of the fourth-order structure
function, as a result of the existence of dangerous operator with $k=2$
in the corresponding OPE. Then the probabilistic inequalities allow one to
show that all the higher-order structure functions are also anomalous, the
leading term of the inertial-range behavior for the function $S_{2n}$ is
given by an operator with $k=n$, the number of dangerous operators is
necessarily infinite, and the spectrum of their dimensions is not restricted
from below.

Let $\Delta_{n}$ be the dimension of the operator that gives the leading
contribution to the OPE for the function $S_{2n}$, so that
$S_{2n} \propto D_{0}^{-n} r^{n(2-\eps)} (mr)^{\Delta_{n}}$.
It is well known in the probability theory that
$|\langle x^{n} \rangle|^{1/n}$ is a nondecreasing function of $n$
for any random variable $x$. Taking
$x= [ \theta_{r} (t,{\bf x}) - \theta_{r} (t,{\bf x'})]^{2}$ we find that
$S_{2n}^{1/n}\propto D_{0}^{-1} r^{(2-\eps)} (mr)^{\Delta_{n}/n}$ is a
nondecreasing function of $n$, and so is the ratio $|\Delta_{n}|/n$
[we recall that $\Delta_{n}$ is negative and $(mr)$ is small].
This proves all the above statements.

In principle, calculation of the critical dimensions related to the
family $(\partial\theta)^{2n}$ for any given $n$ is a purely technical
problem, and the formulas (\ref{Coor})--(\ref{X2}) remain valid in the
general case with obvious alterations. In practice, however, this problem
appears very cumbersome, in particular, because the number of relevant
operators increases rapidly with $n$. The situation simplifies for large
$d$, and below we restrict ourselves with the zeroth and first terms of
the $1/d$ expansion. To avoid possible misunderstandings, it should be
stressed that we deal with the $1/d$ expansion of a critical dimension
$\Delta$ in its $O(\eps)$ approximation, that is, the $1/d$ expansion
of the coefficient $\Delta^{(1)} (d)$ in the representation
$\Delta = \eps \Delta^{(1)} (d) + O(\eps^{2})$.

Despite this simplification, no explicit analytical result is
available for general $n$. Below we only present the results for
the critical dimensions of the families $(\partial\theta)^{2n}$
with $n\le 6$; the detailed derivation is given in Appendix~B.

The first two terms of the $1/d$ expansion for such operators have the form
$\Delta^{(1)} (d) = 2k + \Delta^{(11)} /d  + O(1/d^{2})$, where
$k=0,1, \dots, n$ and $\Delta^{(11)}$ are numerical coefficients independent
of the parameters $\eps$ and $d$. It is clear that for large $d$, dangerous
operators can only be present in the subset with $k=0$.
As already mentioned in Sec. \ref{sec:Operators4},
in the family with $n=2$ there are two such operators with
\begin{mathletters}
\label{d}
\begin{equation}
\Delta^{(11)}= \quad  \pm 2\sqrt{2} .
\label{d2}
\end{equation}
In the family with $n=3$ there are three such operators with
\begin{equation}
 \Delta^{(11)}= \quad  -9.674, \quad -0.973, \quad 7.647.
\label{d3}
\end{equation}
In the family with $n=4$ there are five such operators with
\begin{equation}
 \Delta^{(11)}= \quad -20.617, \quad -7.783, \quad -1.018,
\quad 3.883, \quad 14.534.
\label{d4}
\end{equation}
In the family with $n=5$ there are seven such operators with
\begin{equation}
 \Delta^{(11)}=  \quad  -35.589, \quad -18.660, \quad -8.700, \quad
-2.960, \quad  2.780, \quad 10.674, \quad 23.455.
\label{d5}
\end{equation}
In the family with $n=6$ there are eleven such operators with
\begin{eqnarray}
\Delta^{(11)}=  \quad  -54.572, \quad -33.612, \quad -19.554, \quad
-13.834, \quad  -12.815, \quad -4.908,
\nonumber
\end{eqnarray}
\begin{eqnarray}
3.839, \quad 4.828, \quad 9.681, \quad 19.552, \quad 34.395.
\label{d6}
\end{eqnarray}
\end{mathletters}

\begin{figure}
\centerline{\psfig{file=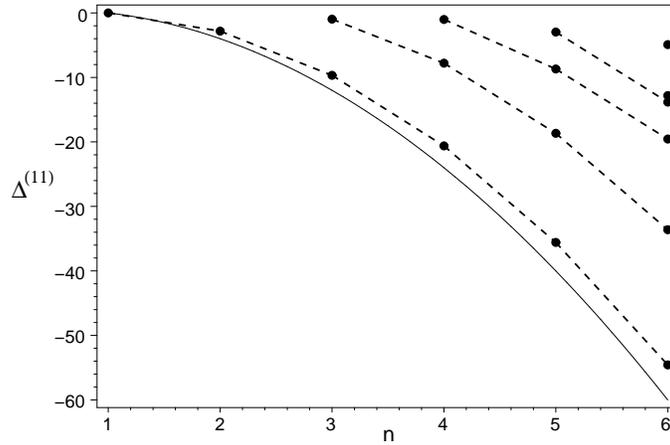,width=0.5\textwidth}}
\caption{Coefficients $\Delta^{(11)}$ in the $O(\eps/d)$ approximation
of the critical dimensions of the operators $(\partial\theta)^{2n}$
in the scalar (solid curve) and vector (thick dots) models. Dashed
lines denote monotonous branches of the critical dimensions in the
vector case.}
\label{Fig4}
\end{figure}

These results confirm and illustrate the general picture outlined above:
in the set of operators with $k\le n$, the most dangerous operator
(that is, the operator with the lowest negative dimension) belongs to the
subset with $k=n$, and its dimension $\Delta_{n}<0$ decreases faster than
linearly with $n$. The results (\ref{d}) are illustrated by Fig.~\ref{Fig4}
(only the negative dimensions are shown). It suggests that the dimensions
form a set of monotonous branches, denoted by dashed lines. The solid
curve corresponds to the well-known expression for the scalar Kraichnan
model \cite{Falk1,GK} in the same $O(\eps/d)$ approximation:
$\Delta_{n} = -2n(n-1)\, \eps/d$. For all $n$, it lies below the lowest
vector branch (the scaling in the scalar model appears ``more anomalous'');
the deviation between the scalar and vector cases becomes stronger as $n$
increases, although the {\it ratio\,} of the dimensions approaches unity.

\subsection{Tensor operators and the scaling of $S_{2}$ in anisotropic
sectors of arbitrarily high orders } \label {sec:Ashells}

In this Section, we apply the RG and OPE approach to the higher anisotropic
sectors of the model (\ref{1})--(\ref{3}). We shall concentrate on the
second-order structure function, for which nonperturbative results can, in
principle, be derived for arbitrarily high values of the parameter $l$
from the exact Dyson--Wyld equation (see Sec.~\ref{sec:Kartina}).
This allows one to identify the solutions of the zero-mode equations,
discussed in Sec. \ref{sec:PAIR}, with definite composite operators and,
in principle, to calculate the corresponding amplitude factors.
Using the OPE technique, we derive explicit analytical expressions
for the leading exponents in all anisotropic sectors to the order
$O(\eps)$ in $d$ dimensions. Furthermore, we present
additional nontrivial exponents which do not appear in the
inertial-range behavior of the model (\ref{1})--(\ref{3}), but will be
activated (and can determine {\it leading} terms in anisotropic sectors!)
if the anisotropy is introduced by the velocity field (like in Ref.
\cite{Novikov}) and not only by the large-scale forcing.

The analysis of the anisotropic sectors for higher-order functions using
the OPE is extremely cumbersome but, in a sense, purely technical problem;
we shall briefly discuss it in Sec.~\ref{sec:hi-fi}.

According to the general rules (see Sec.~\ref{sec:OPE}), the leading terms
in the sector $l=2$ are determined by the second-rank operators built of
two gradients; up to derivatives, there are two such operators:
\begin{equation}
F_{1}={\cal IRP}\,\Bigl[
\partial_{i} \theta_{k} \partial_{j} \theta_{k}\Bigr] , \qquad F_{2}=
{\cal IRP}\,\Bigl[ \partial_{k} \theta_{i} \partial_{k} \theta_{j} \Bigr],
\label{tensors2}
\end{equation}
where ${\cal IRP}$ denotes the irreducible part; cf. Eq. (\ref{tensorN})
in Sec.~\ref{sec:Kartina} (here and below, we use $F_{\alpha}$ and
$\Delta_{\alpha}$ to denote different operators and their dimensions).

We omit the one-loop calculation of the corresponding $2\times2$ matrix of
critical dimensions, which is similar to the calculation discussed in
Sec. \ref{sec:Operators4}  for scalar operators, and present only its
eigenvalues, i.e., the critical dimensions of operators  (\ref{tensors2}):
\begin{equation}
\Delta_{1,2}=\varepsilon \,\frac {d^{3}+5d^2+2d -8 \pm \sqrt { (d+4)
(d^5+2d^4-7d^3-4d^2+8d+16 )}}
{2(d^2-3)(d+4)} +O(\varepsilon^{2}) .
\label{dim2}
\end{equation}
In representation (\ref{OR}), they give rise to exponents
$2-\eps+\Delta_{1,2}$, which agree with the special case $l=2$
of expressions (\ref{slopes3}), (\ref{gam2}), obtained in
Sec.~\ref{sec:Kartina} on the basis of the Dyson--Wyld equation.

In two dimensions, the transverse vector field can be represented in the form
\begin{eqnarray}
\theta_{i} = \epsilon_{ik} \partial_{k} \psi,
\label{Civita}
\end{eqnarray}
where $\epsilon_{ik}$ is the antisymmetric Levi-Civita pseudotensor and
$\psi(x)$ is
some scalar function (stream function). Using the well known identity
$\epsilon_{ik}\epsilon_{js} = \delta_{ij}\delta_{ks} -
\delta_{is}\delta_{kj}$ one can easily check that for $d=2$,
the operators (\ref{tensors2}) coincide. Only one of the dimensions
(\ref{dim2}), namely, $\Delta_{2}= \eps$, corresponds to a nontrivial
basis operator (\ref{2.5}) and remains meaningful, while the other,
$\Delta_{1}= 3\eps$, corresponds to a vanishing basis operator and
should be discarded; cf. Fig.~\ref{EXPO2} for $d=l=2$.

The leading terms in the sector $l=4$ are determined by the fourth-rank
operators, obtained from the monomial $F_{ijkn} \equiv
{\cal IRP}\,\bigl[\partial_{i} \theta_{j} \partial_{k} \theta_{n}\bigr]$
by all possible permutations of its tensor indices. It turns out that there
are only {\it three} different critical dimensions, associated with these
operators. The corresponding basis operators (\ref{2.5}) possess
different symmetries and therefore can be written down without calculation
of diagrams. One of them is the fully symmetrized operator,
\[F_{S} = F_{ijkn}+F_{jink}+F_{ikjn}+F_{kinj}+F_{jikn}+F_{iknj}. \]
The others can be constructed as follows. The monomials can be split
into three groups of two operators each,
\[ F_{1} = F_{ikjn}, \qquad  F_{2} = F_{kinj}, \]
symmetric with respect to the simultaneous exchange of the indices within
the pairs $\{ij\}$ and $\{kn\}$,
\[ F_{3} = F_{ijkn}, \qquad  F_{4} = F_{jink}, \]
symmetric with respect to the exchange of the indices within the pairs
$\{ik\}$ and $\{jn\}$,
\[ F_{5} = F_{jikn}, \qquad  F_{6} = F_{iknj}, \]
symmetric with respect to the exchange within the pairs
$\{in\}$ and $\{jk\}$.

The operators can mix in renormalization only within the groups with the same
symmetry, so that the set $F_{1}$, $F_{2}$, $F_{S}$ is closed with respect
to the renormalization, and so are the sets $F_{3}$, $F_{4}$, $F_{S}$ and
$F_{5}$, $F_{6}$, $F_{S}$. In the first set, the basis operators (\ref{2.5})
are $F_{S}$ (it is fully symmetric and no other operators can admix to it),
$F_{1}-F_{2}$ (it is antisymmetric with respect to the exchange of the
indices within the pairs $\{ik\}$ and $\{nj\}$), and $F_{S}-3F_{1}-3F_{2}$.
The latter operator is symmetric with respect to the exchange of the indices
within the pairs $\{ik\}$ and $\{nj\}$, and for this reason it cannot mix
with the second basis operator; it is not fully symmetric and cannot admix
to $F_{S}$. It remains to note that the contraction of the operators
$F_{1}-F_{2}$ and $F_{S}-3F_{1}-3F_{2}$ with any constant vector with
respect to all four indices gives zero, while analogous contraction
of $F_{S}$ remains nontrivial. This means that $F_{S}$ cannot admix
to those operators in renormalization, and the above construction
indeed gives three independent basis elements of the type (\ref{2.5}).

The explicit one-loop calculation confirms this conclusion and gives three
different critical dimensions, corresponding to the basis operators
$F_{S}$, $F_{S}-3F_{1}-3F_{2}$ and $F_{1}-F_{2}$, respectively:
\begin{mathletters}
\label{tri}
\begin{equation}
\Delta_{1} = \frac {(d+2)(d^2+4d-9)}{(d^2-3)(d+6)}\, \varepsilon,
\label{tri1}
\end{equation}
\begin{equation}
\Delta_{2} = \frac {(d^2-1)}{(d^2-3)}\, \varepsilon,
\label{tri2}
\end{equation}
\begin{equation}
\Delta_{3} = \frac {(d^3+4d^2-d-8)}{(d^2-3) (d+4)} \,\varepsilon,
\label{tri3}
\end{equation}
\end{mathletters}
with corrections of order $O(\varepsilon^{2})$ and higher.

The remaining independent basis operators can be obtained by permutations
of the indices and can be chosen in the form
$F_{3}-F_{4}$, $F_{5}-F_{6}$  and $F_{S}-3F_{3}-3F_{4}$.
At first glance, it seems that one can add another independent element,
$F_{S}-3F_{5}-3F_{6}$, but in fact it is equal to the sum of the
operators $F_{S}-3F_{1}-3F_{2}$ and $F_{S}-3F_{3}-3F_{4}$
up to the minus sign. Therefore the dimension $\Delta_{1}$ in
Eq. (\ref{tri}) is unique, $\Delta_{2}$ has two-fold degeneracy and
$\Delta_{3}$ has three-fold degeneracy.

Although all three dimensions in Eq. (\ref{tri}) make sense, only
$\Delta_{1}$ appears on the right-hand side of expansion (\ref{OR}).
Indeed, both the diagrammatic analysis and dimensional considerations
show that the coefficients in Eq. (\ref{OPE}), corresponding to operators
$F_{ijkn}$, do not involve the function $\langle\theta\theta\rangle_0$
from Eq. (\ref{lines2}). Therefore, they do not depend on the vector
${\bf n}$ and their tensor indices are carried by the Kronecker delta
symbols or vectors ${\bf r}$. The contraction of any {\it irreducible}
operator with the delta symbols always gives zero; the contraction with
the components of a single vector ${\bf r}$ in expansion (\ref{OPE})
``kills'' the basis operators of the form $F_{S}-3F_{1}-3F_{2}$ and
$F_{1}-F_{2}$ (see above).

Thus the only contribution to expansion (\ref{OR}) comes from the
operator $F_{S}$, and the expression $2-\eps+\Delta_{1}$ should be
identified with the leading exponent $\gamma_{4}$ from Eq. (\ref{corre3})
in Sec. \ref{sec:PAIR}; they indeed agree in the order $O(\eps)$.

The remaining dimensions $\Delta_{2,3}$ will be activated (and can
determine {\it leading} terms on the right-hand side of Eq. (\ref{OR})
in the $l=4$ sector) if the vector ${\bf n}$ appears in the corresponding
coefficients $C_{F}$ in expansion (\ref{OPE}). This can happen
if the anisotropy is introduced by the velocity field (like in
Ref. \cite{Novikov}) and not only by the large-scale forcing.
However, it should be noted that in such a case the dimensions
become nonuniversal through their dependence on the anisotropy
parameters, and the expressions like (\ref{tri}) give only the
zeroth order (isotropic) approximations; cf. Ref. \cite{Novikov}
for the scalar case.

Now let us turn to general $l$. The relevant operators are the
$l\,$th rank tensors built of two fields $\theta$ and minimal possible
number of derivatives:
\begin{mathletters}
\label{sorok}
\begin{equation}
{\cal IRP}\, \bigl[\theta_{i_{1}} \partial_{i_{2}} \cdots
 \partial_{i_{l-1}} \theta_{i_{l}} \bigr].
\label{sorok1}
\end{equation}
Up to total derivatives, monomials (\ref{sorok1}) are invariant with respect
to the shift $\theta\to\theta+{\rm const}$, so that their dimensions can
appear in expansion (\ref{OR}). In the above form, all symmetries of
operators (\ref{sorok1}) are obvious: they are symmetric with respect to the
permutation of the indices $\{i_{1},i_{l}\}$ and any permutations within the
subset $\{i_{2} \cdots i_{l-1}\}$. Thus the total number of {\it different}
monomials, obtained from (\ref{sorok1}) by permutations of the indices,
equals $l(l-1)/2$. However, there are only three different dimensions
related to them. Indeed, the counterterm to the monomial (\ref{sorok1})
necessarily has the same symmetries and therefore can include, along with
(\ref{sorok1}) itself, two more structures:
\begin{equation}
Sym\, {\cal IRP}\, \bigl[\theta_{i_{2}} \partial_{i_{1}}\partial_{i_{3}}
\cdots  \partial_{i_{l-1}} \theta_{i_{l}} \bigr], \qquad
Sym\,{\cal IRP}\, \bigl[\theta_{i_{2}} \partial_{i_{1}}\partial_{i_{2}}
\partial_{i_{4}} \cdots
 \partial_{i_{l-1}} \theta_{i_{3}} \bigr],
\label{sorok23}
\end{equation}
\end{mathletters}
where {\it Sym} denotes the symmetrization with respect to the permutation
of the pair $\{i_{1},i_{l}\}$ and any permutations of the indices
$\{i_{2} \cdots i_{l-1}\}$. Thus the set of three operators (\ref{sorok})
is closed with respect to the renormalization, the
corresponding basis operators (\ref{2.5}) are their linear combinations
and determine three different dimensions $\Delta_{1,2,3}$;
all the other basis operators are obtained by permutations
of the indices and give rise only to the same dimensions.

For general $l$, even the one-loop calculation is rather difficult
because individual contributions in the counterterms to the polynomials
(\ref{sorok}) contain powerlike UV divergencies in addition to
logarithmic ones. In contrast with the calculation discussed
in Sec. \ref{sec:Operators4}, one cannot neglect the spacetime inhomogeneity
of the fields $\theta$ in the diagram; in other words, one cannot
neglect the dependence of its integrand on the external
momenta ${\bf p}$ and should expand the integrand up to the terms of
order $p^{2l-2}$. This expansion gives rise to the terms that
diverge for $\Lambda\to\infty$ as some positive powers of the UV
cut-off $\Lambda$. However, all such terms contain delta symbols and cancel
out when all contributions in irreducible structures (\ref{sorok})
are taken into account; the result is finite at $\Lambda\to\infty$,
contains a first-order pole in $\eps$, and reduces to a linear combination
of the three structures (\ref{sorok1}) and (\ref{sorok23}).

We omit the details of this cumbersome calculation and give only the result:
\begin{equation}
\Delta_{\alpha} = (l-4) + \frac{\eps\, x_{\alpha}}
{(d^2-3)(d+2l-6)(d+2l-4)(d+2l-2)} +O(\eps^{2}),
\label{mnogo}
\end{equation}
where
\begin{eqnarray}
x_{1}=d^5+6d^4l+13d^3l^2+10d^2l^3+dl^4-12d^4-53d^3l-60d^2l^2-6dl^3+2l^4+47d^3
\nonumber\\
+92d^2l-29dl^2-44l^3-24d^2 +158dl+214l^2-156d-364l+192,
\nonumber
\end{eqnarray}
\begin{eqnarray}
x_{2}=(d+2l-4)(d+2l-2)(d^3+2d^2l+dl^2-6d^2-5dl+2l^2+3d-16l+30),
\nonumber
\end{eqnarray}
\begin{eqnarray}
x_{3}=(d+2l-2)(d^4+4d^3 l+5d^2l^2+dl^3 -10d^3-25d^2l-5dl^2
+2l^3+27d^2-6dl-26 l^2+30d+92l-96).
\nonumber
\end{eqnarray}
For $l=4$, the results (\ref{tri}) are recovered. Like in the case $l=4$,
the basis operator (\ref{2.5}) that possesses the dimension $\Delta_{1}$
is symmetric with respect to all possible permutations of the full
set $\{i_{1} \cdots i_{l}\}$; only this operator survives the
contraction with the coefficients $C_{F}$ in the operator product
expansion (\ref{OPE}) for the model (\ref{1})--(\ref{3}).
Therefore only $\Delta_{1}$ appears on the right-hand side of
Eq. (\ref{OPE}) and determines the leading exponent for the $l\,$th
anisotropic sector: $\gamma_{l}= 2-\eps+\Delta_{1}$. For all $l$ and $d$,
this recovers the result (\ref{corre3}) obtained in Sec. \ref{sec:PAIR}
on the basis of the Dyson--Wyld equations.

Like in the case $l=4$, the remaining dimensions $\Delta_{2,3}$ in
Eq. (\ref{mnogo}) are activated and appear on the right-hand side of
Eq. (\ref{OR}) when the anisotropy is introduced by the velocity field.

We recall that the general form of the exponent for a given $l\ge 2$ is
$\gamma_{l} = l-2 +2k +O(\eps)$; see Sec.~\ref{sec:Kartina}. From the
OPE viewpoints, $k=0$ corresponds to the  $l\,$th rank irreducible operator
built of two fields $\theta$ and $l-2$ derivatives, symmetric in all indices
(see above). The operators with $k =1,2,\dots$ can be obtained
in two ways: one can add $k$ Laplacians to the operator described above,
or one can add $(k-1)$ Laplacians, two derivatives with free indices, and
contract the indices of the $\theta$ fields. Therefore, for general $d$ the
leading exponent $\gamma_{l} = l-2 +O(\eps)$ is unique, while for all
$k =1,2,\dots$ there are {\it two} correction exponents of the form
$\gamma_{l} = l-2 +2k +O(\eps)$. In two dimensions, these two possibilities
coincide, see the discussion below Eq. (\ref{dim2}), and only one
exponent exists for any $k =0,1,2,\dots$.

For $l=0$, the general form of the exponent is $\gamma = 2k +O(\eps)$; the
first two solutions, $\gamma = 0$ and $\gamma = 2-\eps$, are known exactly
both from the RG and the Dyson--Wyld equation; see Secs.~\ref{sec:Iso}
and~\ref{sec:Operators2}. For any $k$ and $d$, the
solution is unique: for the scalar operators, the indices of the fields are
contracted, and the operator with $\gamma = 2k +O(\eps)$ necessarily reduces
to the unique form $\theta_{i} \Delta^{k} \theta_{i}$.

This picture is in a full agreement with the results obtained
in Sec.~\ref{sec:PAIR} for general $d$ and $l$ on the basis of
the exact Dyson--Wyld equation for the pair correlation function;
see also Ref. \cite{AP} for $d=3$ and $j\le10$.

\subsection{Higher-order structure functions in the higher-order
anisotropic sectors} \label {sec:hi-fi}

Let us briefly discuss the scaling behavior in the anisotropic sectors
for the higher-order structure functions. The RG and OPE analysis given
in Secs. \ref{sec:OPE} and \ref{sec:Ashells} can directly be extended
to the general case. It shows that the leading exponents $\gamma_{nl}$
in the $l\,$th sector of the $2n$-order structure function,
$S_{2n}\propto P_{l}(z)\, r^{\gamma_{nl}}$,
are determined by the $l\,$th rank tensor operators with $k\le 2n$ fields
$\theta$ and minimal possible number of derivatives; the operators
which contain the field $\theta$ without derivative, or reduce to total
derivatives, give no contribution to the expansion (\ref{OR})
and should be discarded.

The practical calculation of the critical dimensions of the operators with
large $l$ or $n$ is a difficult task, as one could see already on the example
with $l=0$ and $n=2$. However, in the zeroth order of the $\eps$ expansion,
some important information can be obtained without
calculation, just by the analysis of the form of relevant operators.

One can easily see that for $l\le 4n$, the leading exponents are determined
by the $l\,$th rank tensor operators of the form $(\partial\theta)^{2n}$,
with $l$ free and $4n-l$ contracted indices. In the $O(1)$ approximation,
the exponents themselves are equal to the number of derivatives entering
into the operators: $\gamma_{nl}=2n+O(\eps)$.

For $l>4n$, the relevant operators necessarily contain more derivatives than
fields, and for the leading exponents one obtains: $\gamma_{nl}=l-2n+O(\eps)$.

One can thus conclude that for higher $n$, the general picture remains the
same as for $S_{2}$: each anisotropic sector possesses its own set of scaling
exponents; the leading exponents obey hierarchy relations at least for
small $\eps$ and $l>4n$; they grow with $l$ without bound. Of course,
there can be several exponents for given $l$ and $n$; we recall that there
are six exponents of the form $\gamma_{20}=4+O(\eps)$ for $l=0$ and $n=2$
(see Sec. \ref{sec:Operators6}). In order to identify the unique
leading exponent within a family with the same zeroth-order value, or to
verify the hierarchy relations for $l\le 4n$, one should perform the
$O(\eps)$ calculation for the relevant families of operators. This
cumbersome task lies beyond the scope of the present paper and will
be discussed elsewhere.

\section{Conclusion} \label{sec:Con}

We have studied the inertial-range scaling behavior in a model of the passive
vector quantity advected by a self-similar white-in-time Gaussian velocity
field, with the large-scale anisotropy introduced by a random forcing.
In two respects, the model is closer to the real Navier-Stokes turbulence
than the famous scalar rapid-change model: nonlocality of the dynamics and
mixing of the composite operators that determine anomalous exponents.

The incompressibility condition for the advected field and the pressure
term in the diffusion-convection equation make the dynamics nonlocal.
This raises the question of realizability of the zero-mode solutions,
that is, convergence of the integrals in the equations for the correlation
functions on powerlike solutions, and consistency of nonlocality and
the existence of infinite families of scaling exponents \cite{AP,WL}. The
detailed analysis of the exact integral equation satisfied by the pair
correlation function has shown that the general picture of the inertial-range
scaling is essentially the same as in the scalar \cite{RG3} and magnetic
\cite{Lanotte2,Lanotte,Arad} variants of the rapid-change model. Namely,
each anisotropic sector is described by an infinite set of scaling exponents,
with the spectrum unbounded from above. The leading exponents in each sector
are organized in the hierarchical order according to their degree of
anisotropy, with the main contribution coming from the isotropic sector in
agreement with the hypothesis on the restored local isotropy of the
fully developed turbulence \cite{Legacy}. The leading exponents themselves
grow without bound with the degree of anisotropy, in disagreement with the
idea of the window of locality \cite{WL}.

The integral operator entering into the
equation for the pair correlation function in the momentum space converges
on the powerlike solutions with the {\it leading} exponents in the $l=0$
and $l=2$ sectors, but formally diverges on powerlike solutions with
{\it subleading} exponents and {\it leading} exponents in the sectors
with $l\ge4$. However, correct analysis of convergence here requires the
knowledge of the behavior of the full solution beyond the inertial range,
where it no longer reduces to a sum of power terms. It turns out, that
natural assumptions about the form of the solution allow one to
perform certain subtractions in the integrals which make them convergent.
One can use the formal rules of analytical regularization and simultaneously
omit the subtracted terms to obtain correct answers for the convergent
integrals with proper subtractions. Moreover, the realizability of these
solutions is also guaranteed by the RG approach, where they are identified
with the contributions of certain composite operators in the corresponding
operator product expansions. Therefore, such exponents indeed appear in
the full solution in the inertial range.

These conclusions are in agreement with the recent analysis performed
in Ref.~\cite{AP} for the model (\ref{1})--(\ref{3}) in the
three-dimensional coordinate space, although the analysis in the momentum
space appears rather different; see also Ref.~\cite{amodel} for the general
vector model. Furthermore, the RG and OPE techniques confirm this picture
and extend it to the higher-order correlation functions.

The second aim of the paper has been the analysis of the anomalous scaling
of the higher-order even structure functions $S_{2n}$. Owing to the
conservation of the ``energy'' $\theta^{2}(x)$, the second-order function
appears nonanomalous with the simple dimensional exponent:
$S_{2}\propto r^{2-\eps}$. The anomalous scaling reveals itself on the level
of the fourth-order structure function. In contrast with the scalar case,
where the leading anomalous exponents were identified with the critical
dimensions of individual composite operators in the corresponding OPE
\cite{RG}, the vector nature of the advected field in our model leads to
mixing of operators. In particular, the inertial-range behavior of the
function $S_{4}$ in $d$ dimensions is given by a set of six close exponents,
determined by eigenvalues of the matrix of critical dimensions for a set of
six operators. One of the dimensions is negative (``dangerous operator'')
and gives rise to anomalous scaling. The number of relevant operators
increases rapidly with the order of the function; they have been calculated
in a controlled approximation (small $\eps$ and large $d$) for the
higher-order functions up to $S_{12}$. The latter involves as many as
sixteen negative exponents, ten of them coming from the lower-order
functions, and a multitude of positive exponents which are small and
therefore close to the negative ones for small $\eps$. The probabilistic
inequalities prove that all the higher-order structure functions are also
anomalous, and the total number of dangerous operators in our model is
infinite, with the spectrum of dimensions unbounded from below.

Since the mixing of operators is a manifestation of the vector nature of
the advected field, there are a few general conclusions regarding the real
NS turbulence, which one can draw from the analysis of model
(\ref{1})--(\ref{3}).

It was demonstrated recently that a careful disentangling of contributions
from different anisotropic sectors is ubiquitous in the analysis of
experimental data on the real turbulence, because it allows one to properly
identify scaling exponents in the situations, where the standard
treatment reveals no scaling behavior at all;
see the discussion in Refs.~\cite{Arad99,Arad991,LP1,KS}
and references therein. The example of the model (\ref{1})--(\ref{3})
shows that even in the isotropic sector, or for the ideal isotropic
turbulence, correlation functions are represented by infinite sums of
powerlike terms, and the number of close terms grows rapidly with the order
of the correlation function. Although these corrections die out in the
formal limit $L\to\infty$, and a pure powerlike behavior with the leading
exponent sets in, in practice it may be obscured by such corrections: the
subleading exponents can be very close to the leading ones and much more
important than the leading terms from the higher anisotropic sectors. This
might result in imaginary nonuniversality of the inertial-range exponents or
deviations from a pure scaling behavior, which increase with the order of
the correlation function. Therefore, reliable analysis of the
inertial-range scaling necessarily requires some theory for the
correction exponents.

In theoretical models, anomalous scaling is usually explained by the
so-called intermittency phenomenon. Within the framework of numerous models,
the anomalous exponents are related to the statistics of the local
dissipation rate or to the dimensionality of fractal structures formed by
small-scale vortices in the dissipative range; the detailed review and
bibliography can be found in Ref.~\cite{Legacy}. As a rule, those theories
predict simple analytic formulas for the dependence of the anomalous
exponents on $n$, the order of the structure function. Although such
formulas can provide a very good fit for the experimental results, the
experience on the rapid-change models suggests that they cannot be
absolutely correct. Even for the scalar model, the $n$-dependence of the
anomalous exponents changes as the order of the $\eps$ increases
\cite{RG,RG1,RG2,AH,cube}. Moreover, for our vector problem
the $n$-dependence of the exponents can hardly be given by a single
explicit formula (except probably for the case $d=2$, where the model
can be mapped onto a nonlocal scalar problem): the relevant families of
operators are completely different for different $n$, so that each function
$S_{n}$ requires special analysis.

Recently, a systematic perturbation theory for the anomalous exponents
in the NS turbulence was proposed, where the role of a formal expansion
parameter is played by the anomalous exponent for $S_{2}$, assumed to be
small but nontrivial \cite{LP}. In contrast with the magnetic variant of
the rapid-change model \cite{V96,RK97} and the general ``${\cal A}$-model''
with a stretching term \cite{amodel}, the second-order structure function
in our vector problem (\ref{1})--(\ref{3}) is nonanomalous, and the
perturbation theory of Ref.~\cite{LP} would be impossible here. Like in the
scalar model \cite{Kraich1,Falk1,GK,VMF}, this is a consequence of the
energy conservation or, in the field-theoretic language,
of vanishing of the critical dimension of the local dissipation rate; see
Sec.~\ref{sec:Operators2} for the vector and Ref. \cite{RG} for the scalar
cases. The vanishing of the critical dimension of the dissipation rate at
the physical value of $\eps$ is also characteristic of the NS case
\cite{Pismak,CL}, which raises serious doubts about the existence of the
second-order anomaly and the possibility of the corresponding perturbation
theory in the NS turbulence.

The analysis of the inertial-range behavior essentially simplifies as
$d\to\infty$. Our model has no finite ``upper critical dimension,''
above which anomalous scaling vanishes (see Ref. \cite{Nelkin} for a
recent discussion of that concept). Like in the scalar case \cite{Infty}
and, probably, in the NS turbulence \cite{FFR,AR}, the anomalous scaling
disappears at $d=\infty$, but it reveals itself already in the $O(1/d)$
approximation. The anomalous exponents can be calculated within the
double expansion in $\eps$ and $1/d$. Along with the results \cite{Falk1}
for the scalar rapid-change model, this confirms the importance of the
large-$d$ expansion for the issue of anomalous scaling in fully developed
turbulence.

Although our analysis has been confined with the linear problem
(\ref{1})--(\ref{3}),
which has only restricted resemblance with the real fluid
turbulence, some of the results can be extended to the case of the vector
passive field advected by the NS field, or the nonlinear NS equation itself
with some classes of random forcing. These questions lie beyond the scope
of the present paper. Detailed exposition of the RG approach to the NS
problem and the bibliography can be found in Refs. \cite{UFN,turbo};
the renormalization of composite operators and the concept of the operator
product expansion are also discussed in
Refs. \cite{JETP,Eight,AR,Pismak,CL,Kim,Triple}. In particular, critical
dimensions of tensor composite operators in the stirred NS problem were
calculated in Ref.~\cite{Triple} (see also Sec. 2.3 of \cite{turbo}); they
demonstrate the same hierarchy as their counterparts from
Sec.~\ref{sec:Ashells}.

We believe that the framework of the renormalization group and operator
product expansion, the concept of dangerous composite operators, exact
functional equations, and the $\eps$ and $1/d$ expansions will become the
necessary elements of the appearing theory of the anomalous scaling in
fully developed turbulence.

\acknowledgments
The authors are thankful to M.~Hnatich, J.~Honkonen, A.~Kupiainen, A.~Mazzino,
P.~Muratore Ginanneschi, S.~V.~Novikov and A.~N.~Vasil'ev for discussions.
The work was supported by the Russian Foundation for Fundamental Research
(Grant No.~99-02-16783) and the Grant Center for Natural Sciences of the
Russian State Committee for Higher Education (Grant No.~E00-3-24).

\appendix

\section{Equations for the exponents in the higher anisotropic sectors
of the pair correlation function}
\label{sec:higher_sectors}

It was shown in Sec.~\ref{sec:Kartina} that the equation for the exponents
$\gamma_{l}$ in $l\,$th anisotropic sector can be written as
$\det |C_{\alpha\beta}^{(l)}|=0$. For $l=2$, the matrix elements
$C_{\alpha\beta}^{(l)}$ were given in Eq. (\ref{coefficients}).
In general, the matrix $C_{\alpha\beta}^{(l)}$ is symmetric and its
elements are finite linear combinations of the integrals
$\widetilde{I}_{1}\equiv I_{1} +2J$ with $J$ from (\ref{J}) and
$I_{n}\equiv I_{n}(\gamma_l-2,\eps+2)$ from (\ref{convolu}),
with $n$ as high as $l/2+2$. In this Appendix, we present the
coefficients $C_{\alpha\beta}^{(l)}$ for higher values of $l$ up to $l=12$.
Then equations for $\gamma_{l}$ can be written in a straightforward way.
Below we denote $d_{k,s} = (d+k)(d+k+2) \dots
(d+s-2)(d+s)$ and $d_{k} = (d+k)$.

\begin{eqnarray}
C_{11}^{(4)}&=& d_{2,4} I_{4} - (d+2) (d^{2}+5d-2) I_{3} +
(d^{2}-1) (2d+5) I_{2} -(d-1)^{2}(d+1) \I,
\nonumber \\
C_{12}^{(4)}&=& -d_{2,4} I_{4} + 2d_{2,4}I_{3} - (d+1)(2d+5) I_{2}
+ (d^{2}-1) \I,
\nonumber  \\
C_{22}^{(4)}&=& d_{2,4} I_{4} -\frac{1}{2} d_{2,4} (d+2) I_{3}
+ \frac{(d+1)}{12}\, (d^3+10d^2+24d+12) I_{2}
- \frac{d}{12} \, (d^2-1)(d+4) \I,
\nonumber  \\
C_{11}^{(6)}&=& -I_{5} + \frac{(d^{2}+10d+1)} {d_{8}} I_{4}  - 3
\frac{d_{3}(d^{2}+6d-5)} {d_{6,8}} I_{3} +
\frac{d_{1,3}(d-1)(3d+13)}{d_{4,8}}I_{2} -
\frac{d_{1,3}(d-1)^{2}}{d_{4,8}} \I,
\nonumber \\
C_{12}^{(6)}&=& I_{5} - \frac{(3d+19)}{d_{8}} I_{4} +
\frac{d_{3}(4d+25)}{d_{6,8}} I_{3} -
\frac{d_{1,3}(3d+13)}{d_{4,8}} I_{2}  +
\frac{d_{1,3}(d-1)}{d_{4,8}} \I,
\nonumber \\
C_{22}^{(6)}&=& -I_{5} + \frac{(d^{2}+14d+43)} {3d_{8}} I_{4} +
\frac{d_{3}(d^{3}+30d^{2}+216d+430)} {30d_{6,8}} I_{3} +
\frac{d_{1,3}(d^{3}+15d^{2}+66d+85)} {15 d_{4,8}} I_{2} -
\nonumber \\
&-& \frac{d_{1,3}(d-1)(d^{2}+8d+10)} {30 d_{4,8}}\I,
\nonumber \\
C_{11}^{(8)}&=& I_{6} - \frac{(d^{2}+15d+8)}{d_{12}} I_{5}+ 2
\frac{d_{5}(2d-1)(d+11)}{d_{10,12}}I_{4}  -2 \frac
{d_{3,5}(3d^{2}+23d-22)}{d_{8,12}}I_{3} +
\frac {d_{1,5}(d-1) (4d+25)}{d_{6,12}} I_{2} -
\nonumber \\
&-&\frac{d_{1,5}(d-1)^{2}} {d_{6,12}}\I,
\nonumber \\
C_{12}^{(8)}&=& -I_{6} + 2\frac {(2d+17)}{d_{12}} I_{5}-
\frac{d_{5}(7d+64)}{d_{10,12}} I_{4} +
\frac{d_{3,5}(7d+58)}{d_{8,12}} I_{3} -
\frac {d_{1,5}(4d+25)}{d_{6,12}} I_{2}+
\frac{d_{1,5}(d-1)} {d_{6,12}}\I,
\nonumber \\
C_{22}^{(8)}&=& I_{6} - \frac {(d^{2}+24d+116)}{4d_{12}} I_{5}+
\frac{d_{5}(d^{3}+58d^{2}+748d+2632)}{56d_{10,12}} I_{4} -
\frac{d_{3,5}(3d^{3}+90d^{2}+788d+2072)}{56d_{8,12}} I_{3} +
\nonumber \\
&+& \frac {d_{1,5}(3d^{2}+50d+196)}{56d_{6,12}} I_{2}-
\frac{d_{1,5}(d-1)(d^{2}+12d+28)} {56d_{6,12}}\I,
\nonumber \\
C_{11}^{(10)}&=&- I_{7} + \frac{(d+1)(d+19)} {d_{16}} I_{6} -
5\frac{d_{7}(d^{2}+15d-4)}{d_{14,16}}I_{5}+ 10
\frac {d_{5,7}(d^{2}+12d-9)}{d_{12,16}} I_{4} -5 \frac
{d_{3,7}(2d^{2}+19d-19)}{d_{10,16}} I_{3}+
\nonumber \\ &+&
\frac{d_{1,7}(d-1)(5d+41)}{d_{8,16}} I_{2} -
\frac{d_{1,7}(d-1)^{2}}{d_{8,16}} \I,
\nonumber \\
C_{12}^{(10)}&=& I_{7} - \frac{(5d+53)} {d_{16}} I_{6} +
\frac{d_{7}(11d+128)}{d_{14,16}}I_{5}- 2\frac{d_{5,7}(7d+81)}{d_{12,16}}I_{4}
+\frac{d_{3,7}(11d+113)}{d_{10,16}}I_{3}-\frac{d_{1,7}(5d+41)}{d_{8,16}}I_{2}+
\nonumber \\ &+& \frac{d_{1,7}(d-1)}{d_{8,16}} \I,
\nonumber \\
C_{22}^{(10)}&=&-I_{7}+\frac{(d^{2}+36d+239)}{5d_{16}}I_{6}-
\frac{d_{7}(d^{3}+94d^{2}+1842d+9432)}{90d_{14,16}}I_{5} +2
\frac{d_{5,7}(d^{3}+49d^{2}+672d+2727)}{45d_{12,16}}I_{4} -
\nonumber \\ &-&
\frac{d_{3,7}(d^{3}+34d^{2}+357d+1167)}{15d_{10,16}}I_{3} +
\frac{d_{1,7}(2d^{3}+53d^{2}+444d+1179)}{45d_{8,16}}I_{2} -
\frac{d_{1,7} (d-1)(d^{2}+16d+54)}{90d_{8,16}}  \I,
\nonumber \\
C_{11}^{(12)}&=&I_{8} - \frac{(d^{2}+25d+34)}{d_{20}}I_{7} +
3\frac {d_{9}(2d^{2}+39d-1)}{d_{18,20}} I_{6} -
5\frac {d_{7,9}(3d^{2}+49d-28)}{d_{16,20}} I_{5} +
5\frac {d_{5,9}(4d^{2}+55d-47)}{d_{14,20}} I_{4} -
\nonumber \\ &-&
3\frac {d_{3,9}(5d^{2}+57d-58)}{d_{12,20}} I_{3} +
\frac {d_{1,9}(d-1)(6d+61)}{d_{10,20}} I_{2}-
\frac {d_{1,9}(d-1)^{2}}{d_{10,20}}\I,
\nonumber \\
C_{12}^{(12)}&=& -I_{8} +2 \frac{ (3d+38)}{d_{20}}I_{7}-
\frac{d_{9}(16d+223)}{d_{18,20}}I_{6} + 5
\frac{d_{7,9}(5d+72)}{d_{16,20}}I_{5} - 5
\frac{d_{5,9}(5d+69)}{d_{14,20}}I_{4} +4
\frac{d_{3,9}(4d+49)}{d_{12,20}}I_{3} -
\nonumber \\ &-&
\frac{d_{1,9}(6d+61)}{d_{10,20}}I_{2} +
\frac{d_{1,9}(d-1)}{d_{10,20}} \I,
\nonumber \\
C_{22}^{(12)}&=& I_{8} - \frac{ (d^{2}+50d+424)} {6d_{20}}I_{7}+
\frac{d_{9}(d^{3}+138d^{2}+3768d+25564)}{132d_{18,20}}I_{6}-5
\frac{d_{7,9}(d^{3}+72d^{2}+1392d+7744)}{132d_{16,20}}I_{5}+
\nonumber \\  &+&
5 \frac{d_{5,9}(d^{3}+50d^{2}+754d+3498)}{66d_{14,20}}I_{4}-
\frac{d_{3,9}(5d^{3}+195d^{2}+2406d+9416)}{66d_{12,20}}I_{3}+
\nonumber \\  &+&
\frac{d_{1,9}(5d^{3}+162d^{2}+1680d+5588)}{132d_{10,20}}I_{2}-
\frac{d_{1,9}(d-1)(d^{2}+20d+88)}{132d_{10,20}} \I.
\nonumber
\end{eqnarray}

\section{Calculation of the critical dimensions of the operators
$(\partial\theta)^{2\lowercase{n}}$ for large \lowercase{$d$} }
\label{sec:largeD}

The critical dimensions of the scalar composite operators of the form
$(\partial\theta)^{2n}$ were presented and discussed in
Sec.~\ref{sec:Operators6} in the first nontrivial orders, $O(\eps)$
and $O(\eps/d)$, of the double expansion in $\eps$ and $1/d$;
see Eqs. (\ref{d}). Below we give the derivation of those results.

The UV divergent part of the diagram from Eq. (\ref{Gamma2}) with
a composite operator $(\partial\theta)^{2n}$ is given by the formulas
(\ref{Gamma2})--(\ref{GammaR}) for $n=2$; see Sec.~\ref{sec:Operators4}.
With obvious alterations, they apply to the case of general $n$ and can be
summarized as [the $O(g)$ approximation in the renormalized variables]:
\begin{equation}
\frac{g\,C_{d}\,(\mu/m)^{\eps}} {2\eps} \,
\frac {\partial^{2} F(a)}
{\partial a_{i_{1} i_{2}}\partial a_{i_{3} i_{4}}} \,
\partial_{j_{1}} \theta_{j_{2}} (x) \,
\partial_{j_{3}} \theta_{j_{4}} (x) \,
T_{i_{1}i_{2}i_{3}i_{4},j_{1}j_{2}j_{3}j_{4}},
\label{Appen1}
\end{equation}
where we have used the relations (\ref{Lambda}) and
(\ref{logint}) and denoted
\begin{equation}
T_{i_{1}i_{2}i_{3}i_{4},j_{1}j_{2}j_{3}j_{4}} \equiv \int d{\bf n}\,
n_{i_{1}}n_{i_{3}}\, P_{i_{2}j_{2}} ({\bf n}) \,
P_{i_{4}j_{4}} ({\bf n}) \, P_{j_{1}j_{3}} ({\bf n}).
\label{Appen2}
\end{equation}
The integration over the unit sphere in $d$-dimensional space
in Eq. (\ref{Appen2}) can be explicitly performed using the isotropy
relations (\ref{isotropy}):
\begin{eqnarray}
T_{i_{1}i_{2}i_{3}i_{4},j_{1}j_{2}j_{3}j_{4}} &=& \biggl[
\frac{1}{d}\, \delta_{i_{1}i_{3}} \, \delta_{i_{2}j_{2}} \,
\delta_{i_{4}j_{4}} \, \delta_{j_{1}j_{3}}  - \frac{1}{d_{0,2}}\,
\Bigl( \delta_{i_{2}j_{2}} \, \delta_{i_{4}j_{4}} \,
\delta_{i_{1}i_{3}j_{1}j_{3}} +
\delta_{j_{1}j_{3}} \, \delta_{i_{2}j_{2}} \,
\delta_{i_{1}i_{3}i_{4}j_{4}} +
\delta_{i_{4}j_{4}} \, \delta_{j_{1}j_{3}} \,
\delta_{i_{1}i_{2}i_{3}j_{2}} \Bigr) +
\nonumber \\
&+& \frac{1}{d_{0,4}}\,
\Bigl( \delta_{j_{1}j_{3}} \, \delta_{i_{1}i_{2}i_{3}i_{4}j_{2}j_{4}} +
\delta_{i_{2}j_{2}} \,
\delta_{i_{1}i_{3}i_{4}j_{1}j_{3}j_{4}} +
\delta_{i_{4}j_{4}} \,
\delta_{i_{1}i_{2}i_{3}j_{1}j_{2}j_{3}} \Bigr)  -
\frac{1}{d_{0,6}}\,
\delta_{i_{1}i_{2}i_{3}i_{4}j_{1}j_{2}j_{3}j_{4}} \biggr],
\label{Appen3}
\end{eqnarray}
where $d_{0,k}=d(d+2)\dots(d+k)$; cf. Appendix~A.

Although the coefficients of the tensor (\ref{Appen3}) behave as $O(1/d)$
for $d\to\infty$, contractions of their indices $i_{k}$ can compensate
the smallness (the indices $j_{k}$ are contracted with the factor
$\partial\theta\partial\theta$ in Eq. (\ref{Appen1}) and cannot be
contracted with each other). Such contractions occur when
the factor $\partial^{2} F(a) / \partial a\partial a$ contains the terms
of the form $\delta_{i_{1}i_{3}} \, \delta_{i_{2}i_{4}} $ or
$\delta_{i_{1}i_{4}} \, \delta_{i_{2}i_{3}} $  (the third term,
$\delta_{i_{1}i_{2}} \, \delta_{i_{3}i_{4}} $, is forbidden by the
transversality of the field $\theta$). In particular,
\begin{equation}
\frac {\partial^{2} \phi_{1}}
{\partial a_{i_{1} i_{2}}\partial a_{i_{3} i_{4}}} =
2 \delta_{i_{1}i_{3}} \, \delta_{i_{2}i_{4}}, \qquad \phi_{1} \equiv
\Phi^{ii}_{jj}.
\label{Appen4}
\end{equation}
Here and below, we use the notation
$ \Phi^{i_{1}\dots i_{k}}_{j_{1}\dots j_{k}} =
\partial_{j_{1}} \theta_{i_{1}} \dots \partial_{j_{k}} \theta_{i_{k}}$;
cf. Eq. (\ref{FF}) in Sec.~\ref{sec:Operators4}.

However, the both such contractions can appear simultaneously only in
the term with $d_{0,4}$ in Eq. (\ref{Appen3}) and give an $O(1/d)$
contribution. The leading $O(1)$ terms arise from the contraction of the
pair $i_{1}i_{3}$ in the first term of (\ref{Appen3}). The
needed term of the form $\delta_{i_{1}i_{3}}$ in the factor
$\partial^{2} F(a) / \partial a\partial a$ appears in every
differentiation of the block of $F$ that contains the contraction of
the indices of derivatives:
\begin{equation}
\frac {\partial^{2} \Phi^{bc}_{kk}}
{\partial a_{i_{1} i_{2}}\partial a_{i_{3} i_{4}}} = \delta_{i_{1}i_{3}} \,
(\delta_{i_{2}b}\delta_{i_{4}c} + \delta_{i_{2}c}\delta_{i_{4}b} ).
\label{Appen5}
\end{equation}
Substituting Eq. (\ref{Appen5}) into Eq. (\ref{Appen1}) gives the
contribution
\[\frac{g\,C_{d}\,(\mu/m)^{\eps}} {2\eps} \,
\partial_{j_{1}} \theta_{b} \partial_{j_{1}} \theta_{c} =
\frac{g\,C_{d}\,(\mu/m)^{\eps}} {2\eps} \, \Phi^{bc}_{j_{1}j_{1}},\]
that is, the block $\Phi^{bc}_{kk}$ is reproduced, and the counterterm
to the operator $F_{\alpha}$ in the order $O(1)$ is proportional to the
same monomial $F_{\alpha}$. The number of such contributions equals to
the number $\widetilde n_{\alpha}$ of the contractions between the
derivatives in the monomial $F_{\alpha}$, and we obtain:
\begin{mathletters}
\label{Appen6}
\begin{equation}
A_{\alpha\beta} = \frac{1}{2} \widetilde n_{\alpha} \delta_{\alpha\beta} +
O(1/d),
\label{Appen61}
\end{equation}
\begin{equation}
\gamma^{(0)}_{\alpha\beta}(g_{*}) =
- \widetilde n_{\alpha} \delta_{\alpha\beta} + O(1/d),
\label{Appen62}
\end{equation}
\begin{equation}
\Delta_{\alpha}^{(0)} = (n-\widetilde n_{\alpha}) \eps + O(1/d),
\label{Appen63}
\end{equation}
\end{mathletters}
where $0\le\widetilde n_{\alpha}\le n$ for operators from the family
$(\partial\theta)^{2n}$. The minimal possible value, $\Delta_{\alpha}=0$,
is reached for $\widetilde n_{\alpha}=n$, that is, for the operators
where all the derivatives are contracted only with each other.
In the family (\ref{FF}), one has $\widetilde n_{\alpha}=n=2$ for
$F_{2}$ and $F_{5}$; $\widetilde n_{\alpha}=1$ ($\Delta_{\alpha}=\eps$)
for $F_{3}$, $F_{4}$ and $F_{6}$; $\widetilde n_{\alpha}=0$
($\Delta_{\alpha}=2\eps$) for $F_{1}$.

Let us turn to the calculation of the $O(1/d)$ correction to the results
(\ref{Appen6}). We write
\begin{equation}
\Delta_{\alpha}=\Delta_{\alpha}^{(0)} + C_{\alpha} /d
\label{Det11}
\end{equation}
with $\Delta_{\alpha}^{(0)}$ from (\ref{Appen63}) and numerical coefficients
$C_{\alpha}$ determined by the relation
\begin{equation}
\det \Bigl[ \Delta - \Delta_{\alpha}^{(0)} - C_{\alpha} /d\Bigr] =0,
\label{Det}
\end{equation}
where $\Delta$ is the matrix (\ref{32B})
for the family $(\partial\theta)^{2n}$ with a given $n$.

Since in the $O(1)$ approximation the matrix $\Delta$ is diagonal with the
diagonal elements $\Delta_{\alpha}^{(0)}$, all the nondiagonal elements of
the matrix in Eq. (\ref{Det}) are of order $O(1/d)$, as well as its
diagonal elements that correspond to the (degenerate) eigenvalue
$\Delta_{\alpha}^{(0)}$. The diagonal elements that correspond to the
eigenvalues different from $\Delta_{\alpha}^{(0)}$ are of order $O(1)$.
It then follows that the determinant in Eq. (\ref{Det}) is of order
$O(1/d^{N})$, where $N$ is the degeneracy of the eigenvalue
$\Delta_{\alpha}^{(0)}$, and the vanishing of the full determinant in the
leading approximation is equivalent to the vanishing of its
$N\times N$ subdeterminant that corresponds to $\Delta_{\alpha}^{(0)}$.

This means that in the $O(1/d)$ approximation, the equations for the
coefficients $C_{\alpha}$ corresponding to different values of
$\Delta_{\alpha}^{(0)}$ or, equivalently, $\widetilde n_{\alpha}$,
are independent, and these coefficients can be sought separately.

It is clear from Eq. (\ref{Appen63}) that for small $1/d$, dangerous
operators ($\Delta_{\alpha}<0$) can be present only among the operators
with $\Delta_{\alpha}^{(0)}=0$, and below we confine ourselves with this
family. For such operators, $\widetilde n_{\alpha}=n$ (tensor indices of
the derivatives are contracted only with each other, and the same holds
for the indices of the fields), and they always can be represented in the
form
\begin{equation}
F = (\phi_{1})^{n_{1}}(\phi_{2})^{n_{2}} \dots (\phi_{q})^{n_{q}},
\label{Form}
\end{equation}
where $\sum_{k=1}^{q} kn_{k} =n $ and $\phi_{k}$ is the scalar operator
that contains $2k$ fields $\theta$ and cannot be represented as a product.
This operator necessarily reduces to the form [cf. Eq.
(\ref{Appen4}) for $k=1$]
\begin{equation}
\phi_{k} = \Phi^{l_{1}l_{1}l_{2}l_{2}l_{3}l_{3}\cdots l_{k}l_{k}}
_{s_{k}s_{1}s_{1}s_{2}s_{2}s_{3}\cdots s_{k-1}s_{k}} \equiv
[s_{k}s_{1},s_{1}s_{2},s_{2}s_{3},\dots, s_{k-1}s_{k}].
\label{Form2}
\end{equation}

Now let us collect all possible contributions of order $O(1/d)$.

(i) As already mentioned above, an $O(1/d)$ contribution appears from the
term with $d_{0,4}$ in Eq. (\ref{Appen3}), when the both derivatives in the
vertex (\ref{Vertex}) act on $\phi_{1}$. In (\ref{Appen1}), this gives the
contribution of the form
\[  \frac{gC_{d}} {2\eps d}\left(\frac{\mu}{m}\right)^{\eps} \,
\partial_{j_{1}} \theta_{j_{2}} \, \partial_{j_{1}} \theta_{j_{2}} =
\frac{g{C_{d}}}{2\eps d}\left(\frac{\mu}{m}\right)^{\eps} \, \phi_{1} , \]
that is, the operator $F$ in Eq. (\ref{Form}) reproduces itself, and the
number of such terms is equal to $n_{1}$. Therefore, the corresponding
contributions to the matrices $A_{\alpha\beta}$ and
$\gamma^{*}_{\alpha\beta}$ in Eq. (\ref{Appen6}) are diagonal and have
the forms
\begin{equation}
\delta_{1} A_{\alpha\beta} = \frac{n_{1}}{2d}\, \delta _{\alpha\beta},
\qquad \delta_{1} \gamma^{*}_{\alpha\beta} = - \frac{n_{1}}{d}\,
\delta _{\alpha\beta}.
\label{Form11}
\end{equation}

(ii) An $O(1/d)$ contribution appears from the term with $d_{0,2}$ in Eq.
(\ref{Appen3}), when the both derivatives in the vertex (\ref{Vertex})
act on any one of the $n$ factors $\Phi^{bc}_{kk}$ in $F$ and produce
the delta symbol $\delta_{i_{1}i_{3}}$. Each of these differentiations
gives into Eq. (\ref{Appen1}) the contribution
\[ - \frac{g{C_{d}}}{2\eps d}\left(\frac{\mu}{m}\right)^{\eps} \times 3
\partial_{j_{1}} \theta_{b} \, \partial_{j_{1}} \theta_{c}, \]
and the corresponding contributions into the functions (\ref{Appen6}) are
again diagonal:
\begin{equation}
\delta_{2} A_{\alpha\beta} =- \frac{3n}{2d}\, \delta _{\alpha\beta},
\qquad \delta_{1} \gamma^{*}_{\alpha\beta} =  \frac{3n}{d}\,
\delta _{\alpha\beta}.
\label{Form12}
\end{equation}

(iii) The contributions into Eq.~(\ref{Appen1}) from the first term of
Eq.~(\ref{Appen3}):
\begin{equation}
\frac{g{C_{d}}}{4\eps}\left(\frac{\mu}{m}\right)^{\eps} \,
\frac {\partial^{2} F(a)}
{\partial a_{i_{1} i_{2}}\partial a_{i_{1} i_{4}}} \,
\partial_{j_{1}} \theta_{i_{2}}  \,
\partial_{j_{1}} \theta_{i_{4}}.
\label{Form13}
\end{equation}
The operation $\partial_{j_{1}} \theta_{i_{2}} \,
\partial/\partial a_{i_{1} i_{2}}$
breaks the chain of contractions in $\phi_{k}$:
\begin{equation}
\partial_{j_{1}} \theta_{i_{2}} \, \partial \phi_{k}/
\partial a_{i_{1} i_{2}} = 2k \,
[j_{1}s_{1},s_{1}s_{2},s_{2}s_{3},\dots, s_{k-1}i_{1}]
\label{Form14}
\end{equation}
in the notation of (\ref{Form2}). In the following, the field with the
index $j_{1}$ is not differentiated, since it does not belong to the
operator $F(a)$ in the vertex (\ref{Vertex}).

The operation
$\partial_{j_{1}} \theta_{i_{4}} \,\partial/\partial  a_{i_{1}i_{4}}$
acts either onto
the factor (\ref{Form14}) or onto some other factor $\phi_{s}$ in the
operator (\ref{Form}). In the latter case, another broken chain of the
form (\ref{Form14}) appears; along with the first broken chain it gives
rise to the unbroken chain with $k+s$ elements, that is, $\phi_{k+s}$.
Therefore, this process gives rise to the counterterm
\begin{equation}
\frac{g{C_{d}}}{2\eps d}\left(\frac{\mu}{m}\right)^{\eps} \times
4ks \phi_{k+s}
\label{Form15}
\end{equation}
for any pair of factors $\phi_{k}$, $\phi_{s}$ in the operator (\ref{Form}),
and they determine nondiagonal contributions to the matrices (\ref{Appen6}).

The action of the operation
$\partial_{j_{1}} \theta_{i_{4}} \, \partial / \partial a_{i_{1} i_{4}}$
onto the chain (\ref{Form14}) produces a number of different terms. One
possibility is the breakdown of the chain of contractions of two kinds:
\begin{mathletters}
\label{susik}
\begin{equation}
[j_{1}s_{1},s_{1}s_{2},\dots, s_{p-1}j_{1}, i_{1}s_{p+1}, \dots,
s_{k-1}i_{1}]
\label{Form16}
\end{equation}
\begin{equation}
[j_{1}s_{1},s_{1}s_{2}, \dots, s_{p-1}i_{1},j_{1}s_{p+1},
\dots, s_{k-1}i_{1}].
\label{Form17}
\end{equation}
\end{mathletters}
The first variant is obviously $\phi_{p}\phi_{k-p}$, that is, the ``decay''
\begin{equation}
\phi_{k} \to \frac{gC_{d}}{2\eps d} \left(\frac{\mu}{m}\right)^{\eps} .
\sum_{p=1}^{k-1} \, k\, \phi_{p}\phi_{k-p}
\label{Form18}
\end{equation}
The second variant gives $(k-1)$ factors $\phi_{k}$, and they give the
diagonal contribution to the matrices (\ref{Appen6}):
\begin{equation}
\delta_{3} A_{\alpha\beta} = \frac{1}{2d}\, \sum_{k=2}^{q} n_{k} k(k-1)\,
\delta_{\alpha\beta}
\label{Form19}
\end{equation}
with $n_{k}$ and $q$ from Eq.~(\ref{Form}).

Finally, the differentiation of the rightmost factor in Eq. (\ref{Form14})
gives $d \phi_{k}$. This is the leading $O(1)$ contribution, but it also
gives the $O(1/d)$ term after the substitution \[g\to g_{*} =
\frac{2\eps} {C_{d}}  \left(1+\frac{2}{d}\right) + O(1/d^{2}).\]
The contribution to the matrices (\ref{Appen6}) is also diagonal:
\begin{equation}
\delta_{4} A_{\alpha\beta} = \frac{n}{d}\, \delta_{\alpha\beta}, \qquad
\delta_{4} \gamma^{*}_{\alpha\beta} = -\frac{2n}{d}\, \delta_{\alpha\beta}.
\label{Form20}
\end{equation}

Collecting the contributions (\ref{Form11}), (\ref{Form12}),
(\ref{Form19}) and (\ref{Form20}) gives
\begin{mathletters}
\label{susik2}
\begin{equation}
\delta A_{\alpha\beta} = \frac{1}{2d}\, \Bigl(n_{1}-n +\sum_{k=2}^{q} n_{k}
k(k-1) \Bigr)\, \delta _{\alpha\beta},
\label{Form21}
\end{equation}
\begin{equation}
\delta \gamma^{*}_{\alpha\beta} =  \frac{1}{d}\,
\Bigl(n-n_{1} -\sum_{k=2}^{q} n_{k} k(k-1) \Bigr)\, \delta _{\alpha\beta}
\label{Form22}
\end{equation}
\end{mathletters}
for the total diagonal $O(1/d)$ contribution to the matrices (\ref{Appen6}),
while their nontrivial nondiagonal elements are determined by Eqs.
(\ref{Form15}) and (\ref{Form18}), with the summation in the former over
all pairs of the factors $\phi_{k}\phi_{s}$ in the operator (\ref{Form}).

Consider a few examples that illustrate the above algorithm and lead to
the results announced in Sec.~\ref{sec:Operators6}.

For $n=2$, the family of operators with $\widetilde n_{\alpha}=n=2$
consists of two elements,
\[ F= \{ \phi_{1}^{2}, \quad \phi_{2} \}, \]
and the equation (\ref{Det}) for the coefficients $C_{\alpha}$
in the representation (\ref{Det11}) has the form
(here and below we omit the subscript $\alpha$ and  change the signs of the
matrix elements so that they all become positive):
\begin{equation}
\left| \matrix { C & 4 \cr  2 & C \cr } \right| =0 .
\label{Mat2}
\end{equation}
The solutions are $C= \pm 2\sqrt{2}$, as announced in Eq.~(\ref{d2}).

For $n=3$, the relevant family consists of three elements,
\[ F= \{ \phi_{1}^{3}, \quad \phi_{1}\phi_{2}, \quad \phi_{3}  \}, \]
the equation has the form
\begin{equation}
\left| \matrix { C & 12 & 0 \cr 2 & C & 8 \cr 0 & 6 & 3+C \cr  } \right| =0
\label{Mat3}
\end{equation}
with the solutions given in Eq.~(\ref{d3}).

For $n=4$, the relevant family consists of five elements,
\[ F= \{ \phi_{1}^{4}, \quad \phi_{1}^{2}\phi_{2}, \quad
\phi_{2}^{2}, \quad , \phi_{1}\phi_{3}, \quad \phi_{4}  \}, \]
the equation has the form
\begin{equation}
\left| \matrix { C & 24 & 0 & 0 & 0 \cr
                 2 & C & 4 &16 & 0 \cr
                 0 & 4 & C & 0 & 16 \cr
                 0 & 6 & 0 & 3+C & 12 \cr
                 0 & 0 & 4 & 8 & 8+C \cr } \right| =0
\label{Mat4}
\end{equation}
with the solutions given in Eq.~(\ref{d4}).

For $n=5$, the relevant family consists of seven elements,
\[ F= \{ \phi_{1}^{5}, \quad \phi_{1}^{3}\phi_{2}, \quad
\phi_{1}\phi_{2}^{2}, \quad  \phi_{1}^{2}\phi_{3}, \quad
\phi_{1}\phi_{4}, \quad  \phi_{2}\phi_{3}, \quad \phi_{5}  \}, \]
the equation has the form
\begin{equation}
\left| \matrix { C & 40 & 0 & 0 & 0 & 0 & 0 \cr
                 2 &  C & 12 & 24 & 0 & 0 & 0 \cr
                 0 &  4 & C & 0 & 16 & 16 & 0 \cr
                 0 &  6 & 0 & 3+C & 24 & 4 & 0 \cr
                 0 &  0 & 4 & 8 & 8+C & 0 & 16 \cr
                 0 &  0 & 6 & 2 & 0 & 3+C & 24 \cr
                 0 &  0 & 0 & 0 & 10 & 10 & 15+C \cr } \right| =0
\label{Mat5}
\end{equation}
with the solutions given in Eq.~(\ref{d5}).

For $n=6$, the relevant family consists of eleven elements,
\[ F= \{ \phi_{1}^{6}, \quad \phi_{1}^{4}\phi_{2}, \quad
\phi_{1}^{2}\phi_{2}^{2}, \quad   \phi_{2}^{3}, \quad
\phi_{1}^{3}\phi_{3}, \quad   \phi_{1}\phi_{2}\phi_{3}, \quad
\phi_{3}^{2}, \quad \phi_{1}^{2}\phi_{4}, \quad
\phi_{2}\phi_{4}, \quad  \phi_{1}\phi_{5}, \quad \phi_{6} \}, \]
the equation has the form
\begin{equation}
\left| \matrix { C & 60 & 0 & 0 & 0 & 0 & 0 & 0 & 0 & 0 & 0 \cr
                 2 &  C & 24 & 0 & 32 & 0 & 0 & 0 & 0 & 0 & 0 \cr
                 0 &  4 & C & 4 & 0 & 32 & 0 & 16 & 0 & 0 & 0  \cr
                 0 & 0 &  6 &  C & 0 & 0 & 0 & 0 & 48 & 0 & 0 \cr
                 0 &  6 & 0 & 0 & 3+C & 12 & 0  & 36 & 0 & 0 & 0  \cr
                 0 & 0 & 6 & 0 & 2 & 3+C & 8 & 0 & 12 & 24 & 0 \cr
                 0 & 0 & 0 & 0 & 0 & 12 & 6+C & 0 & 0 & 0 & 36 \cr
                 0 & 0 & 4 & 0 & 8 & 0 & 0 & 8+C & 4 & 32 & 0 \cr
                 0 & 0 & 0 & 4 & 0 & 8 & 0 & 2 & 8+C & 0 & 32 \cr
                 0 & 0 & 0 & 0 & 0 & 10 & 0 & 10 & 0 & 15+C & 20 \cr
        0 & 0 & 0 & 0 & 0 & 0 & 6 & 0 & 12 & 12 & 24+C \cr } \right| =0
\label{Mat6}
\end{equation}
with the solutions given in Eq.~(\ref{d6}).


\begin{table}
\caption{Canonical dimensions of the fields and parameters in the
model (\protect\ref{action}).}
\label{table1}
\begin{tabular}{cccccccc}
$F$ & $\btheta$ & $\btheta'$ & $ {\bf v} $ & $\nu$, $\nu _{0}$
& $m$, $\mu$, $\Lambda$ & $g_{0}$ & $g$ \\
\tableline
$d_{F}^{k}$ & 0 & $d$ & $-1$ & $-2$ & 1& $\eps $ & 0 \\
$d_{F}^{\omega}$ & $-1/2$ & $1/2$ & 1 & 1 & 0 & 0 & 0 \\
$d_{F}$ & $-1$ & $d+1$ & 1 & 0 & 1 & $\eps $ & 0 \\
\end{tabular}
\end{table}

\end{document}